# In pursuit of the Sun, from Jules Janssen to the present day

*Jean-Marie Malherbe, emeritus astronomer*

*Observatoire de Paris (OP), 24 January 2024*

**Abstract**

The Sun has been observed through a telescope for four centuries. However, its study made a prodigious leap at the end of the nineteenth century with the appearance of photography and spectroscopy, then at the beginning of the following century with the invention of the coronagraph and monochromatic filters, and finally in the second half of the twentieth century with the advent of space exploration (satellites, probes). This makes it possible to observe the radiations hidden by the Earth's atmosphere (Ultra Violet, X-rays, γ) and to carry out "in situ" measurements in the solar environment. This article retraces the major stages of this fantastic epic in which renowned scientists such as Janssen, Deslandres, d'Azambuja, Lyot and Dollfus entered the scene, giving the Paris-Meudon Observatory a pioneering role in the history of solar physics until 1960. After this golden age, space exploration required large resources shared between nations, which could no longer be implemented within teams or even individual institutes. The development of numerical simulation, a new research tool, also required the pooling of supercomputers.

**Introduction**

Section 1 of this article presents everything you need to know about the Sun, allowing you to understand the path of knowledge and associated technical progress. Section 2 is dedicated to the beginnings of solar observation, from 1610 to 1860, and section 3 describes the major advances of the period 1860-1960. During these 100 years, the Paris-Meudon Observatory played a major, innovative and world-renowned role, with the first photographs of the Sun, the introduction of spectroscopy, the invention of the coronagraph and monochromatic imaging, the birth of radio astronomy and the beginnings of balloon exploration. Section 4 summarizes the advances of the space era in solar astronomy after 1960, and section 5 mentions the appearance of a new complementary activity to observations, numerical simulation, which was able to develop thanks to the rapid evolution of supercomputers from 1970 onwards.

**1 - What is the Sun ?**

The Sun is an unremarkable star in our galaxy. It is a sphere of hot gas with a diameter 110 times that of the Earth, and a mass 300000 times that of our planet. The Sun is composed of 90% hydrogen, the rest being helium. The majority of Earth's elements, such as metals, exist in trace amounts in the Sun, but in a gaseous state due to the surface temperature of 6000 degrees. The internal structure of the Sun is opaque, so it is unobservable. The solar core, at high temperature and pressure, is a "nuclear reactor" that fuses hydrogen into helium, the corresponding mass loss (4 million tons/s) producing the power radiated by the star throughout space ($3.86 \; 10^{26}$ W). The Earth, outside the atmosphere, receives $1.75 \; 10^{17}$ W, i.e. a flux of 1365 W/m². The age of the Sun is estimated to be 5 billion years old, and it would be at mid-life. Only the surface, called the photosphere, and its atmosphere can be observed through a telescope. The photosphere, at 5800 degrees, reveals spots (Figure 1) which are regions of concentrated magnetic field (0.1 to 0.25 T). The chromosphere, just above, is partially ionized and more tenuous. It is only visible in spectroscopy, where dark filaments and bright areas are visible (Figure 1). Its temperature is 8000 degrees. The Sun is surrounded by a vast corona, a faint glory that appears only during total eclipses. It is made up of charged particles (electrons, protons, ions) that are constantly escaping from the Sun, making up the solar wind. It bathes the interplanetary medium and corresponds to a second mass loss of a few million tons/s. The corona is very hot (more than a million degrees), so totally ionized, but very tenuous. Electrons (very light particles) escape the solar gravity because their thermal agitation exceeds the release velocity (600 km/s) ; as they move outwards, they create an electric field that accelerates the protons (1860 times heavier) and together they form the solar wind that travels at 500 km/s (or more during active events).

The Sun is a variable star, in the sense that its radiative flux varies by one thousandth with a periodicity of 11 years. This variation in brightness is related to the cyclical appearance of magnetized spots and faculae. The covered area is small or non-existent during minima, but it can be more than one thousandth of the disk (Figure 1) during maxima. Active regions (spots and faculae) are often unstable at maxima, releasing excess magnetic energy that they store, in the form of violent eruptions and ejections of matter, which can spread into the interplanetary medium and cross the Earth. It is from the Sun-Earth interaction that the aurora borealis originate, which also exist on planets (Jupiter, Saturn, Uranus) with atmospheres and magnetic fields.

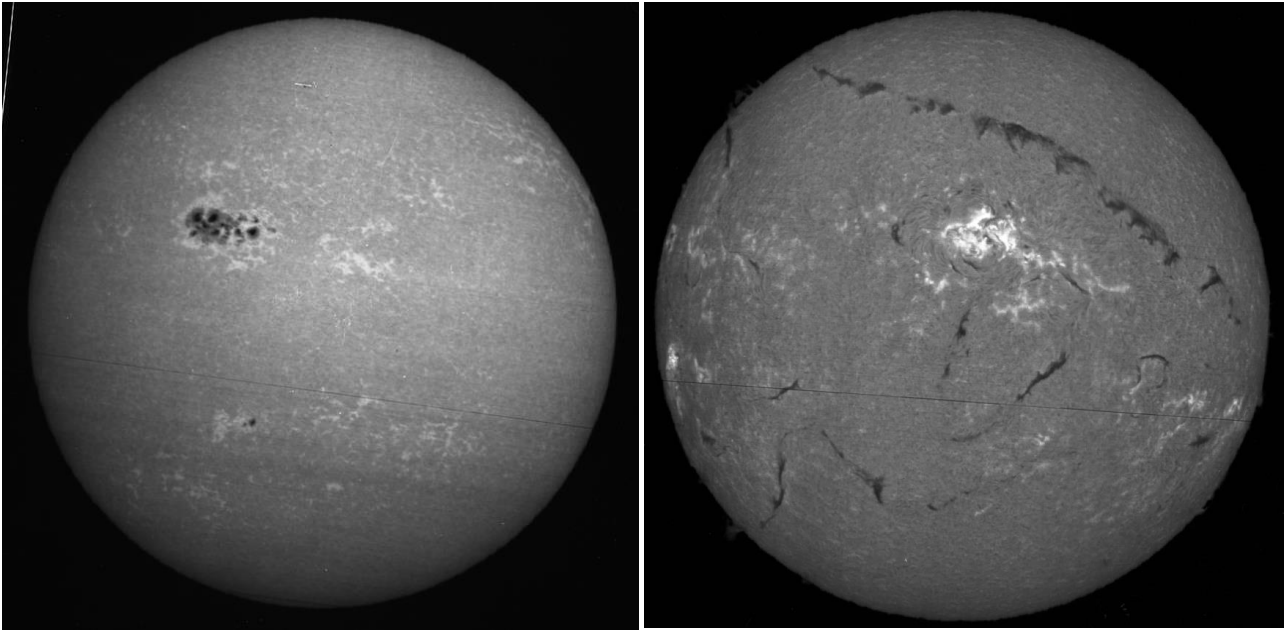

*Figure 1 : The surface of the Sun or photosphere with its spots (left), and the chromosphere, the layer just above the photosphere, visible only in spectroscopy, with its dark filaments and bright facular areas (right). All these structures are more or less magnetized. Courtesy OP.*

## 2 – The beginnings of solar observations with a telescope

The largest sunspots are at the limit of the resolving power of the human eye (about 1 arc minute). Some of them would have been seen for a millennium, especially on the horizon during sunrises and sunsets. It was the invention of Galileo's telescope in 1609 (figure 2) that made it possible to observe them with regularity, most often by projecting the Sun onto a screen through the eyepiece to avoid glare on the one hand, and to draw the outline of the dark spots on paper with a pencil on the other hand. At the beginning of the seventeenth century, the principal observers were, besides Galileo in Italy, Harriot in England, Scheiner and Fabricius in Germany. Scheiner recorded numerous surveys (Figure 2) in Rosa Ursina (1635). Hevelius then made more than 4000 observations over forty years from 1640 to 1680 in Poland, and is credited with numerous drawings (Figure 3). Very soon, the rotation of the spots (linked to the rotation of the Sun on its axis in 26 days) was noticed by the observers.

Astronomy was structured in France under the auspices of Louis XIV with the foundation of the Paris Observatory in 1667, a few years after that of the Greenwich Observatory in England, and that of the French Academy of Sciences. The first director was Jean Dominique Cassini, to whom we got the first solar surveys around 1675 (Figure 3). Then solar observation developed in Paris with measurements of the Sun's angular diameter (it fluctuates with the seasons) by Abbé Jean Picard and his pupil Philippe de la Hire. It then fell into disuse, in favour of celestial mechanics, until the end of the nineteenth century when it was revived and then transferred to Meudon by Henri Deslandres (see below).

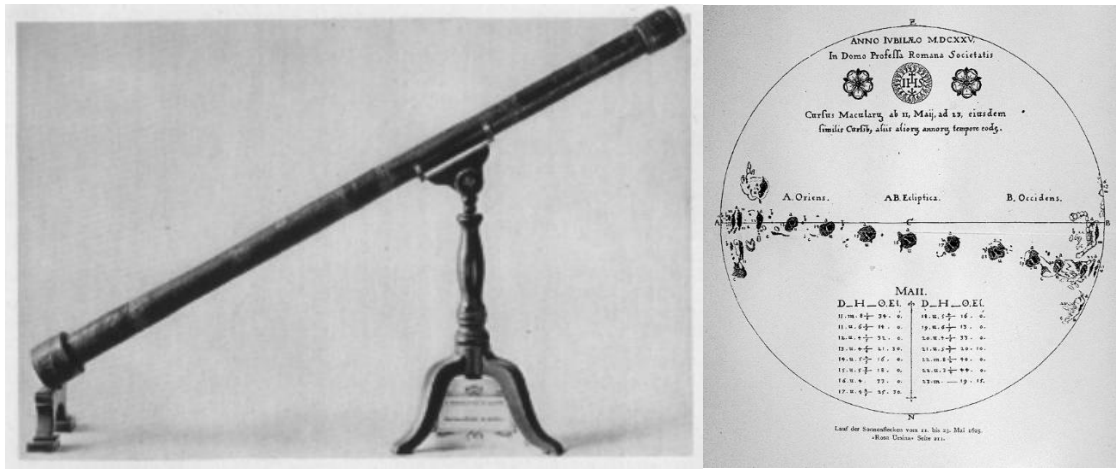

*Figure 2 : Galileo's telescope and Christoph Scheiner's first spot surveys in 1625 (Rosa Ursina)*

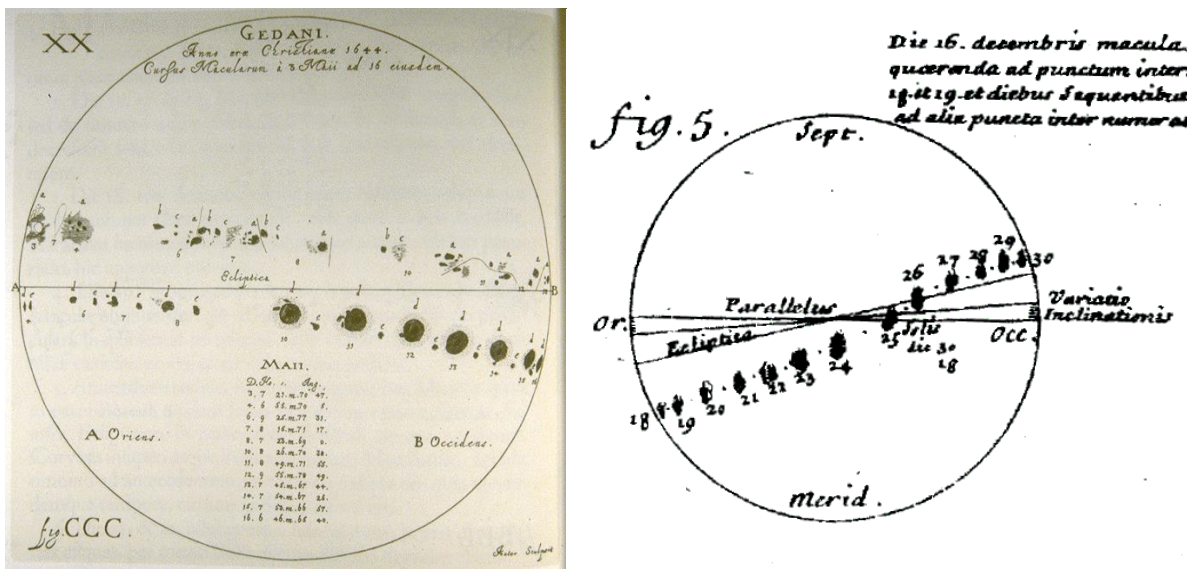

*Figure 3 : Drawings of the rotation of the spots by Hevelius (left) and Cassini (right)*

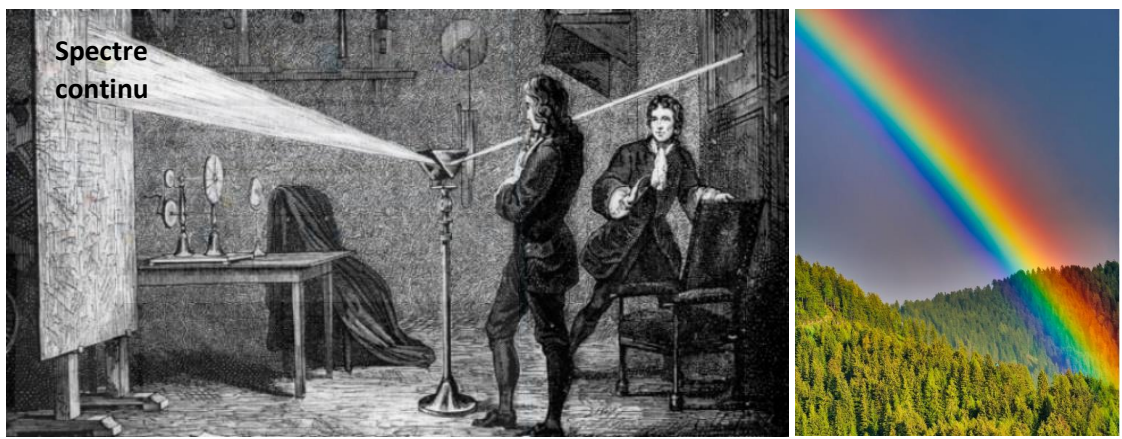

*Figure 4 : Newton studies the sunlight with a prism (1666) and discovers the continuous spectrum of the Sun; the dispersion of light by raindrops (the rainbow) has the same effect, because the refractive index of water and glass depends on the colour (or wavelength).*

Newton decomposed sunlight in 1666 using a prism and found that white light is the superposition of an infinite number of monochromatic radiations characterized by their wavelength (figure 4); this is what water droplets do when they form the rainbow, because the refractive index of glass and water depends on the colour (blue rays are more deflected than red ones by the dispersion phenomenon). This is the continuous spectrum of the Sun, quite close to that of a blackbody at the temperature of 5800 degrees, whose maximum emissivity is located in the yellow-green range. The superposition of the coloured radiations results in the white light. It was not until 1817 (figure 5) that dark lines were distinguished in the continuous spectrum: this is the line spectrum highlighted by Fraunhofer, where we can distinguish a few broad lines (the lines of Hydrogen, Calcium, Magnesium, Sodium) and thousands of very fine lines, which will only be resolved and identified much later. The lines of the spectrum are the signature of the chemical elements present in the solar atmosphere. They are dark because they absorb light from the surface. On the contrary, structures above the limb (such as prominences) give bright lines because they emit against the background of the black sky.

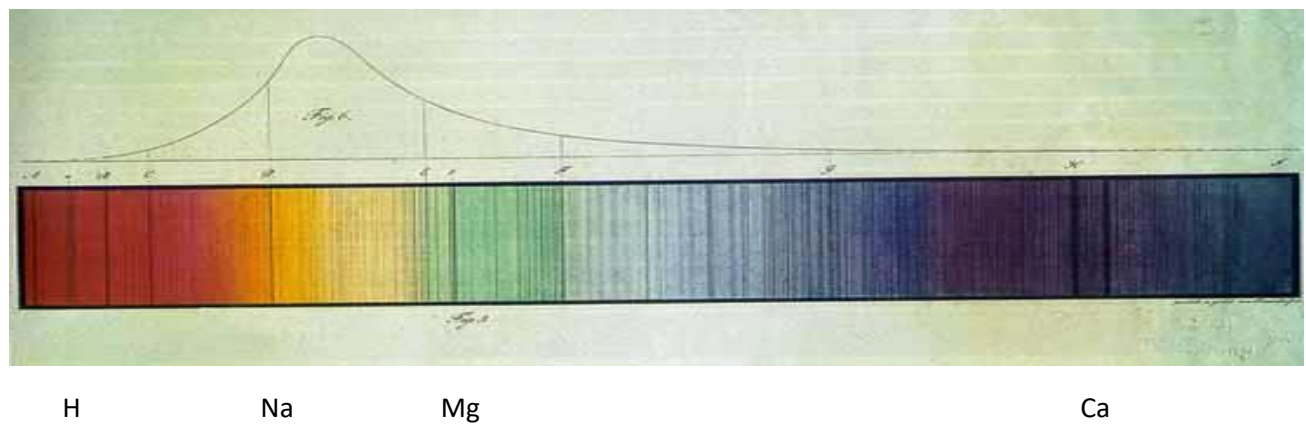

H            Na            Mg                                                            Ca

*Figure 5 : Fraunhofer's line spectrum (1817) with strong lines of the elements H, Na, Mg, Ca, etc…*

## 3 – Period 1860-1960: fast advances in solar physics with Meudon as a major actor

The observation of the Sun, from Galileo until the middle of the nineteenth century, consisted essentially of studying the movements of the spots. Their survey became very regular from 1750 onwards, which allowed the German astronomer Heinrich Schwabe to discover the existence of the 11-year solar cycle around 1850, by compiling previous data. However, the more sporadic surveys of the seventeenth century showed the existence of an abnormally prolonged minimum, known as the Maunder minimum (named after its discoverer) between 1650 and 1700; other deep minima, but less extended in time, were noticed by Gleissberg with a very approximate periodicity of around 100 years. The 11-year activity cycle is therefore modulated, there are successions of strong and weak cycles. The magnetic cycle lasts 22 years (reversal of the polarity of the poles every 11 years during the maxima), it was discovered by George Hale in the USA, at the beginning of the twentieth century, about ten years after he understood the magnetic nature of spots (1908). Hale recognized the Zeeman effect, evidenced earlier by the physicist Pieter Zeeman (this is a special effect which perturbs atomic energy levels when magnetic fields are present).

Until the middle of the nineteenth century, drawing was the only tool to preserve and archive solar observations. The first white light photograph was produced in 1845, it is attributed to Léon Foucault and Hippolyte Fizeau in Paris with the daguerreotype process (Figure 6). The birth of photography was a first revolution in the history of solar physics, because it introduced an objective way for the recording and conservation of observations, unlike drawings in which a personal and inevitable interpretation of the observer appears (for example, the pastels of Etienne Trouvelot, astronomer from Meudon, figure 7). At the same time, the Englishman Warren de la Rue took the first photograph of a solar eclipse (1860) in Spain (Figure 8), using his photoheliograph, a nice instrument specially designed for imaging the Sun in white light on a photographic plate (including the disk outside eclipses).

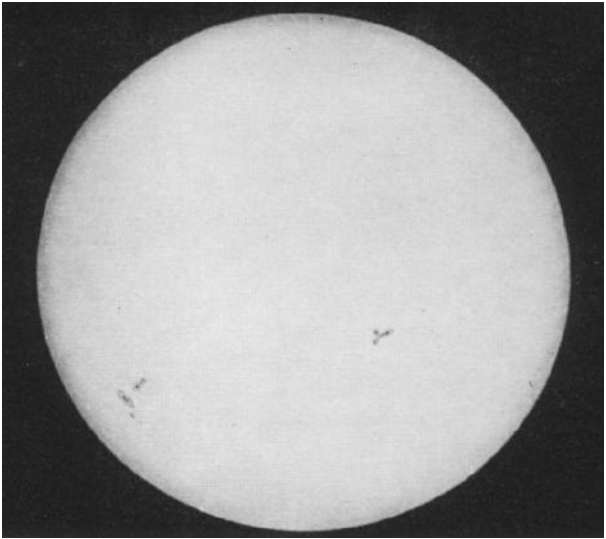

*Figure 6 : First photograph of the Sun (white light) by Foucault and Fizeau (1845), at the Paris Observatory (daguerreotype process).*

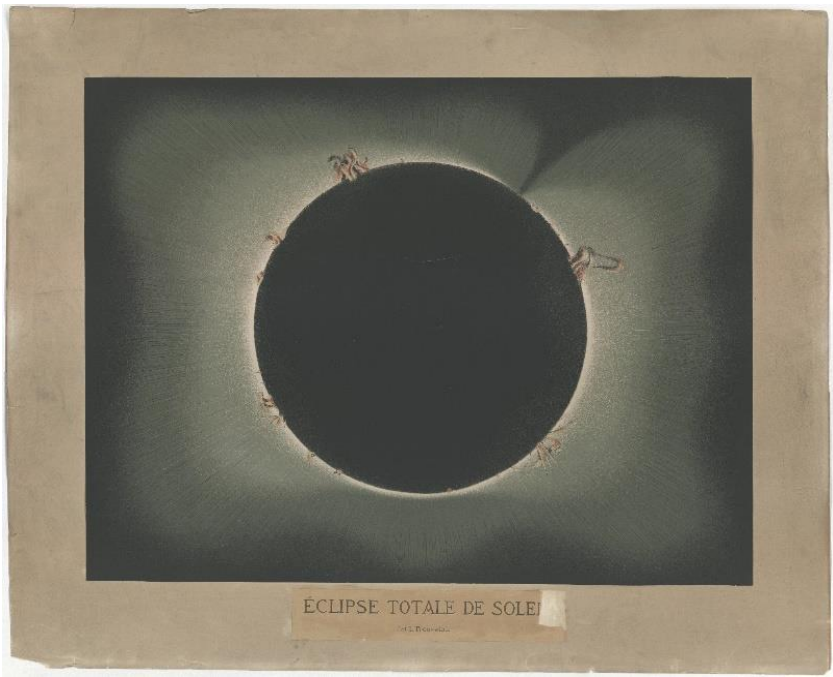

*Figure 7 : a solar eclipse pastel by Etienne Trouvelot (Meudon), probably 1860, showing the hot white corona (1 million degrees), scattering (through electrons) the light from the disk obscured by the Moon; and the prominences around the limb, which are colder (8000 degrees) and denser hydrogen structures, red coloured, emitting in the Hα line. The corona and prominences are invisible outside of eclipses. Credit: OP.*

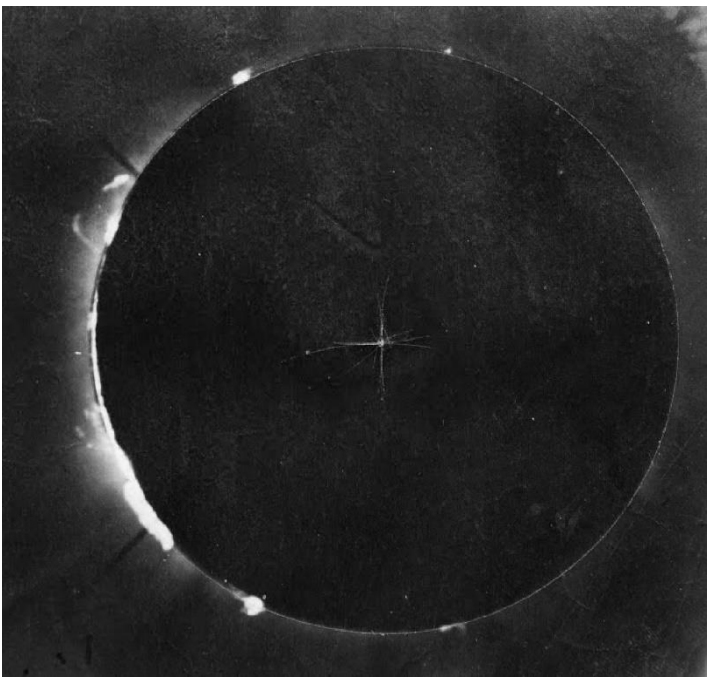

*Figure 8 : First photograph of a solar eclipse (1860) by the British Warren de la Rue, in Spain, with his photoheliograph, specially designed for photographing the Sun. We can see the prominences around the limb, which are Hydrogen structures whose natural colour is red, because they emit in the Hα line. They remain invisible outside of eclipses.*

We owe to Jules Janssen (figure 9) the introduction of physical astronomy in France, a new science now called astrophysics, and whose goal is the physico-chemical study of the stars and celestial objects (the constitution and physical nature). Indeed, the Paris Observatory was essentially devoted to celestial mechanics (the description and study of the movement of the stars and planets). Janssen, inspired by the work of early spectroscopists such as Kirschoff and Bunsen, applied the technique to astronomy, with the aim of isolating light from the lines of the solar spectrum, and thus studying the medium that forms them, i.e. the atmospheric layers that are inaccessible to observation in white light. Janssen thus created a second revolution during the eclipse of 18 August 1868 in India, at Guntoor, when he demonstrated, with Norman Lockyer, the possibility of observing prominences at any time, i.e. outside eclipses, thanks to spectroscopy.

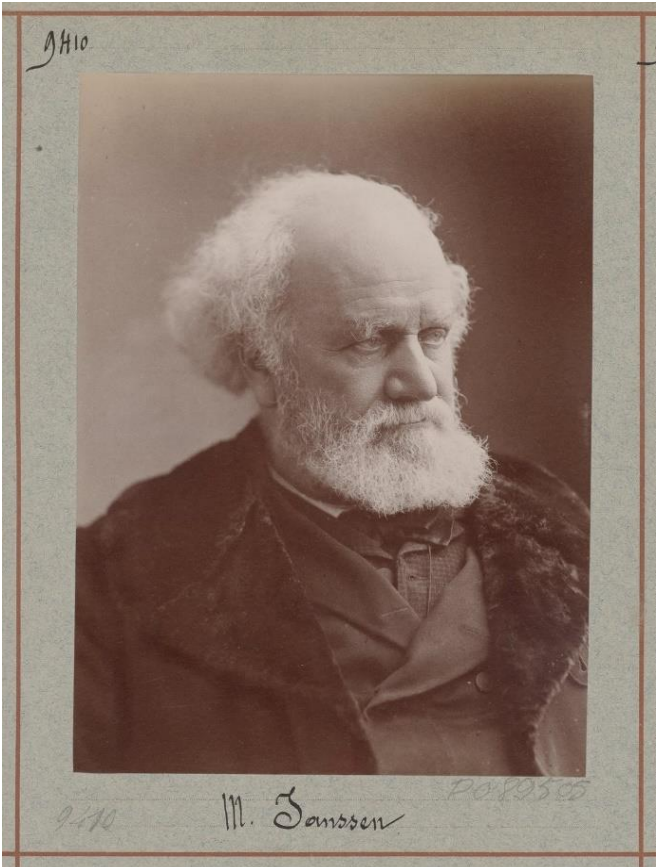

*Figure 9 : Jules Janssen (1824-1907), photographed by Nadar, credit Gallica/BNF. Janssen founded the Meudon Observatory in 1875. A true "adventurer" or "globetrotter" of celestial physics, as Françoise Launay explains in her book published in 2008 (Vuibert editor).*

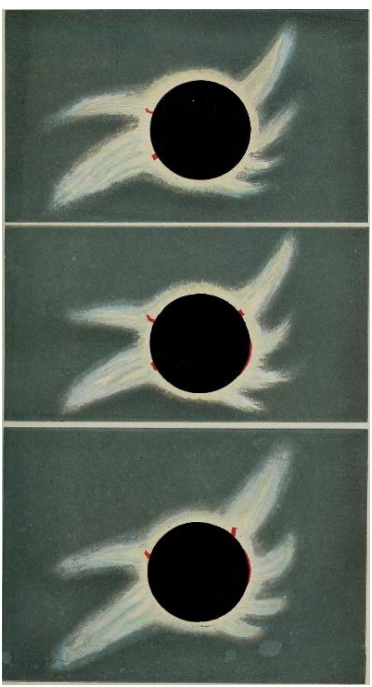
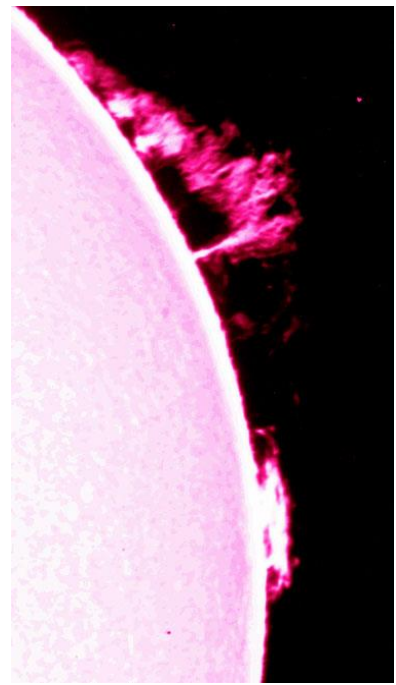

*Figure 10*

*On the left, drawings of the 1868 eclipse, showing the elongated white corona and red prominences of the solar limb. Corona and prominences are only visible to the eye during eclipses.*

*On the right, a protuberance seen in the light of the red Hα line of Hydrogen, outside eclipses, thanks to thin slit-scanning spectroscopy.*

*Courtesy OP.*

Thus, on the day of the eclipse, Janssen (Figure 10) visually spotted a prominence at the limb of the Sun. These hydrogen structures naturally appear red to the eye during a total eclipse because they are made of Hydrogen. The corona, which is 1000 times less dense, is also revealed, but by scattered white light from the disk (Thomson scattering by electrons). But outside of eclipses, we no longer see these structures, because their luminosity is far too faint compared to that of the disk (a million times less for the nearby corona). The day after the eclipse, Janssen pointed his spectroscope at the location spotted during the eclipse, and it appeared in the red line of the spectrum, known as the Hα of Hydrogen. Great discovery! Janssen has found a method for observing the structures of the solar atmosphere, that are hidden from vision outside of eclipses, using spectroscopy. This advance had a world wide impact, as it is at the origin of all the techniques for the study of spectral lines and monochromatic imaging allowing the sounding of atmospheres at various altitudes. Spectroscopy will therefore make it possible to do much more physics. Janssen was admitted to the French Academy of Sciences in 1873.

During this same eclipse, an unknown yellow line (Figure 11) appeared in the spectrum of the solar limb. Janssen noticed this but paid little attention to it, it was Lockyer who found a new element called Helium, still unknown on Earth; this was the second major discovery of spectroscopy !

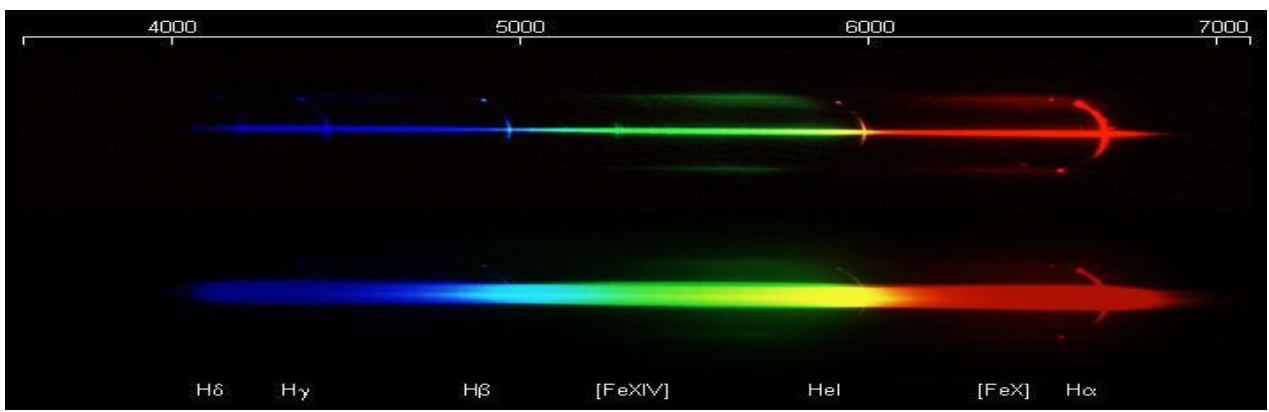

*Figure 11 : spectrum of the solar limb, where the coronal green line of the 13-times ionized iron (Fe XIV), the yellow line of neutral helium and several lines of hydrogen appear during eclipses (blue for Hβ, red for Hα). Credit: AstroSurf.*

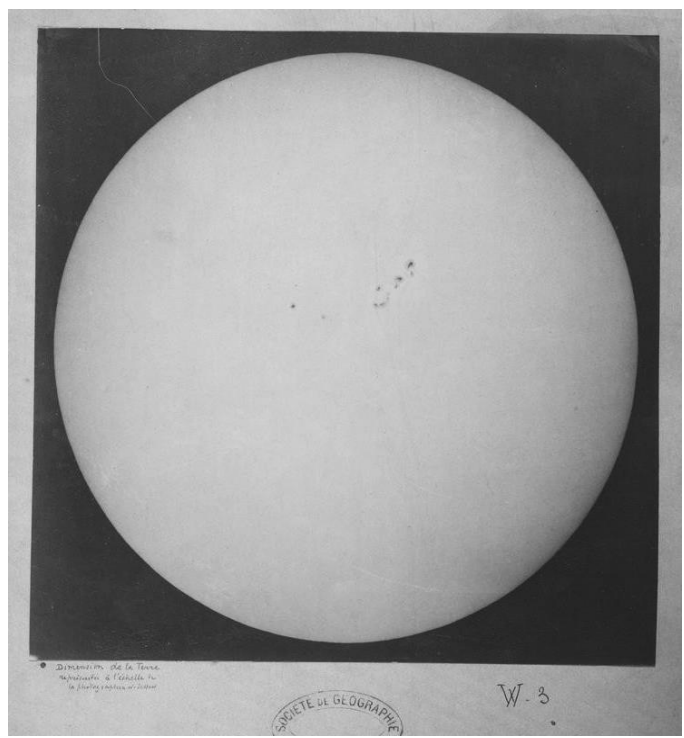

*Figure 12 : image taken by Janssen on July 20, 1874, well before the move to Meudon. Credit: Gallica/BNF.*

Janssen then developed a telescope for observing the Sun in white light (Figures 12 and 13) and designed a special apparatus (Figure 13, left) for observing the transit of Venus in front of the Sun, in December 1874, visible from Japan, in order to improve the calculation of the distance from the Sun to the Earth. It was the first chrono-photography camera, which can be considered as one of the ancestors of the cinematograph, that would come later. This device, called a photographic revolver, had a sensitive surface in the form of a rotating circular corona with 48 sectors, thus able to take 48 successive frames of a fast phenomenon such as the points of contact between Venus and the solar limb.

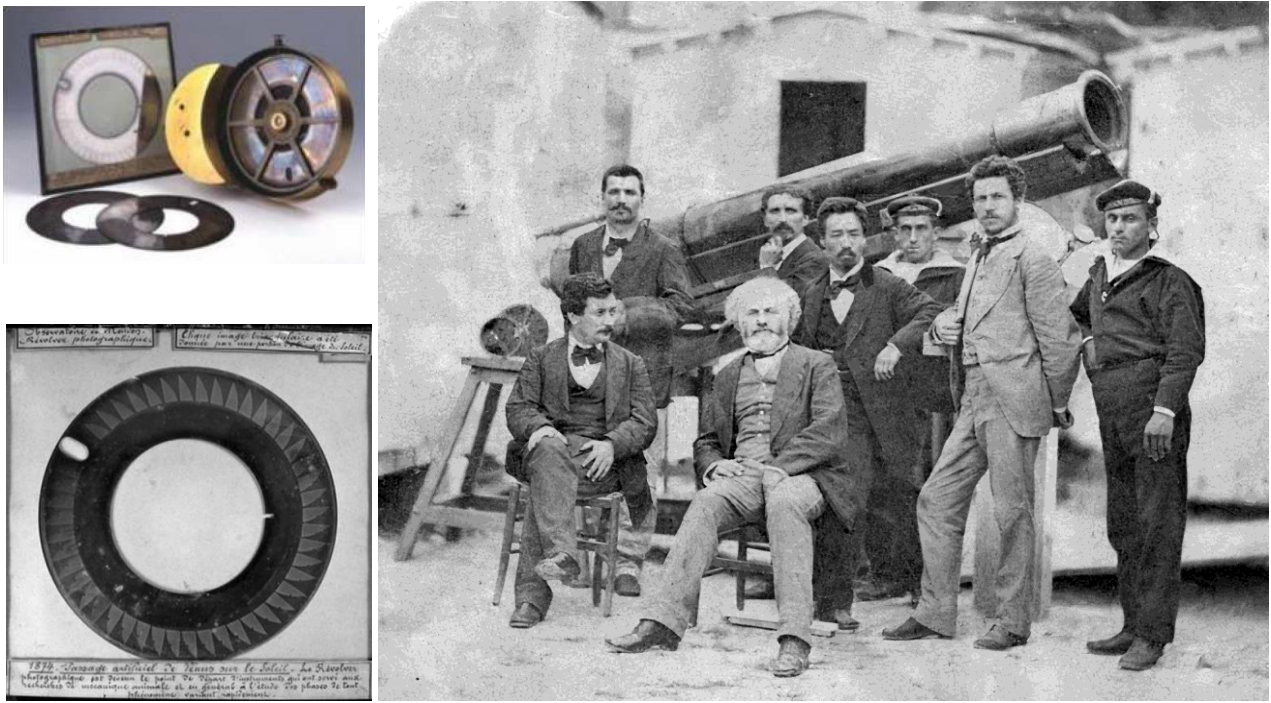

*Figure 13 : Janssen, the photographic revolver (credit OP) and the telescope for observing the transit of Venus, late 1874. It should be noted that the wooden support was recovered to carry a much better optical quality telescope after 1875 in Meudon.*

Happy with these great successes in spectroscopy and imaging, Janssen considered that the need for a new observatory entirely dedicated to physical astronomy was becoming imperative, and so he obtained from the government the allocation of the former royal domain of Meudon in 1875 (he had also thought of another possibility, the Malmaison). In this place overlooking Paris, in the countryside (the city has since devoured it), occupied by the army, Janssen found the new castle devastated in 1870 by fire (figure 14). It undertook its transformation into an observatory, a project that lasted 20 years. The wings were razed and the central part was preserved and surmounted by a large 18 metres diameter dome (Figure 15). Since 1893, it has housed the largest telescope in Europe, consisting of two superimposed telescopes of 67 cm and 83 cm (Figure 16).

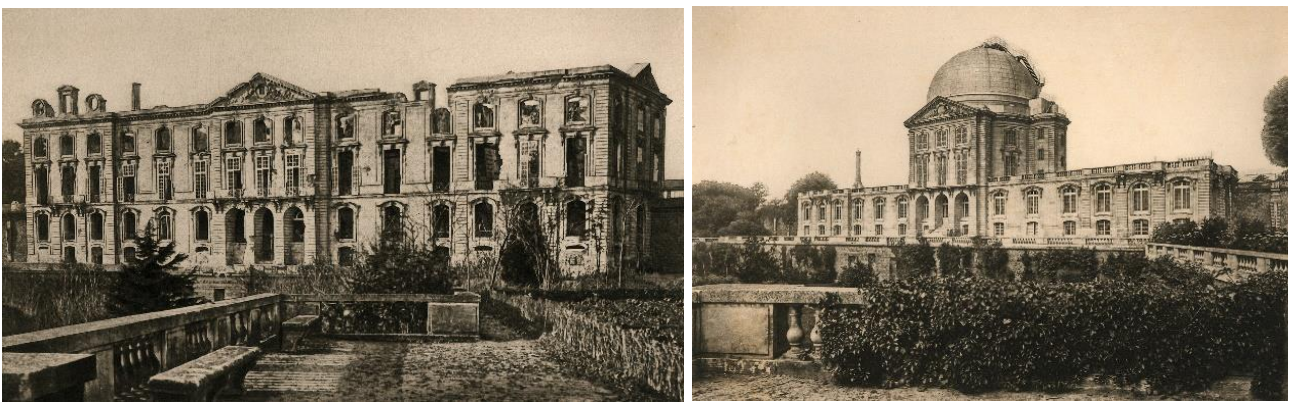

*Figure 14 : the devastated Meudon castle (left), then transformed into an observatory (right). Credit: OP.*

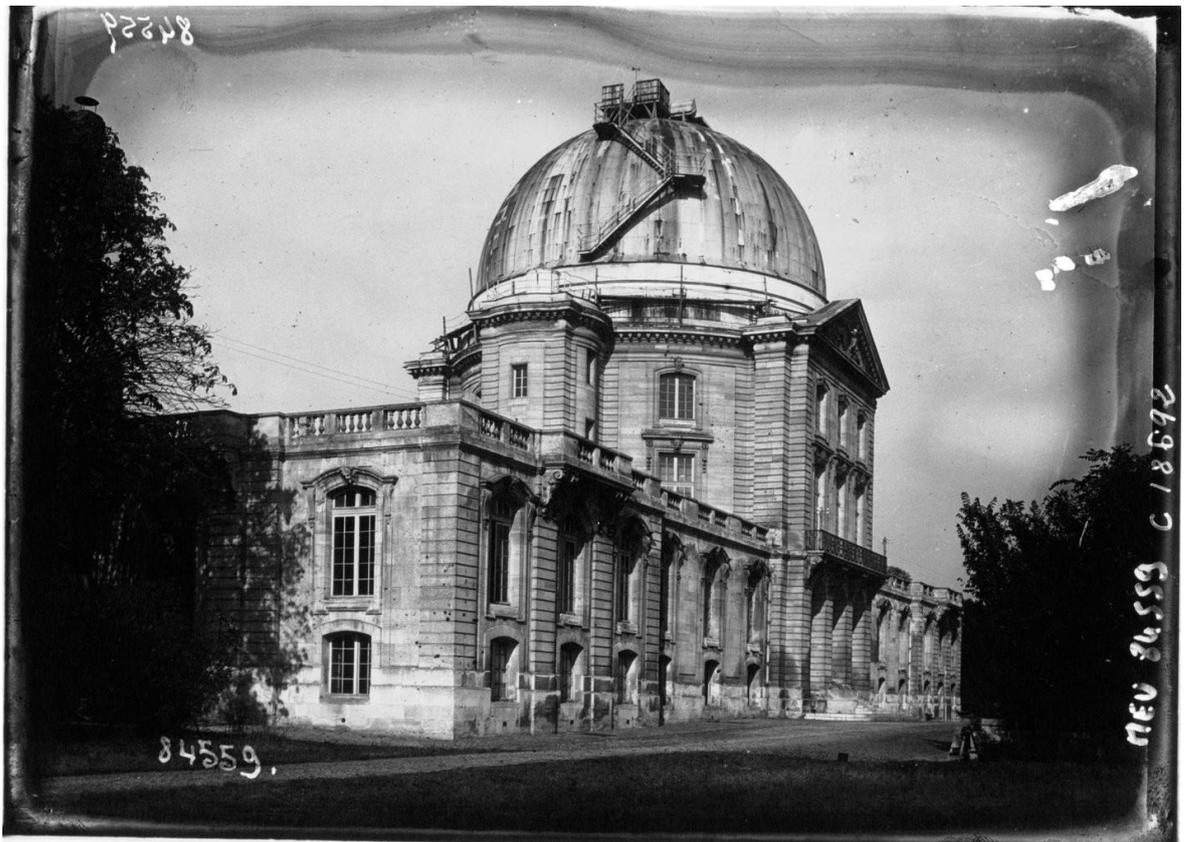

*Figure 15 : Great dome of Meudon in 1920 built on the ruins of the "Château neuf". Credit Gallica/BNF.*

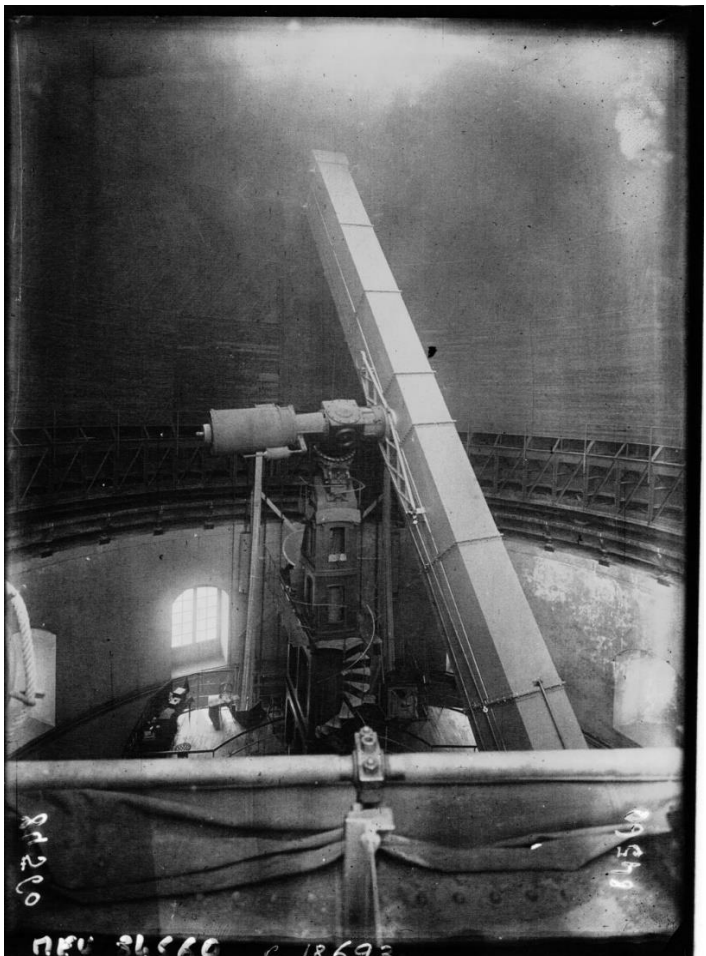

*Figure 16 : The large Meudon telescope in 1920. The instrument is a dual refractor, with a visual lens with a diameter of 83 cm and a photographic lens with a diameter of 67 cm. Visual optics are optimized for the sensitivity of the eye, while photographic optics are optimized for the sensitivity range of the plates of that epoch (blue). The focal length of the instrument is 16 m. A movable floor was subsequently installed. The two largest refractors are American, with a slightly larger diameter. They have not been built for a long time, because mirror telescopes have much more advantages. Agence Meurisse, credit Gallica/BNF.*

During the restoration work on the "Château neuf" and the construction of the great dome, Janssen was busy with the installation of a telescope specially dedicated to photographing the solar surface in white light (Figure 17), with optics optimized by Adam Prazmowski. The primary image is magnified by a projection eyepiece on large 30 x 30 cm² glass photographic plates. Janssen developed an associated photographic laboratory, using a wet collodion process, and invented a curtain shutter (the photographic trap made by Gautier) to minimize exposure times (1/3000 s), and freeze atmospheric turbulence as much as possible.

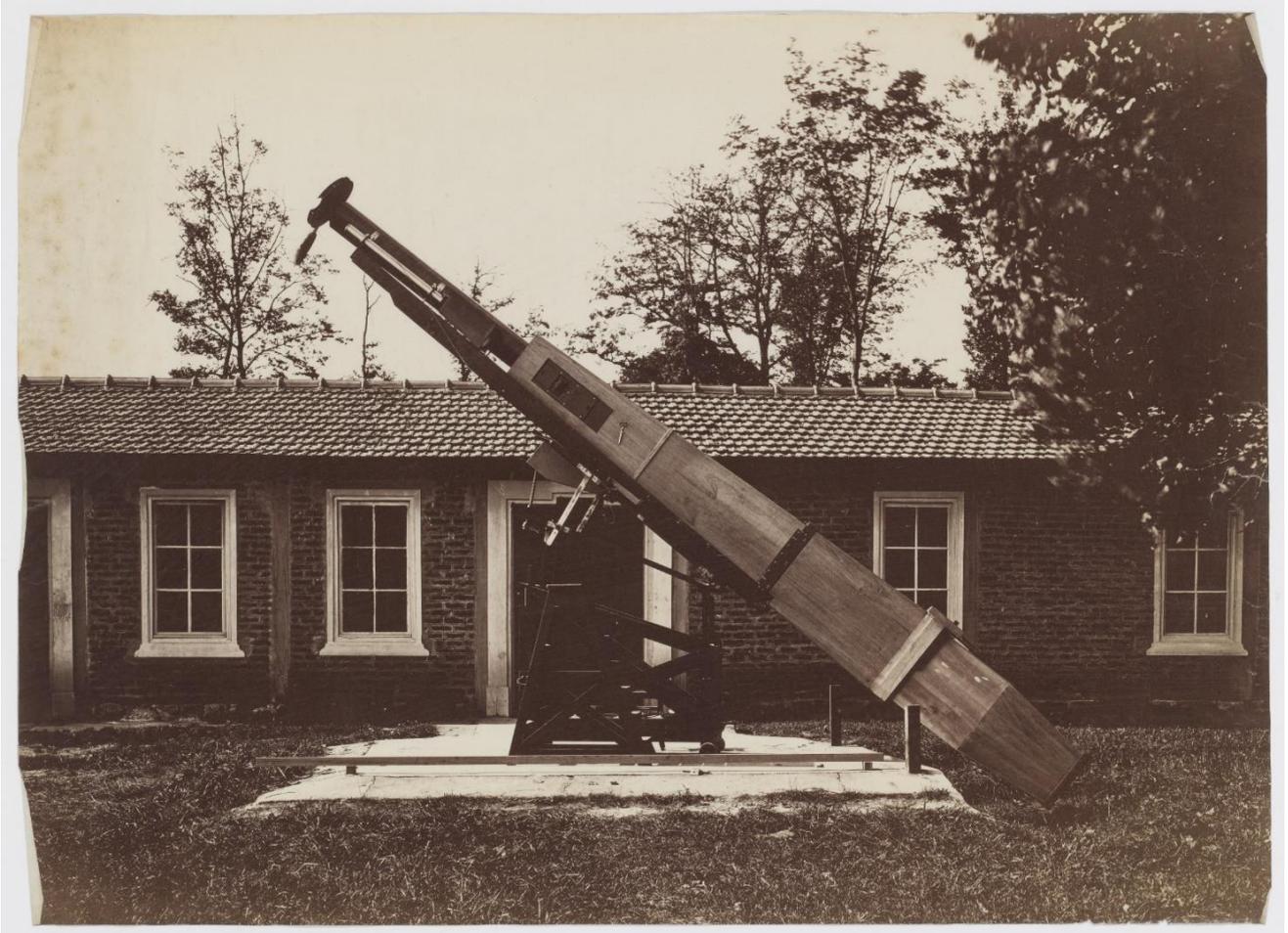

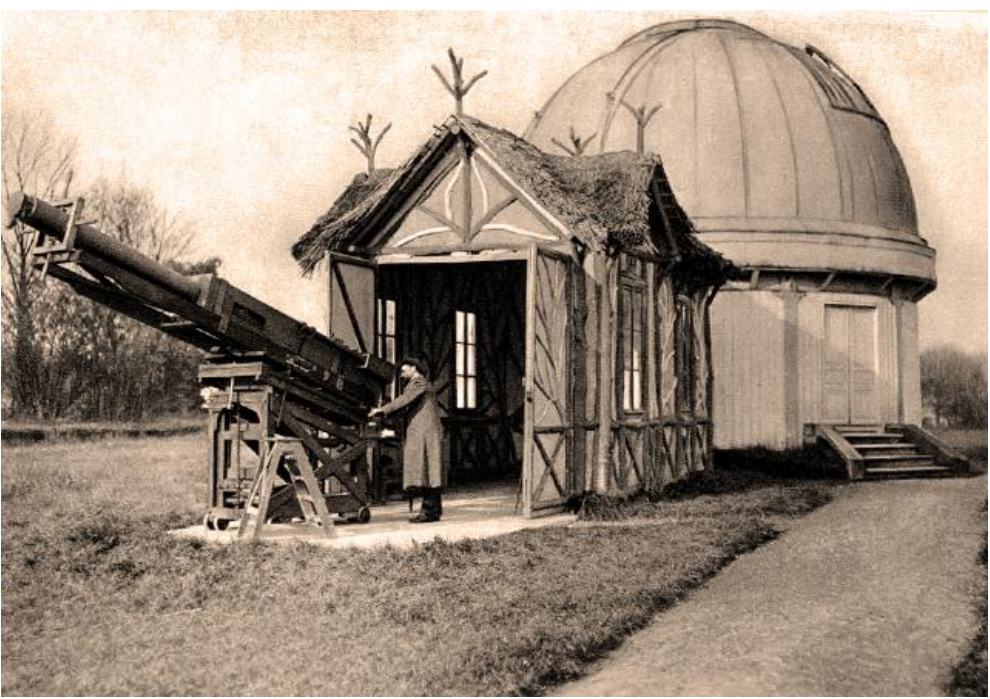

*Figure 17 : Janssen's refractor in the park of Meudon at the end of the nineteenth century. In its early days, it was installed on a wooden rolling mount. The objective lens was 14 cm in diameter and 2 m in focal length, but an eyepiece greatly enlarged the images to use 30 x 30 cm² plates. Credit: OP.*

In practice, the photographic plates were mostly sensitive in the blue, so despite the observation in white light, the result was images from the blue part of the spectrum, around the G-band of the Fraunhofer spectrum (Figures 18 and 19), close to the 430 nm wavelength. For this reason, Prazmowski had optimized the lens of Janssen's refractor for the blue.

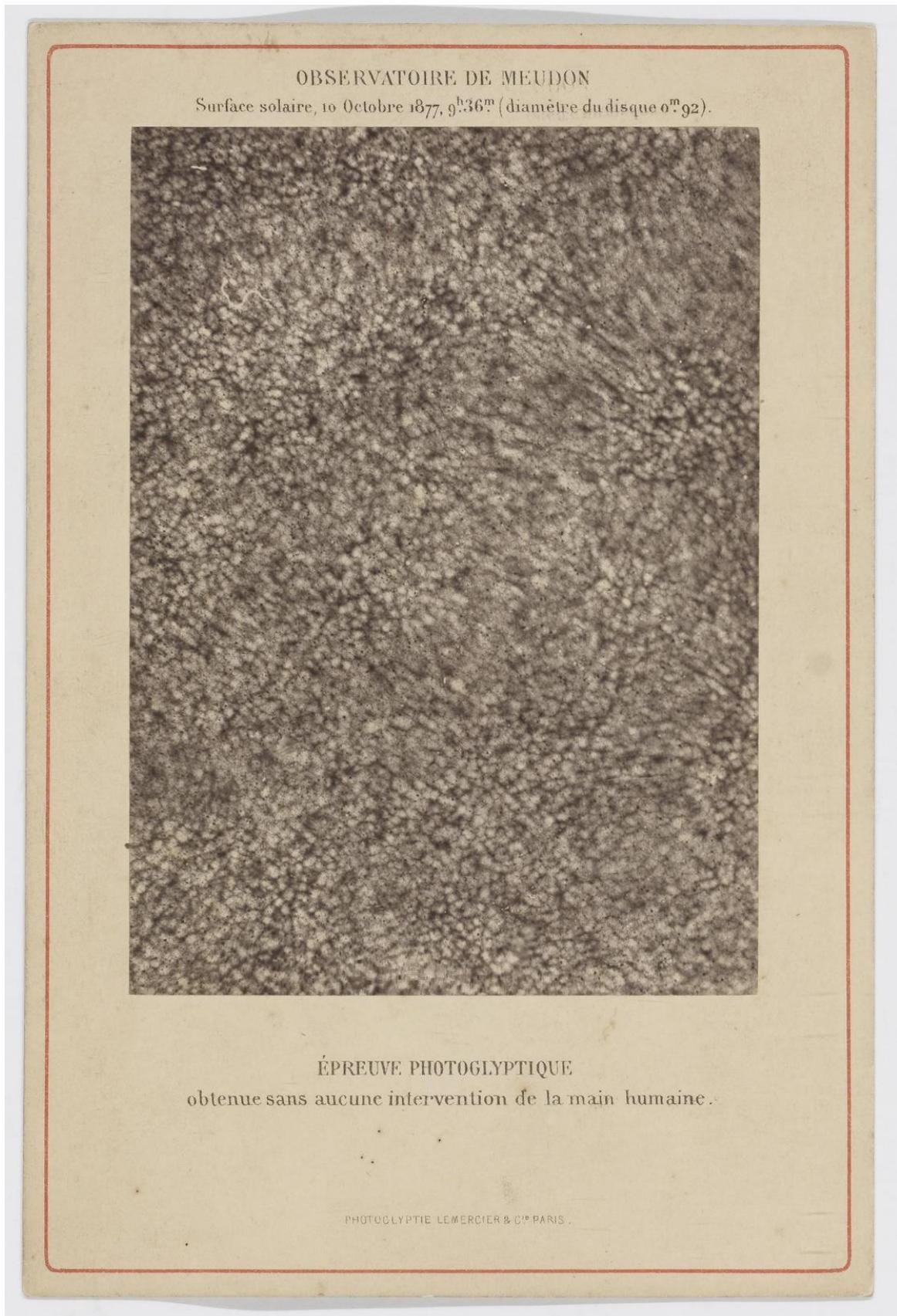

*Figure 18 : A fine example of a photograph of the solar granulation, October 1877. Courtesy OP.*

Janssen and his assistants took 6000 photographs, of which unfortunately only 1% have survived, the rest having been lost or inadvertently destroyed. These are the first detailed, and therefore high-resolution, photographs of the solar surface, in particular of the granulation, that is the signature of the underlying convection, which is difficult to observe in ordinary sites due to the small size of its structures (about 1 second of a degree). Janssen published a sample of 50 plates in his atlas of solar photographs (observatory's library), it is digitized and online at Gallica/BNF.

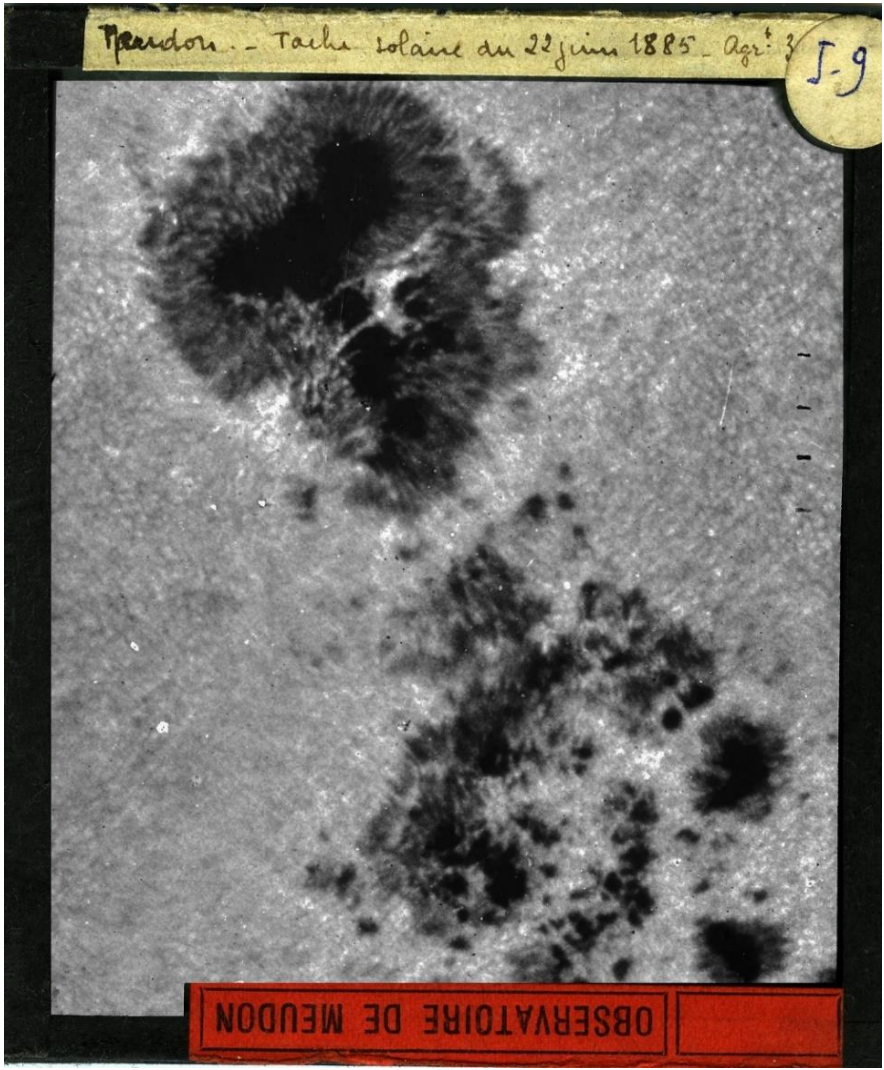
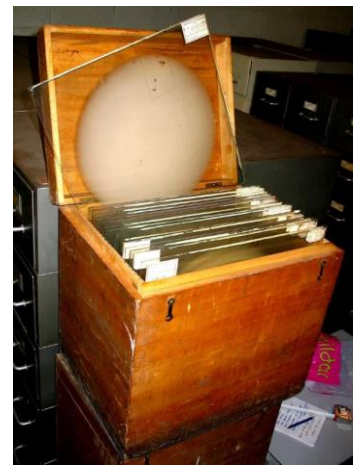

*Figure 19 : a group of sunspots, photographed in 1885 with the Janssen's refractor in the park of Meudon. Around the spot, we can see the granulation, present everywhere on the solar surface (there are 5 million granules in perpetual renewal on the Sun, the lifetime of each granule is 10 minutes).*

*Below: a plate box.*

*Credit: OP.*

After this pioneering work in high-resolution imaging, Janssen became interested in the presence (or absence) of oxygen in the solar spectrum. It is known that the Earth's atmosphere, which contains water vapour, oxygen and nitrogen, contaminates the solar spectrum with the lines of these molecules, which are often found in the red part. The question was whether the oxygen lines could also be of solar origin. The first idea that one may have to solve such a question, consists in ascending high mountains, in order to decrease the thickness of the Earth's atmosphere above the observer, and thus cause the gradual fading of the telluric lines as one ascends (those of solar origin do not vary). Janssen therefore undertook to transport his spectroscope to the mountains, in particular to the Pic du Midi (2870 m). However, he got fascinated by the 4810 m of Mont Blanc, on the slopes of which Joseph Vallot was already doing geophysics with his observatory located at 4350 m on the rocks of "Les Bosses". Thus, a first incursion into the massif (1888) brought Janssen to the "Grands Mulets", a refuge at 3050 m. In 1890, he made his first ascent, under epic conditions. His difficulties of locomotion and the vicissitudes of the ascent through complex and crevassed glaciers did not affect him: he considered travelling seated in a ladder chair or by sleigh with the assistance of 12 guides and porters (Figure 20). Stopping at the Vallot Observatory, Janssen reached the top, was struck by its fantastic horizon, and decided to build an observatory. It could have been permanent if it had been able to benefit from rocky foundations, but it was not the case, because the summit is totally covered by a very thick layer of ice.

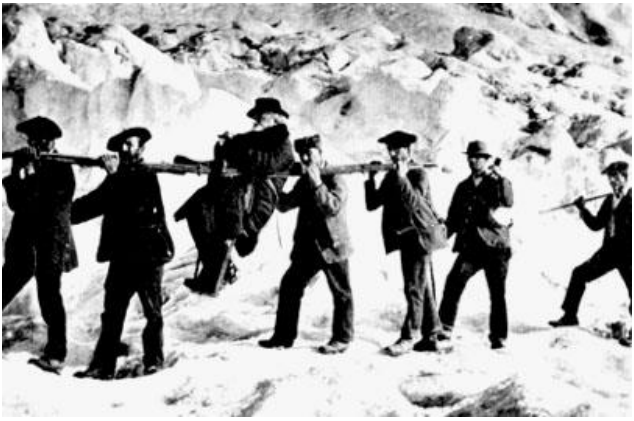
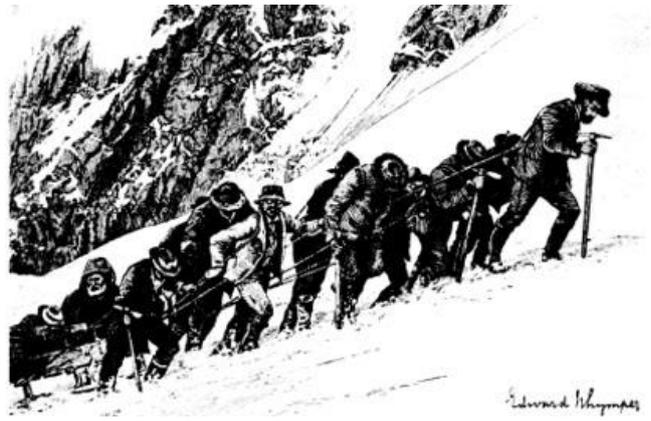
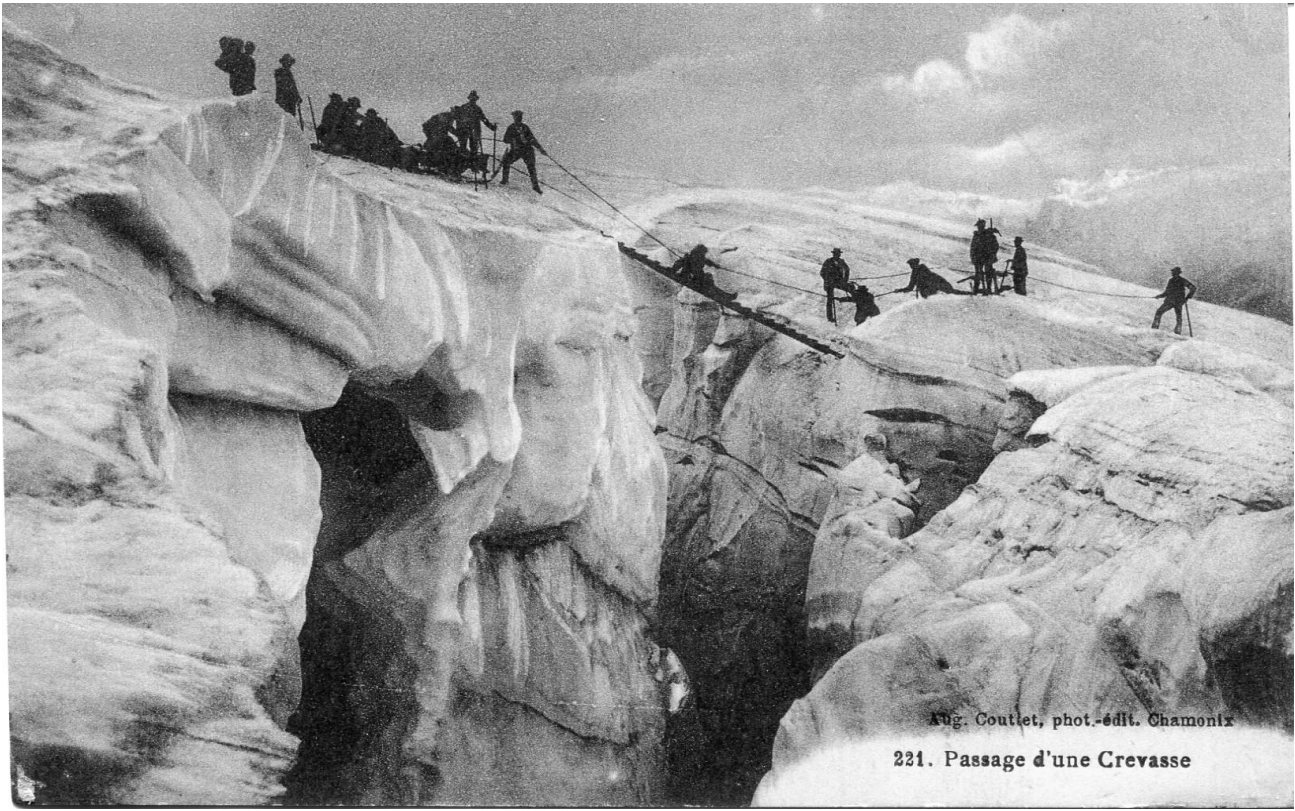

*Figure 20 : epic ascents of Jules Janssen to Mont Blanc, in ladder chair or sleigh. Credit: OP.*

Facing the impossibility of a solid foundation, Janssen decided to place his observatory directly in the ice of the summit. The building (Figure 21) was designed and built in Meudon before being transported in spare parts adapted to the loads that are possible for a carrier or a sleigh. There were two levels, with the low level to be buried in the snow (Figure 22). Actuators were used to adjust the base of the building. The turret was used for weather observations (a long-march meteorograph designed by Jules Richard was permanently placed there). The large refractor (with the Henry brothers' lens) had a diameter of 30 cm and a focal length of 5 m. It was not put on an astronomical mount; alternatively, it was supplied with light by a polar siderostat, a 60 cm flat mirror, installed in front of its lens, with the axis of the telescope remaining fixed and parallel to the Earth's axis of rotation. The refractor protruded from one of the upper walls and astronomers observed from the lower level. Janssen inaugurated it in 1893 and returned in 1895. In reality, he did not use the large telescope himself for his observations of the solar spectrum, but a Duboscq spectroscope that he always carried with him. He found that the intensity of the oxygen lines decreased with altitude. We now know that there is oxygen in the solar atmosphere, but not in the form of diatomic gas $O_2$. Space observations in Ultra Violet have shown since 1960 that there are ionized atoms of oxygen and nitrogen in the solar corona, as well as ions of many other elements known on Earth such as metals. The Mont Blanc telescope was fitted with a 1 m spectrograph in 1904. The co-founder of the Czechoslovak Republic, Milan Stefanik, visited the place several times.

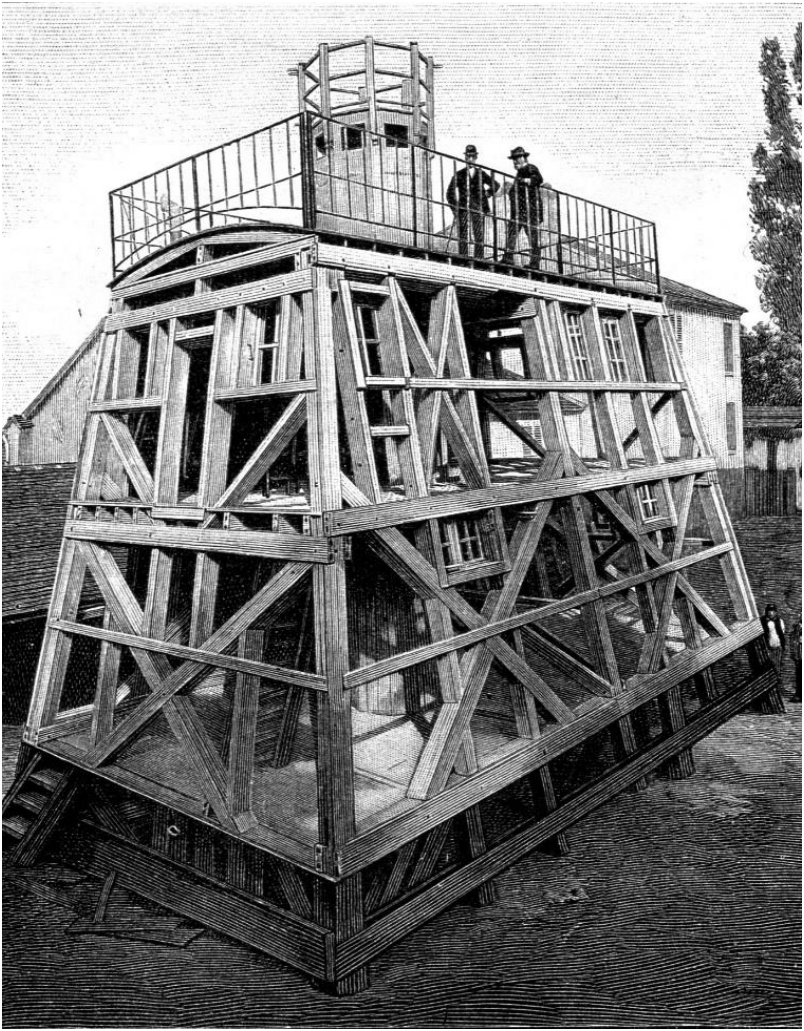
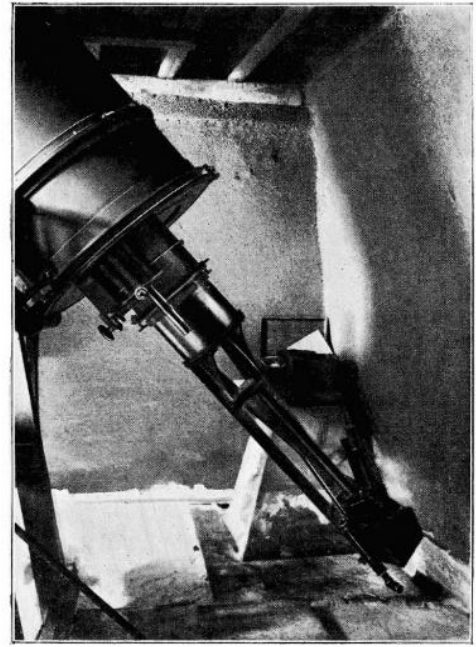

Fig. 1. — Partie oculaire de la lunette, portant le spectroscope à monture d'aluminium.

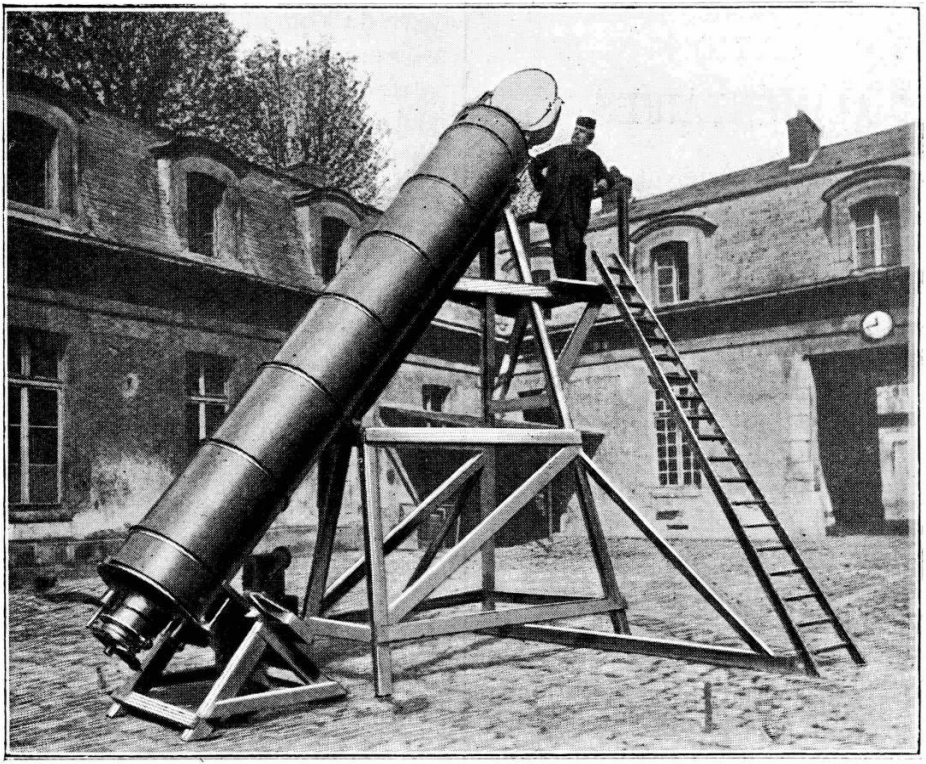

Fig. 3. — La lunette montée, dans la cour de l'Observatoire de Meudon, avant son transport au Mont-Blanc.

*Figure 21* : *the Mont Blanc observatory built and tested in Meudon before being transported on the back of carriers or by sleigh to the summit, with its large polar refractor with a diameter of 30 cm and a focal length of 5 m (1893), and a small spectrograph of 1 m attached to the focus (1904). Credit: OP.*

The Mont Blanc Observatory operated until 1909 (two years after Janssen's death), with numerous expeditions, often astronomical, but sometimes multidisciplinary. Progressively damaged by the pressure and the movement of the ice, it ended its life dislocated, which caused it to be abandoned. A short film (from a 9.5 mm Pathé film) tells the story of an ascent around 1900 at the time of the observatory (Figure 23).

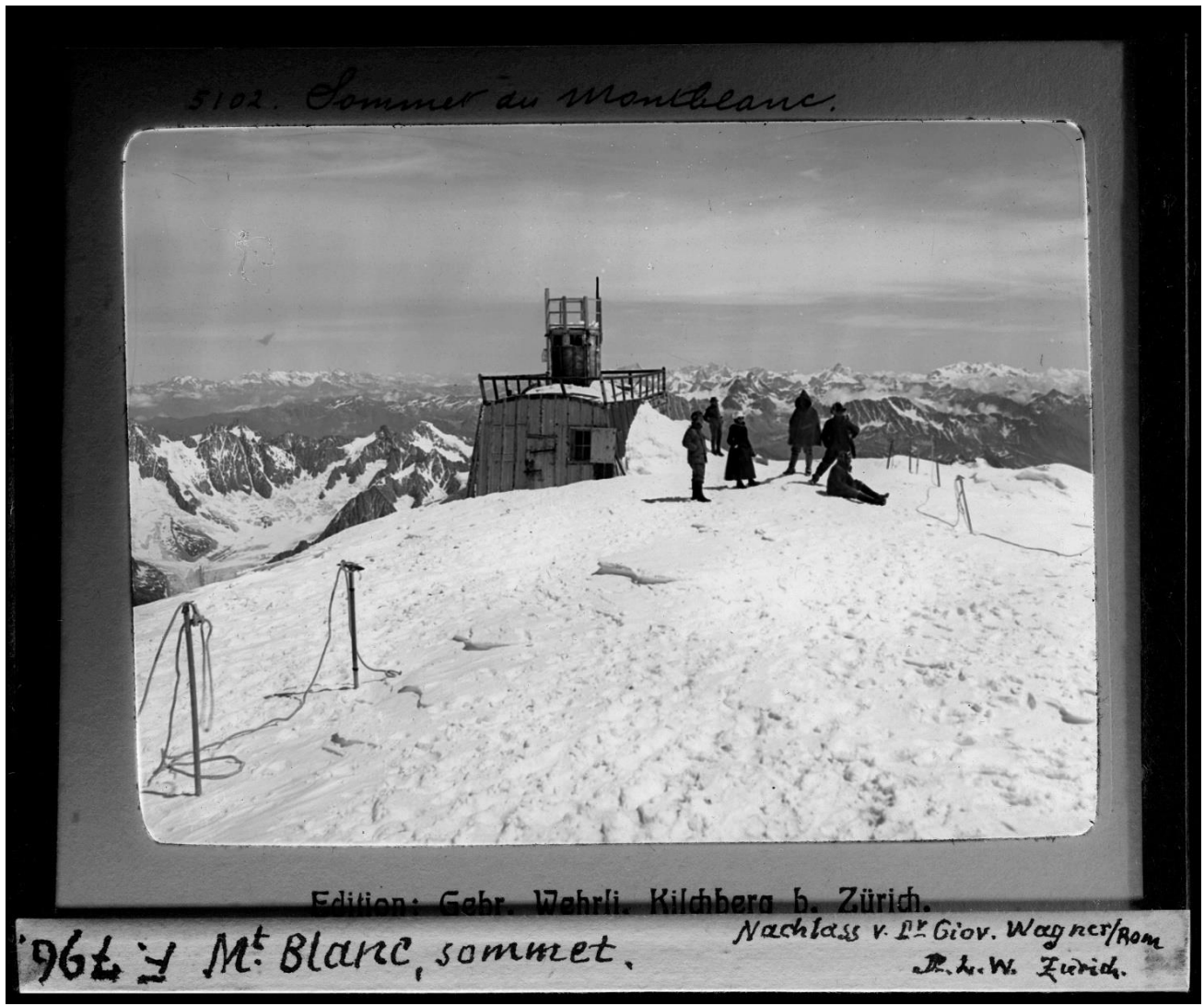

*Figure 22 : the Mont Blanc Observatory. Credit ETH/Zürich.*

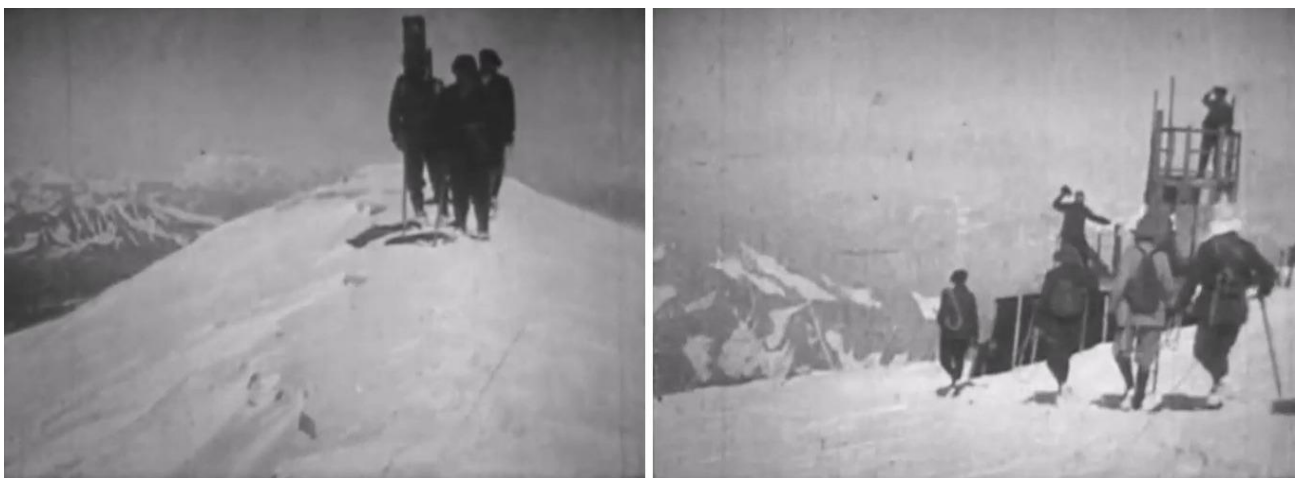

*Figure 23 : arrival at the summit of Mont Blanc, MP4 clip from a Pathé 9.5 mm film. Download there:*

*https://www.lesia.obspm.fr/perso/jean-marie-malherbe/HY/Media1.mp4*

While Janssen was busy at Mont Blanc, Henri Deslandres (a future academician, figure 24) was hired in 1888 at the Paris Observatory by Admiral Ernest Mouchez, its director, to found a spectroscopy laboratory. It was natural that he went to solar spectroscopy, where everything had to be developed. He recalled that Janssen, after the eclipse of 1868, had enunciated the principle of the spectrohelioscope, or the monochromatic slit-scanning imaging spectrograph: "by rotating the spectrograph on its axis, the input slit scans the solar surface, and by placing a second slit in the spectrum isolating a well-chosen spectral line, a monochromatic image is formed on the retina through an eyepiece". To do this, you have to benefit from retinal persistence, so you must turn the spectroscope fast enough (Janssen's instrument had a crank). If the eye is replaced by a light-sensitive plate, a recording spectroheliograph has been made. This was Janssen's idea, enunciated in 1869, taken up and improved by Deslandres in Paris, but also by George Hale in the USA, independently. This is how the spectroheliograph was born, giving in Paris the first monochromatic images of the Sun as early as 1893 (called spectroheliograms), and the first recordings of prominences as early as 1894 (Figure 25). Hale's instrument at Mount Wilson (the Snow telescope) started observations in 1892, using similar principles, and is displayed in figures 26 and 27.

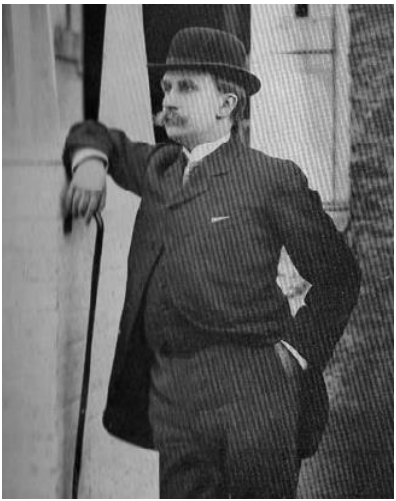

*Figure 24* : *Henri Deslandres (1853-1948). Co-inventor (with Hale) of the spectroheliograph, he set up the first instrument in 1893 in Paris, and he moved to Meudon in 1898 where he started working on a large quadruple spectroheliograph, which he built with his student Lucien d'Azambuja. The latter was in charge of organizing a service of daily observations of the Sun from 1908, that he developed and which continues today. Deslandres returned to Paris in 1926 to be appointed director of the newly merged observatories of Paris and Meudon. An anecdote: "When Deslandres was appointed to Meudon, he came to make contact and entered through the concierge's door. Barring the corridor for him, he saw the little figure of father Bardiot, who said to him: "Are you the new astronomer?" Don't come, sir, they're all crazy here... »… Credit: OP.*

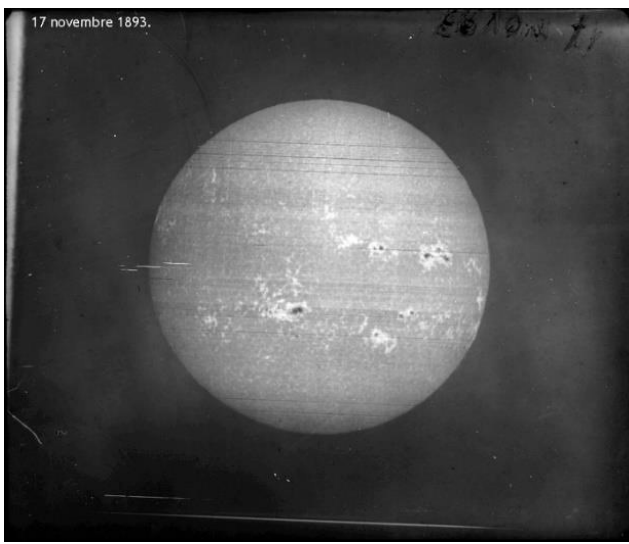 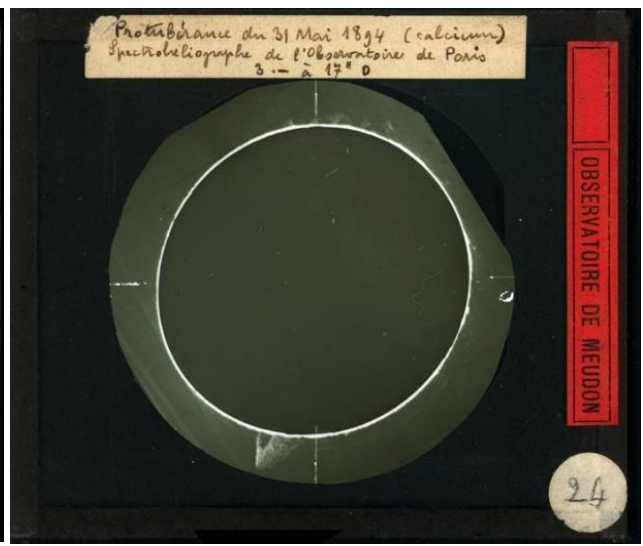

*Figure 25* : *First spectroheliograms or monochromatic images made in Paris in 1893 and 1894 by Henri Deslandres. The line chosen is the purple "K" line of calcium. The image on the left shows the active regions surrounded by bright regions (faculae). The one on the right (with a simple mask on the sun's disk to avoid overexposure) reveals the prominences at the limb. Credit: OP.*

Thinking he could find more resources in Meudon, Deslandres moved there in 1898 with his instruments. He hired an assistant, Lucien d'Azambuja (Figure 28), who was only 15 years old, and with him began work on a large quadruple spectroheliograph (Figures 29 and 30), which was completed in 1908 and was intended for both daily observations and advanced scientific research.

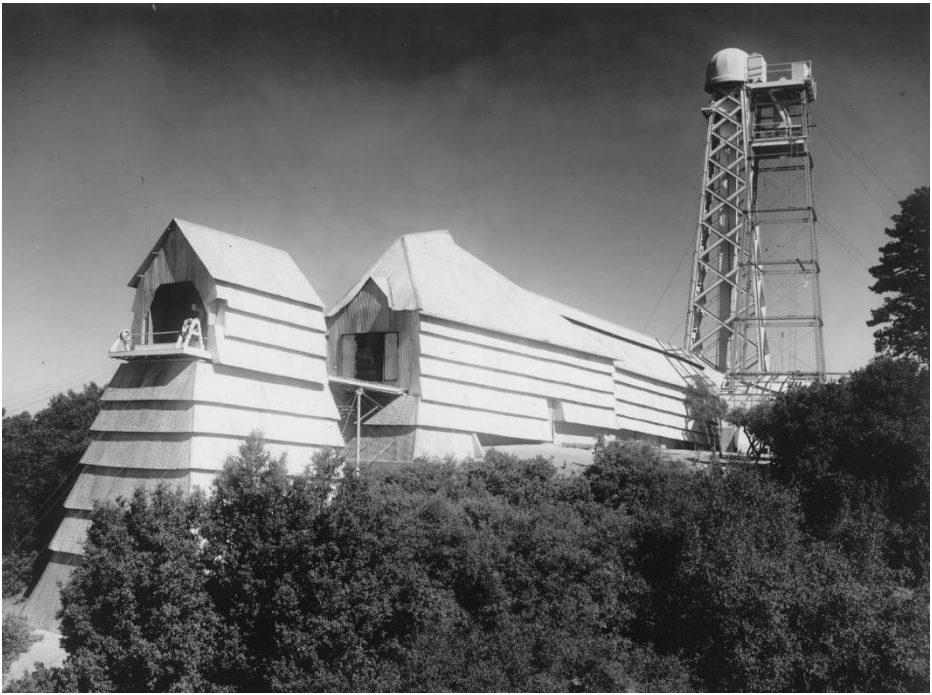

*Figure 26 : The Snow telescope at Mount Wilson in the USA. George Hale started there a huge collection of monochromatic images (the spectroheliograms) in 1892. Courtesy Carnegie institution.*

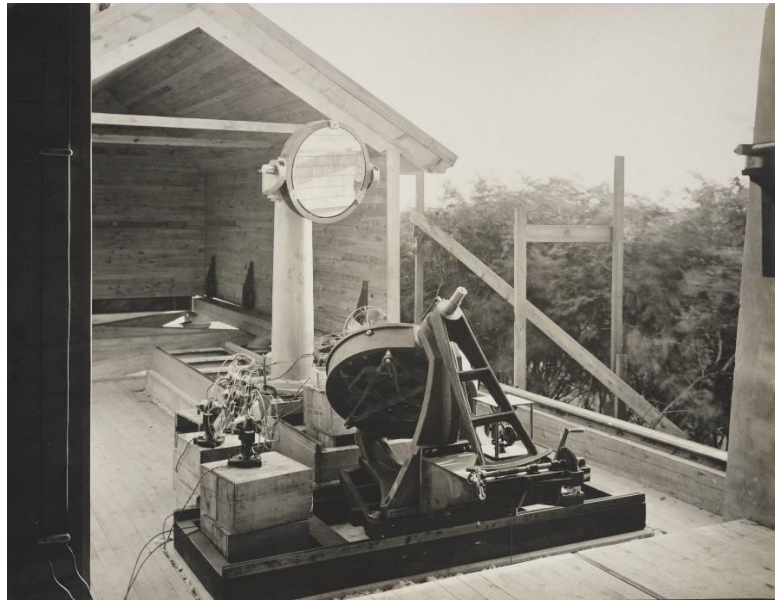

*Figure 27 : The two flat mirror coelostat of the Snow telescope at Mount Wilson in the USA. This device catches the solar light to feed the spectroheliograph. Courtesy Carnegie institution.*

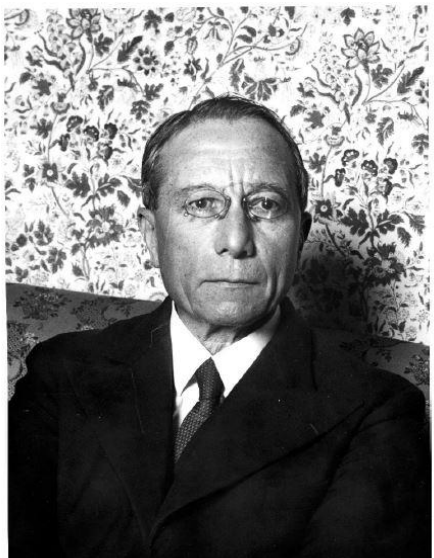

*Figure 28 : Lucien d'Azambuja (1884-1970), "hired by Deslandres at the age of 15, he was a man not very tall, always dressed to the nines, with a three-piece suit, a hard-necked shirt, and eyeglasses. An excellent instrumentalist, but he was intractable on the service... ». Hired extremely young (15 years old in 1899), d'Azambuja remained in Meudon for 60 years, until his wife's retirement in 1959. He was 75 years old. Amazing career today! Credit: OP.*

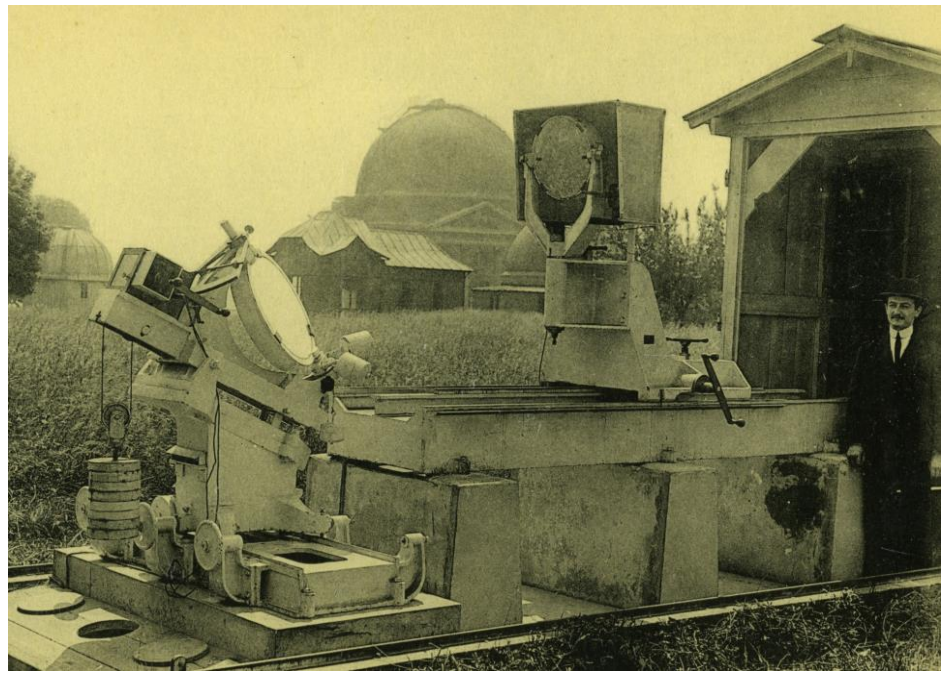

*Figure 29* : The two flat mirror coelostat catching the Sun and feeding the large quadruple spectroheliograph of Deslandres and d'Azambuja at Meudon, after 1908. Courtesy OP.

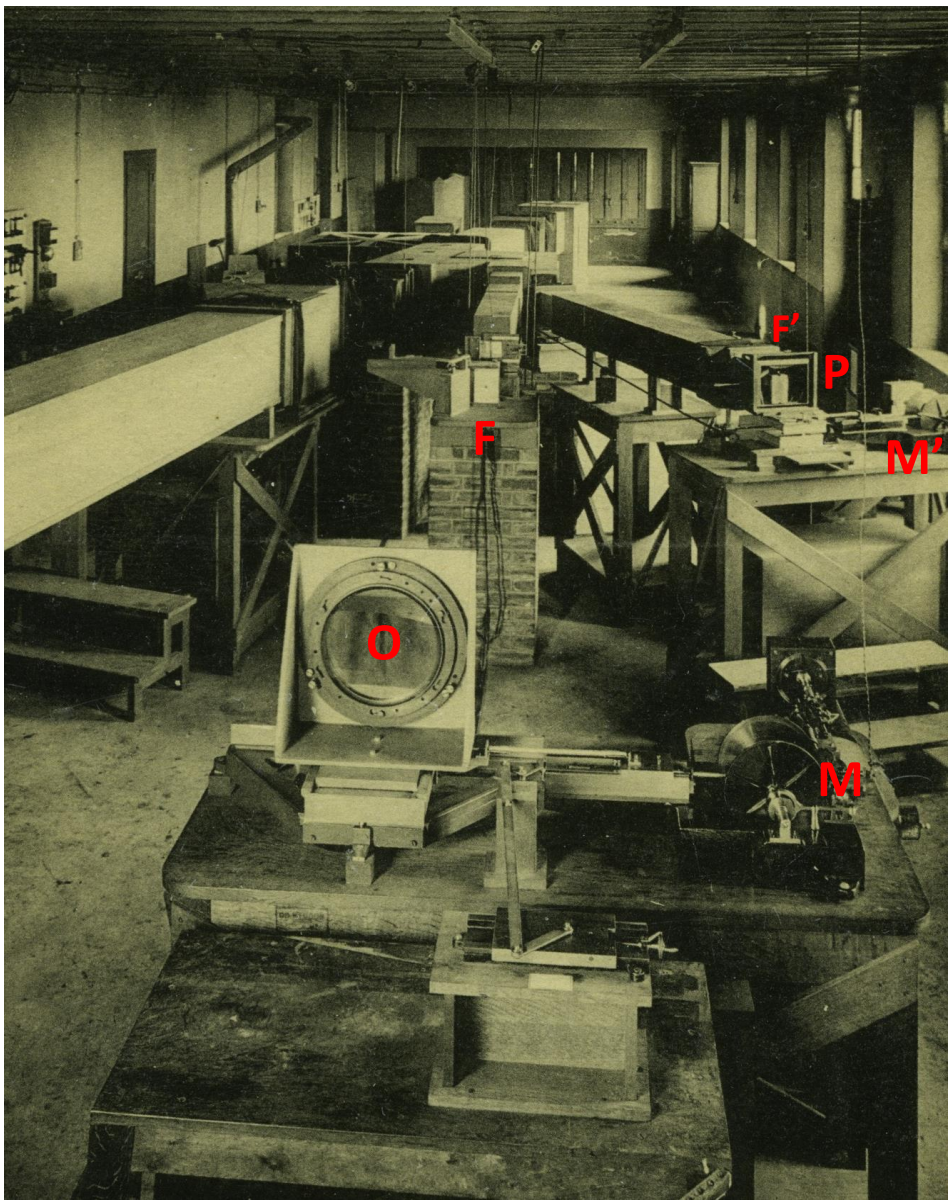

*Figure 30* : the large quadruple spectroheliograph of 1908. It is a versatile instrument with two 3 m chambers for daily observations in Hydrogen and Calcium; for research observations, a multi-purpose, multi-line 3 m chamber and a 7 m high spectral resolution chamber. There were two dispersive elements, a train of 3 prisms and a plane Rowland grating. In the figure, we see the imaging lens O, common to all chambers, with a focal length of 4 m. It was in right/left translation to move the Sun at a constant speed on the spectrograph slit F (M motor). We see the 13 x 18 cm² plate holder (P and motor M') for Hα and the associated output slit F' in the spectrum. Today, the instrument retains the basic principles but has been renovated several times. Credit: OP.

Lucien d'Azambuja was in charge of organizing a service of daily and systematic observations as early as 1908 (Figures 31 and 32). Images of the solar atmosphere were recorded in two lines, Calcium and Hydrogen, which give a synthetic view of the phenomena in the Sun's atmosphere. Today his service is still active, but it has been digitized since the year 2000. The observers recorded more than 100000 photographs, and the old glass plates and 13 x 18 cm² films having been scanned (Figure 33). They can be accessed via the http://bass2000.obspm.fr portal. Marguerite d'Azambuja (figure 34) succeeded Lucien, her husband, as head of the service when he retired in 1954; she herself left in 1959, and it was Marie-Josèphe Martres (figure 34) who continued, along with Gualtiero Olivieri; then Elisabeth Nesme-Ribes; Isabelle Bualé is currently in charge, which shows that solar observation is as much a matter for women as it is for men.

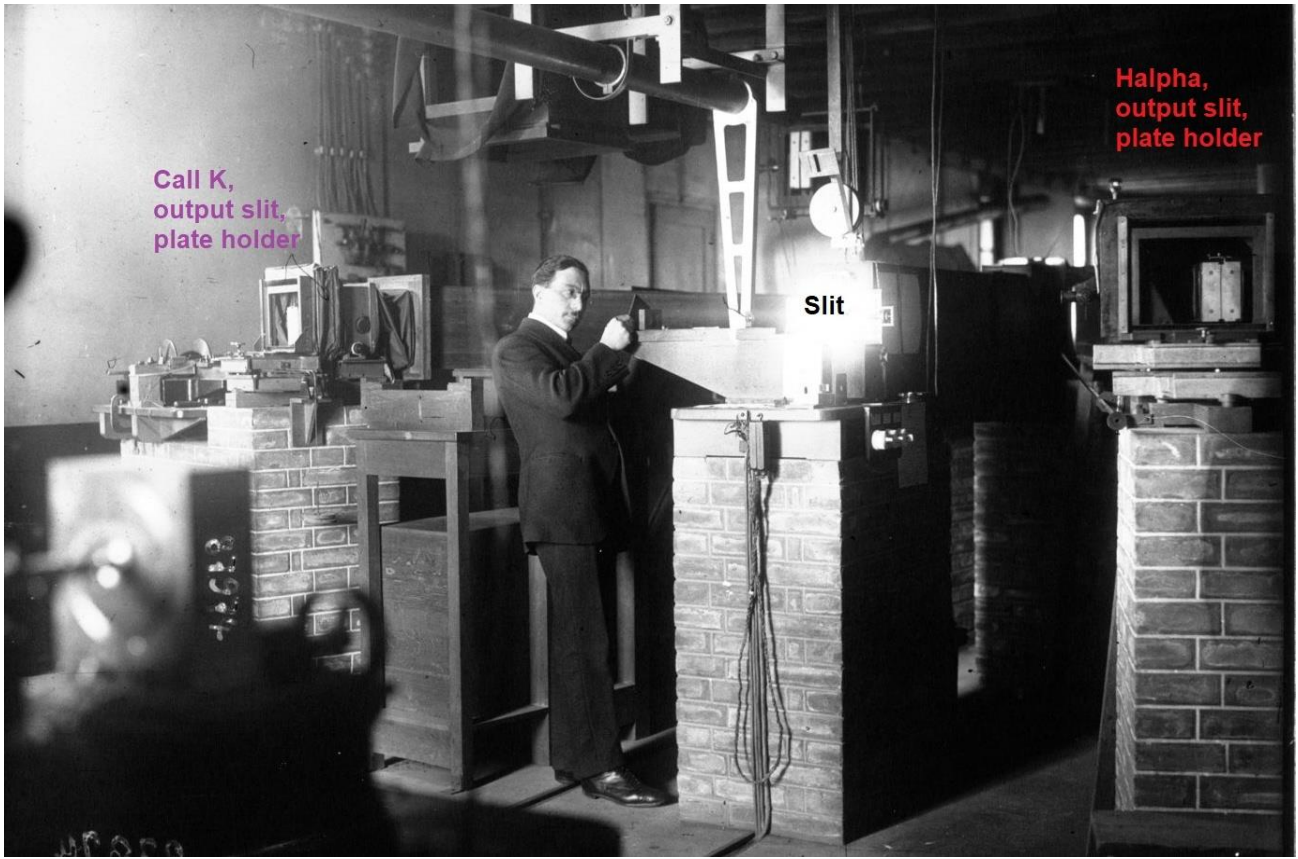

*Figure 31* : *Lucien d'Azambuja in observations in 1921. On the left, the Calcium chamber, on the right the Hydrogen Hα chamber; we can distinguish the associated slits in the spectrum and the motorized plate carriers, and the entrance slit of the two-arm spectrograph. Credit: Gallica/BNF.*

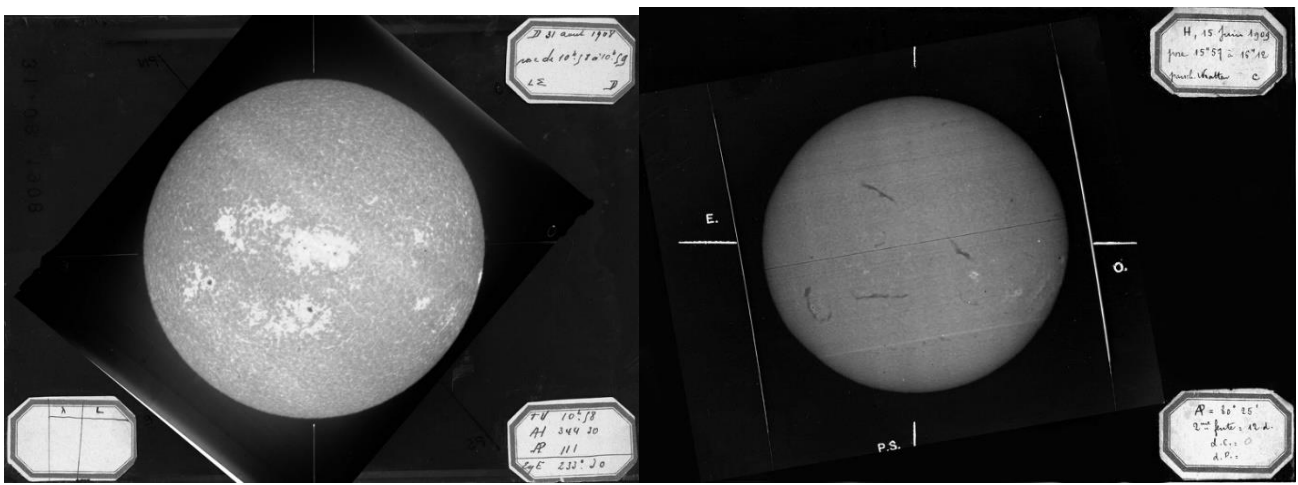

*Figure 32* : *The first spectroheliograms from the huge collection of the Meudon spectroheliograph (1908-2024), on the left in Calcium (1908) and on the right in Hydrogen Hα (1909). Credit: OP.*

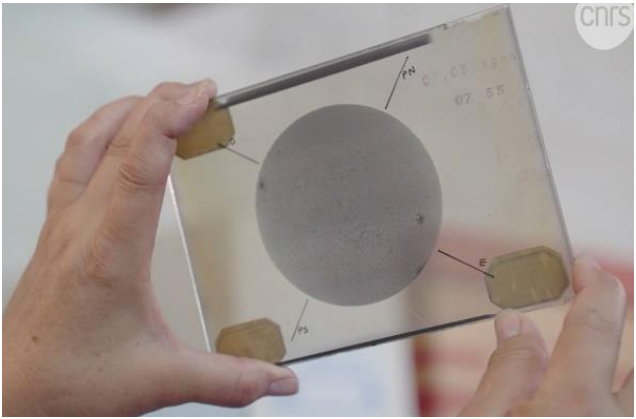

*Figure 33* : Typical glass photographic plate of the spectroheliograph, format 13 x 18 cm², negative, Sun 86 mm in diameter. Films (same format) followed the glass plates before a digital camera took over in the year 2000. The plates and films were digitized with a scanner.

Credit: OP.

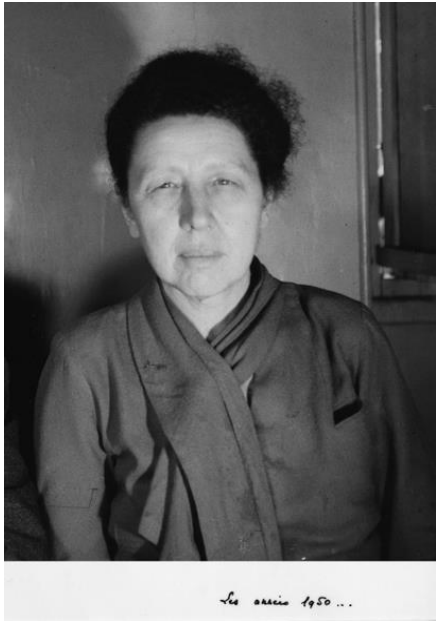
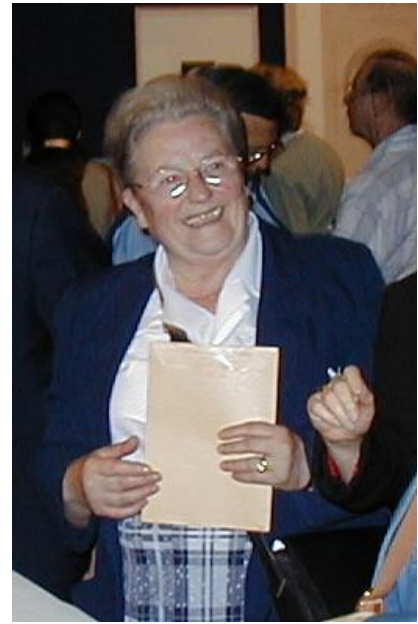

*Figure 34* : Marguerite d'Azambuja (left) and Marie-Josèphe Martres (right).

Credit OP.

The history of solar observations is punctuated by a number of rare events and exceptional phenomena, such as the great flare of Figure 35. At that time, it was said about the Meudon director (André Danjon): "observation was, along with mushroom picking, a favourite hobby of the director; he was there during the large eruption of July 25, 1946, and the orders he gave did not help to keep the observers cool"!

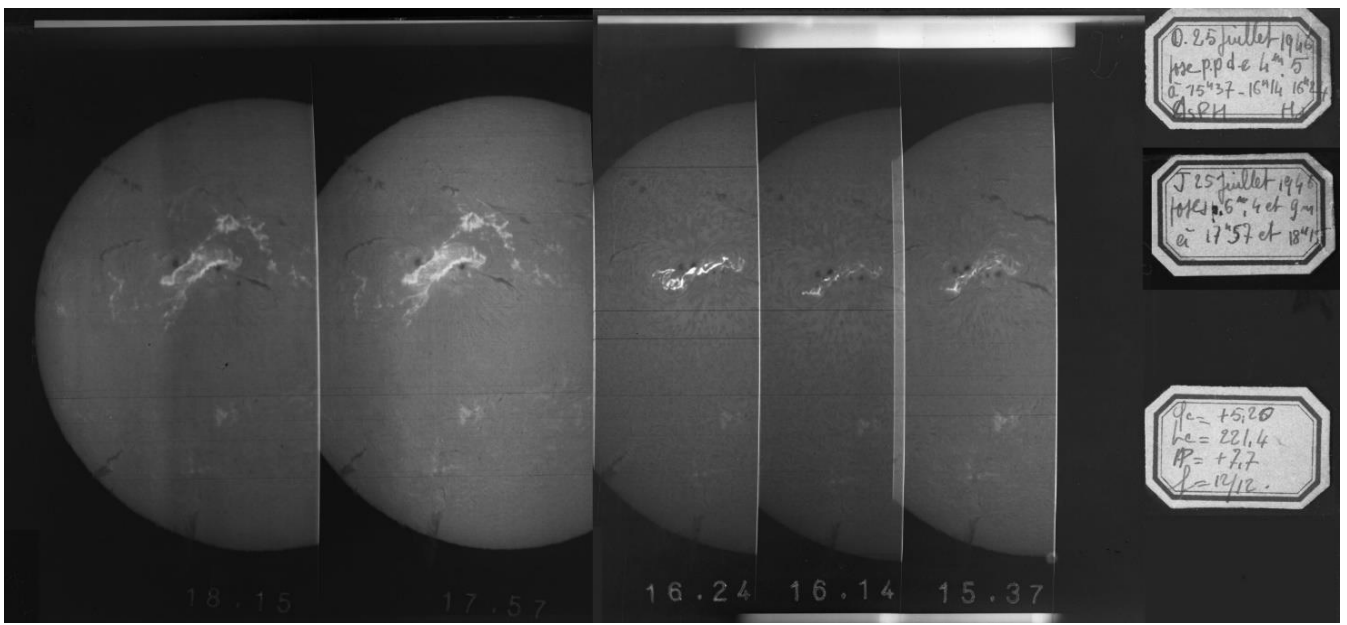

*Figure 35* : Large solar flare of July 25, 1946 at 5 successive moments (from right to left). Credit OP.

The number of sunspots on the surface of the Sun has been counted since 1750 by the Royal Observatory of Brussels (Figure 36), it is known with precision since 1700 and estimated since 1650, allowing us to see the 11-year cycle in action (and its centennial modulation). More recently, since the beginning of the twentieth century, spectroheliograms bring a new and rich contribution to the survey of solar structures (spots, active regions, filaments, prominences) thanks to a faithful photographic record (Figure 37).

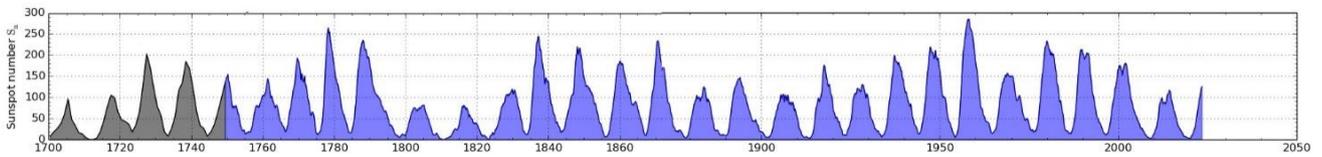

*Figure 36 : Counting spots since 1700. Credit: SIDC, Royal Observatory of Brussels.*

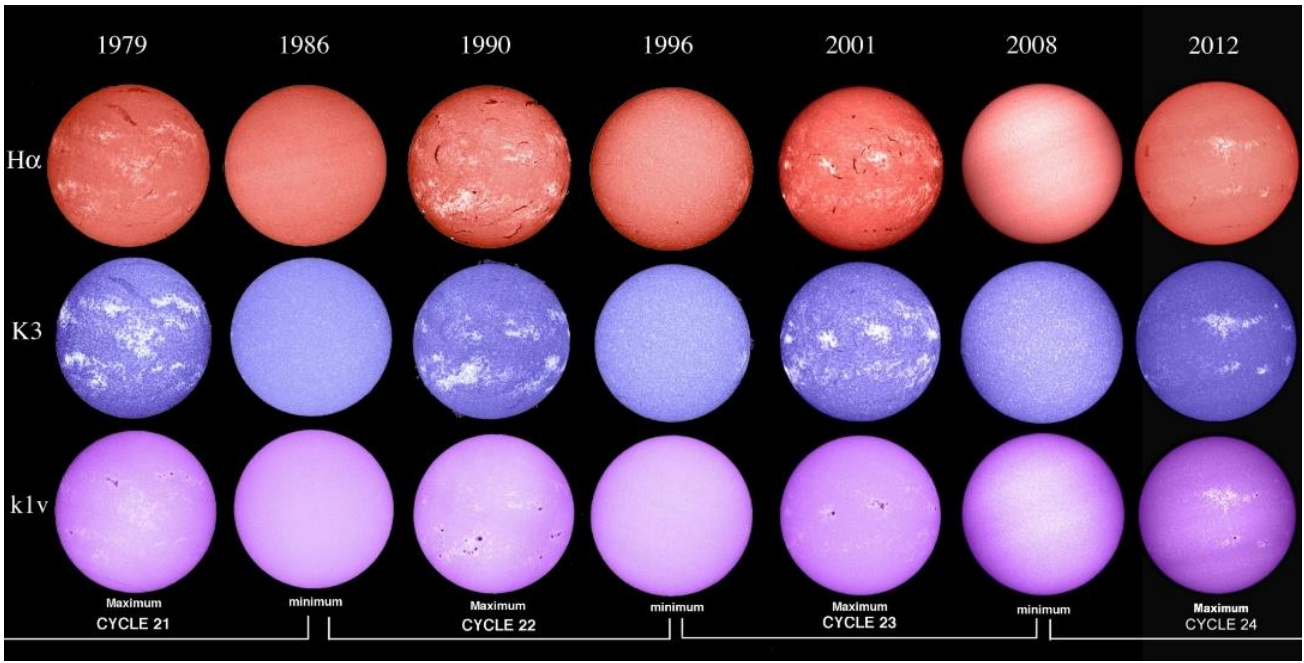

*Figure 37 : extract from the collection of monochromatic images of Meudon in the lines of Hydrogen and Calcium, vision at three altitudes, over three solar cycles. Courtesy OP.*

Between 1930 and 1950, it was Bernard Lyot (1899-1952, figure 38) who distinguished himself with two remarkable inventions that toured the astronomical world and which still equip many instruments on the ground and in space today: the coronagraph and the birefringent monochromatic filter.

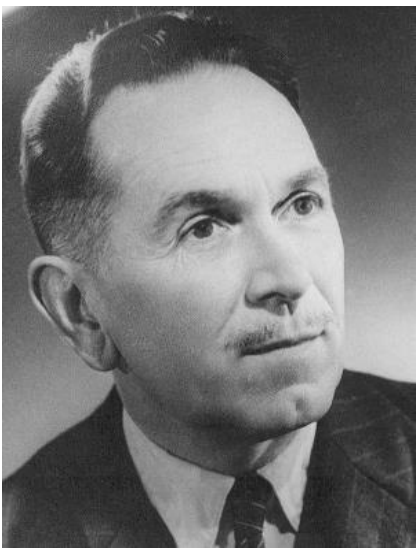 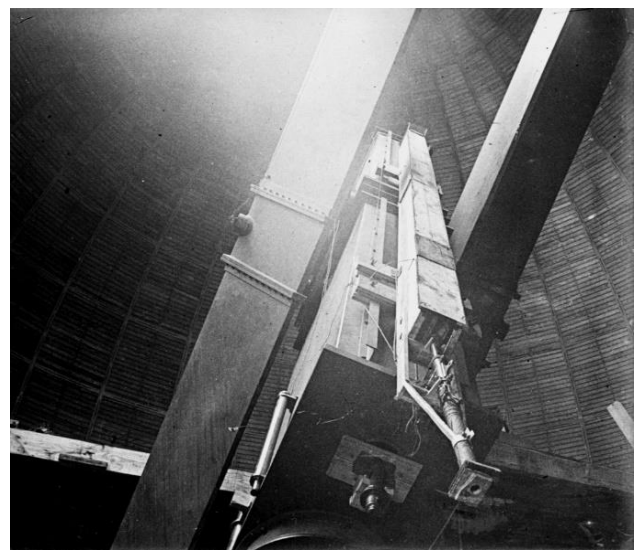

*Figure 38*

*Bernard Lyot and his first coronagraph, fixed here on the mount of the "carte du ciel" dome in Paris. Credit OP.*

Lyot is said to have been "a charming and shy man despite his celebrity (he became an academician). He came to work on a velosolex or in an old Citroën B12 when he brought his pair of poodles. His distraction was proverbial, if he borrowed something from you, it was better to follow him and retrieve the object before he made the trip to the Pic du Midi...". The coronagraph was born in 1930 in Meudon. It is well known that the solar corona is only observable during total eclipses. The coronagraph is an instrument whose optical formula is specially designed and optimized to observe the lower solar corona (one million times less luminous than the disk), outside of eclipses, provided that the sky is pure, a situation that occurs in the high mountains. This is the reason why the device was mounted in Meudon, but tested and used at the Pic du Midi at 2870 m. However, in 1950, Lyot supplemented the coronagraph with a coronameter capable of isolating the polarized light component of the corona, allowing it to be observed even in diffusing sky or at sea level. The second invention dates back to 1933, this is the monochromatic birefringent filter (Figure 39). If the coronagraph allows the low corona to be observed in white light under good conditions, a filter must be added to isolate the spectral lines emitted by solar structures visible at the limb, resulting from cold prominences (hydrogen lines, 8000 degrees) or hot magnetic loops (highly ionized iron lines, 1 million degrees or more, Figure 40), otherwise it is not possible to discern them. The filter is also capable of isolating the absorption lines of the solar disk that reveal dark filaments and active regions, but remains less selective than the spectroheliograph of Deslandres and d'Azambuja. These two inventions by Lyot have revolutionized solar physics, they have been adopted by the majority of observatories in the world, and they can even be found on modern satellites or space instruments (such as the coronagraphs of SOHO, 1996, and those of STEREO A and B, 2006; or the filter of HINODE, 2006...).

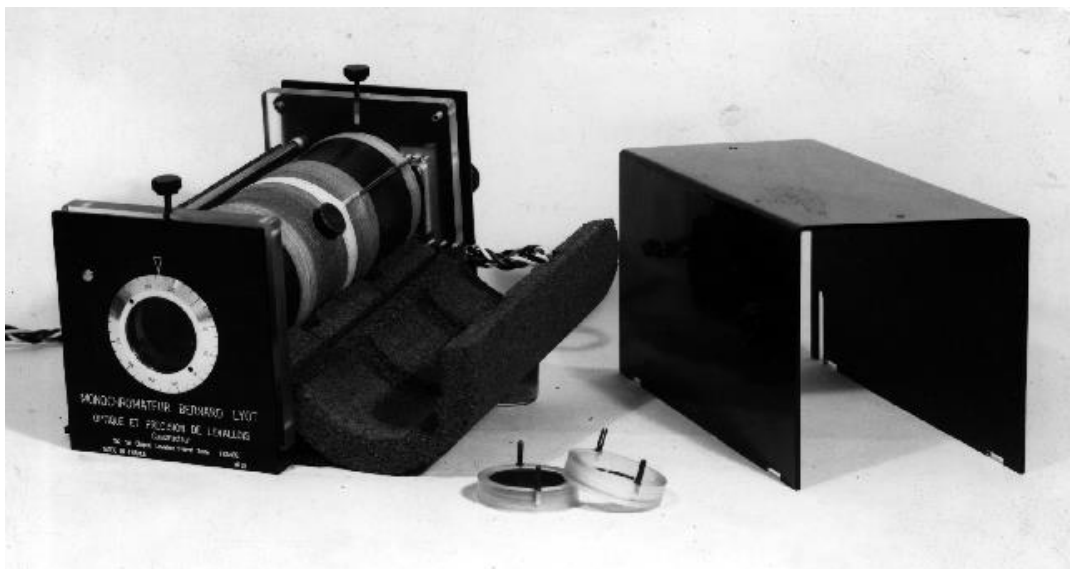

*Figure 39 : Example of a Lyot monochromatic filter; this model was "mass-produced" by the firm "Optique de Précision Levallois" (OPL) to equip many observatories around the world. Credit: OP.*

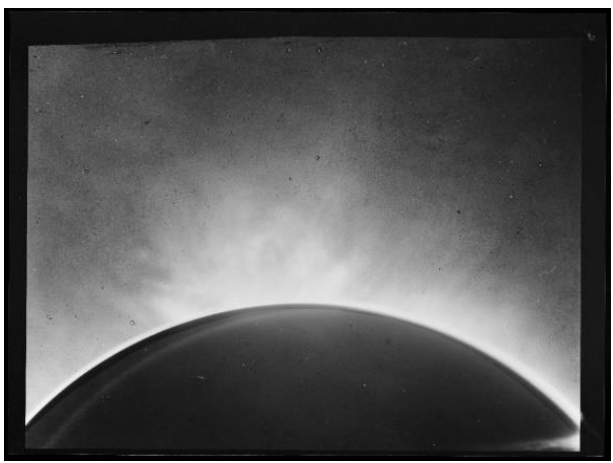

*Figure 40*

*Monochromatic image, in the green line of the 13 times ionized Iron (Fe XIV), of the hot magnetic loops in the lower solar corona, which was observed with a Lyot filter installed on the coronagraph of the Pic du Midi in 1941. The Sun's disk is obscured by an occulting cone, this is the centre-piece (along with the non-scattering sophisticated optics) of the instrument. Credit OP.*

Lyot's new filter had exceptional performance, as it was able to produce both monochromatic and instantaneous images of regions of the Sun much faster than the spectroheliograph (because the latter operates in spectroscopy and has to scan slowly the surface of the Sun with a slit, but in return it has better spectral selectivity). Indeed, at a time t, a filter provides an image (x, y) for a value of the wavelength λ, while a spectrograph gives a spectrum (λ, x), so that the Sun must be scanned along the y-axis with the entrance slit. Lyot was able to exploit the speed of his instrument by starting in 1937, at the Pic du Midi, a program of cinematography of prominences in the red line of Hydrogen (Figure 41, and associated MP4 film). Lyot associated the professional film-maker Joseph Leclerc with this work.

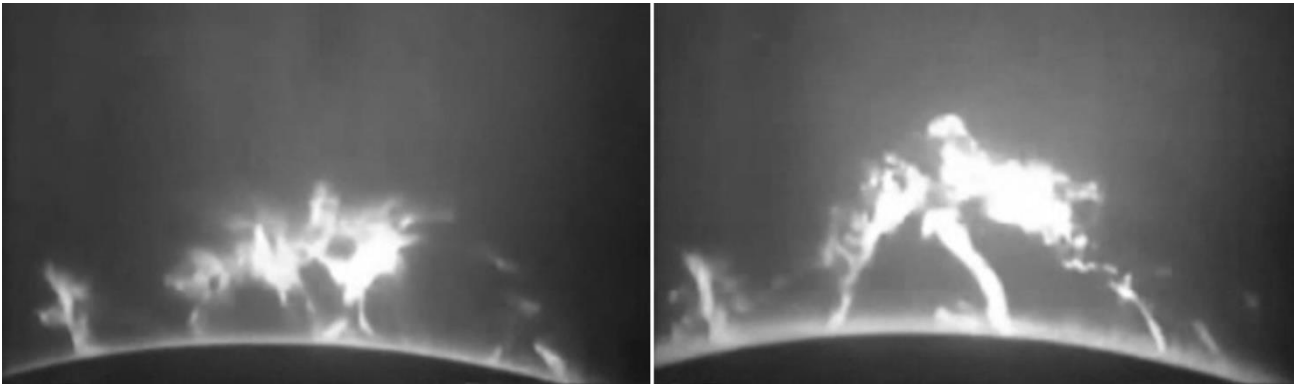

*Figure 41 : a fine example of cinematography of prominences with the coronagraph and the monochromatic filter of Lyot, in the red line Hα of Hydrogen, with an extract of a digitized film (MP4 format). Courtesy OP.*

https://www.lesia.obspm.fr/perso/jean-marie-malherbe/HY/Media2.mp4

The cinematographic possibilities of the Lyot filters were then intensively used in the context of the International Geophysical Year (IGY) of 1957. The aim was to do a survey of the Sun and film the flares on the disk in the vicinity of the maximum of the solar activity cycle. Lyot unfortunately died in 1952, but in 1954 Henri Grenat and Gérard Laborde developed a monochromatic heliograph (Figure 42) based on a Lyot filter (Figure 39) and operating in the red line of Hydrogen. This instrument recorded the images on 35 mm film reels of 45 m long (Figure 42, and associated MP4 film of Figure 43).

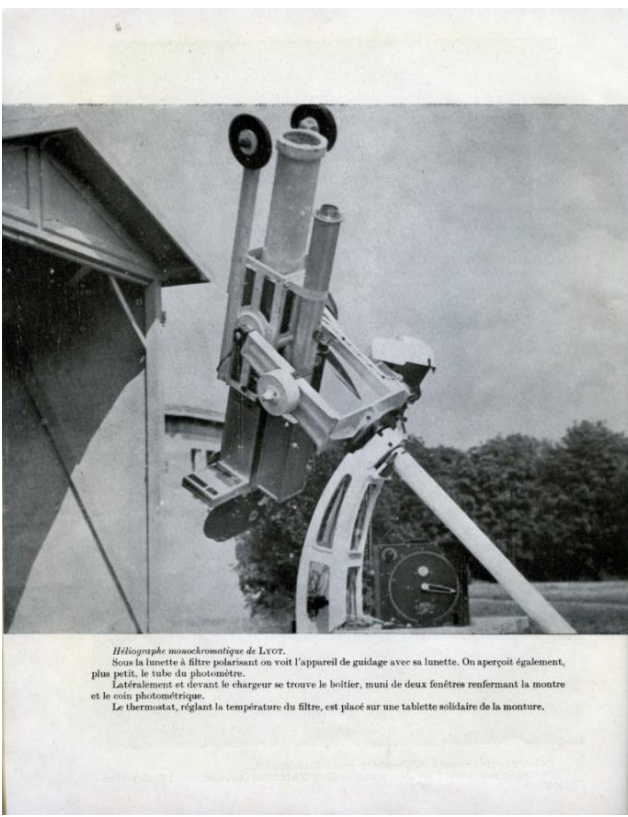

*Figure 42*

*Monochromatic heliograph by Grenat and Laborde equipped with a Lyot filter to film the eruptions on the occasion of the International Geophysical Year of 1957. Below is a reel of 35 mm film. About 3000 of 45 m reels were used (130 km) during the solar activity survey from 1956 to 1994. Courtesy OP.*

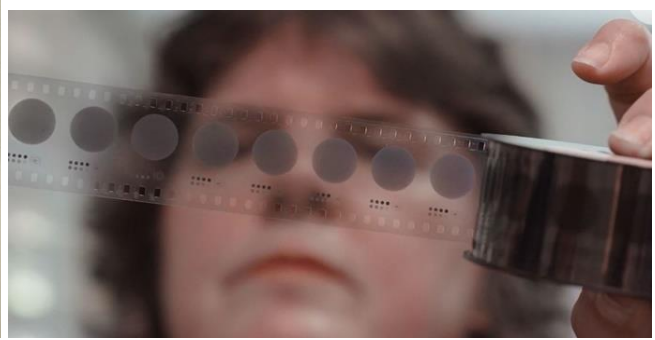

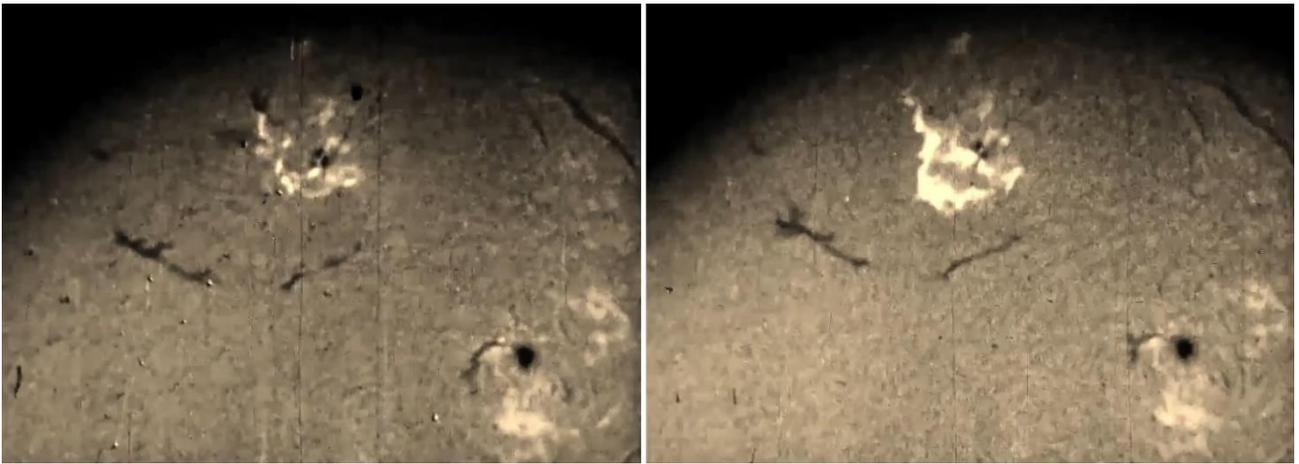

*Figure 43 : example of cinematography of eruptions with a Lyot filter, in 1957 (credit OP). Film :*

*https://www.lesia.obspm.fr/perso/jean-marie-malherbe/HY/Media3.mp4*

It was decided, as part of the monitoring of eruptions and material ejecta in the frame of the 1957 International Geophysical Year program, to duplicate the original Lyot filter and heliograph to equip many observatories around the world. This task was entrusted to the companies SECASI (Bordeaux), which made 6 telescopes and mounts, and OPL (Levallois), which made more than 10 monochromatic filters for the red line of Hydrogen. One of these instruments (Figure 44) was placed at the Haute Provence Observatory (OHP) where it operated from 1958 to 1994. The original Meudon instrument was renovated several times, with increasingly sophisticated filters, and ran from 1956 to 2004. In particular, tunable filters to record not only the line core, but also the line wings, were built, in order to estimate the motions of plasma by the Doppler effect. A total of 7 million images was taken in Meudon and at the OHP, including 6 million on film (3000 reels of 45 m, or 130 km) and 1 million with a digital CCD detector.

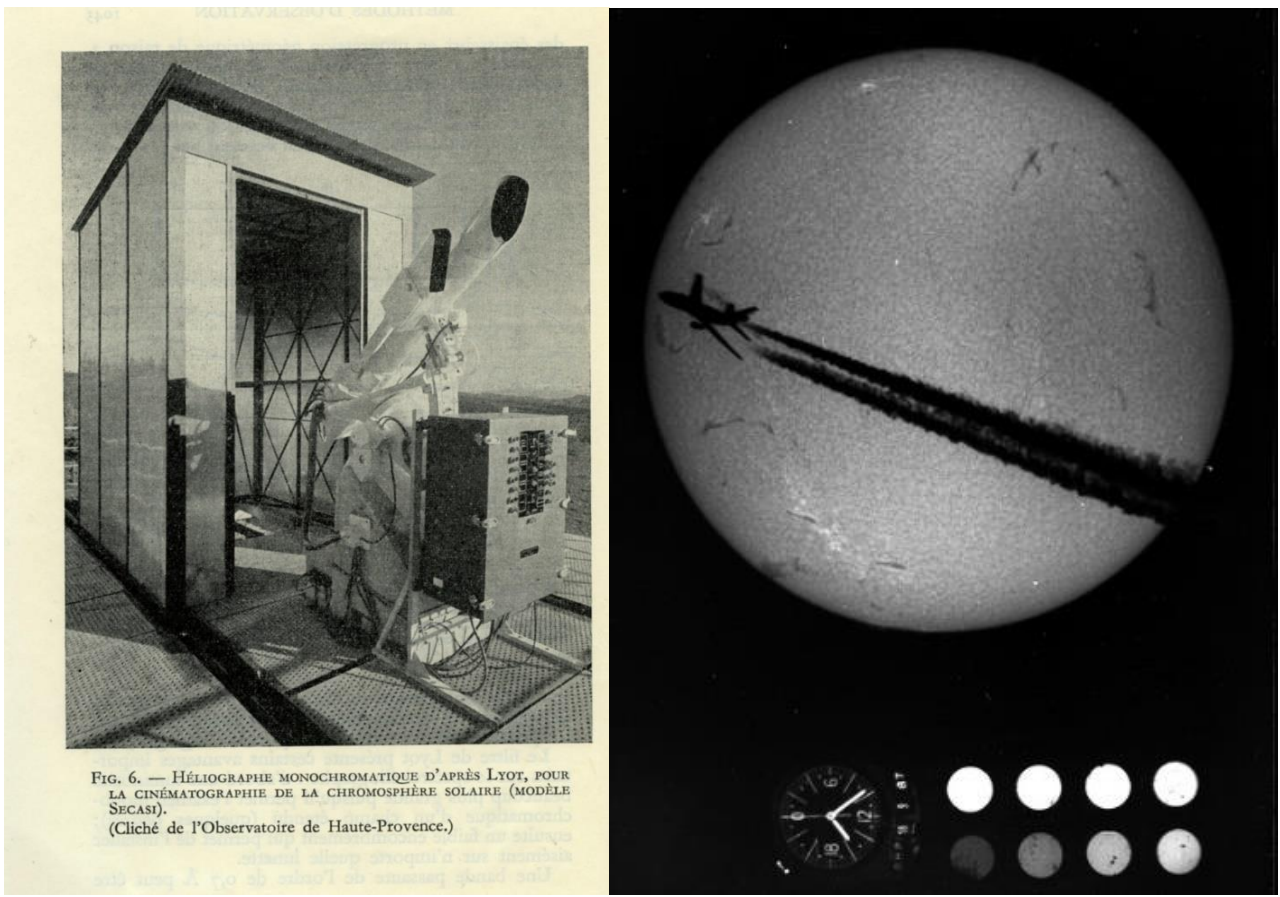

*Figure 44 : heliograph Hα (SECASI/OPL) of the Observatoire de Haute Provence (courtesy OP).*

Let us now turn to radio astronomy. The first solar radio signal was identified in 1942 by James Hey when analyzing signals picked up by the British military radars during the Second World War. The Germans built 1500 Würzburg-Riese giant radars with a diameter of 7.50 m during the war, and a few of them (Figure 45) were recovered after 1945 for scientific purposes.

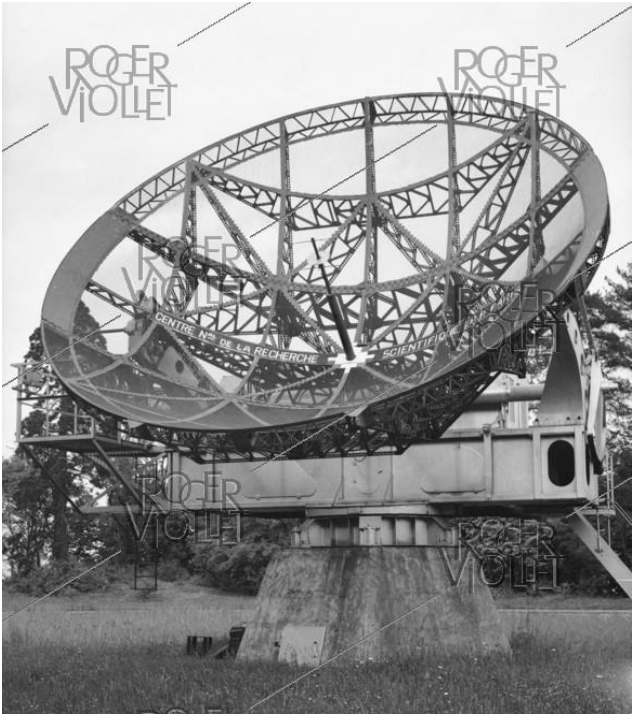

*Figure 45*

*On the left, Würzburg radar installed in Meudon in 1949. Credit: Roger-Viollet.*

*Below, the two Würzburg radars installed in Nançay in 1953 on rails, allowing their distance to be varied for interferometric purpose. Credit: OP.*

*The Meudon radar was later transferred to the Bordeaux Observatory. One antenna remains in Nançay, the second has been donated to the radar museum at Douvres la Délivrande in Normandy, near Caen.*

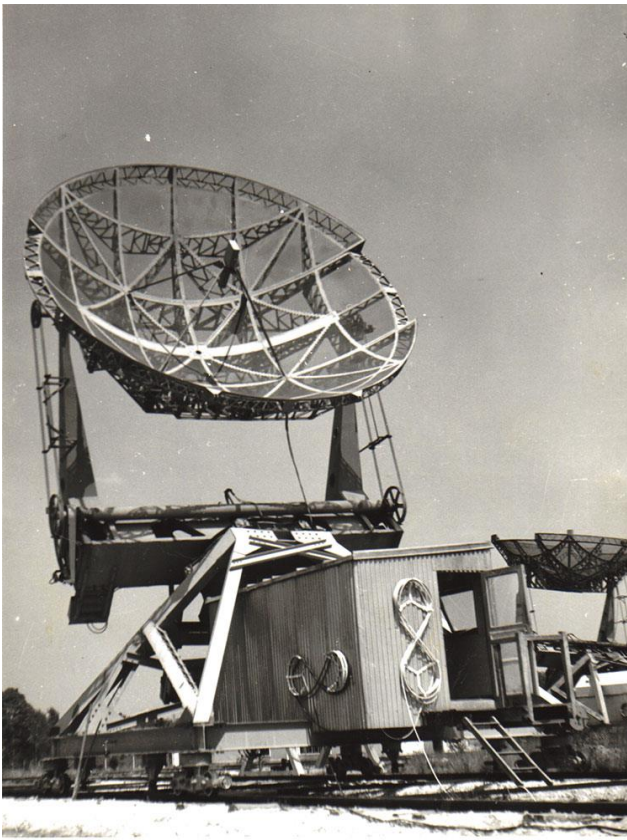
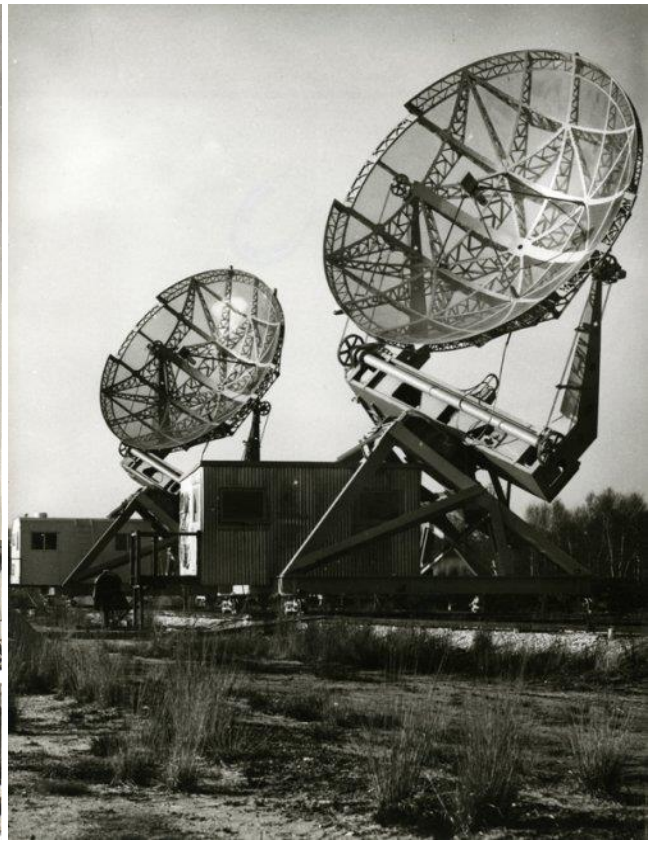

These radars operated at the 55 cm wavelength. The one in Meudon was transformed into a solar radio antenna by Marius Laffineur in 1949. The Nançay radio astronomy station was created shortly afterwards, in 1953, at the instigation of Yves Rocard ("Ecole Normale Supérieure"), who supported Jean-François Denisse, Jean-Louis Steinberg and Jacques-Emile Blum in this innovative task to open a new window on the universe.

In parallel with ground-based observation, the appearance of balloons encouraged astronomers to try new and outstanding experiments. The balloon can be used to rise to high altitudes, above most of the Earth's atmosphere, making it possible to capture radiation that does not reach the ground, or to improve the quality of images by eliminating seeing effects (this last goal can be complicated by the vibrations of the balloon). The first flight undertaken by an astronomer dates back to December 2, 1870. Paris being besieged by the Prussians, Jules Janssen obtained from the government the provision of a balloon to escape the enemy lines and to observe an eclipse of the Sun in Algeria, at Oran (Figure 46). This was the only scientific expedition of the 70 balloons during the siege of Paris.

*Figure 46*

*Below, the railway station "Gare d'Orléans" in Paris is transformed into a balloon assembly hangar during the siege of Paris in 1870.*

*On the right, the stele of the statue of Janssen on the first terrace of Meudon, located below the observatory. Janssen takes off aboard the balloon "le Volta" with his instruments for the observation of a total eclipse in Oran, Algeria.*

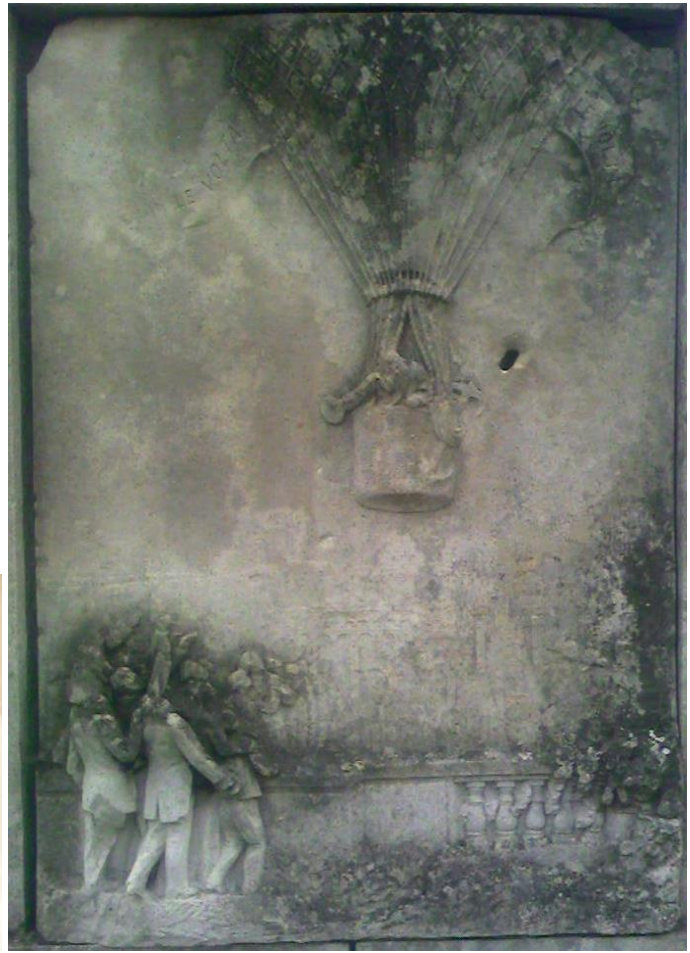
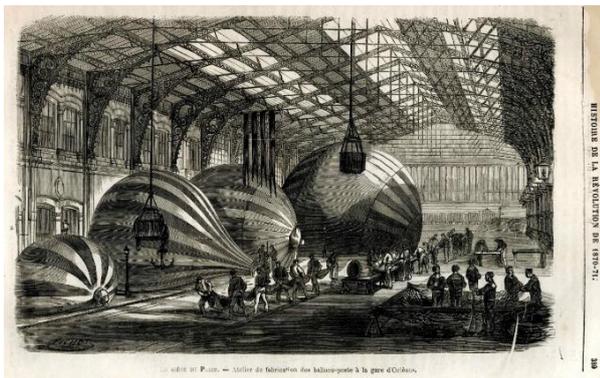

Mounted balloons are not dirigible, but subject to winds. Janssen took off at night, discreetly, accompanied by a military pilot and his scientific equipment, he made an perfect flight driven by the wind that blew in the direction of the Atlantic Ocean, and sometimes was driving the balloon himself. He landed in the vicinity of Saint Nazaire. Here is what he later wrote about his expedition: "A special train took me to Nantes and from there I went to Tours, where I arrived at 11 o'clock in the evening. I had left Paris at 6 a.m. From Tours I travelled to Bordeaux and Marseille where I embarked for Oran, where the rumour soon spread that a great French gastronome had just arrived in Algeria" !  Unfortunately, this extraordinary adventure remained unsuccessful due to cloud coverage on the day of the eclipse... The balloon "Le Volta" was later kept in Meudon, then donated by Mrs. Janssen to the Army Museum after the death of her husband in 1907.

A few years later, Aymar de la Baume Pluvinel, a distinguished amateur astronomer and assiduous collaborator of Jules Janssen, instrumented a balloon without any passenger (Figure 47) which reached the altitude of 9000 m and was retrieved in the countryside after the descent. The aim was to deepen the research carried out at Mont Blanc on the oxygen lines of the solar spectrum. The balloon carried an automatic spectroscope with a photographic plate that corroborated the observations of Mont Blanc, namely the decrease in intensity of these lines, which could now be attributed with certainty to the Earth's atmosphere.

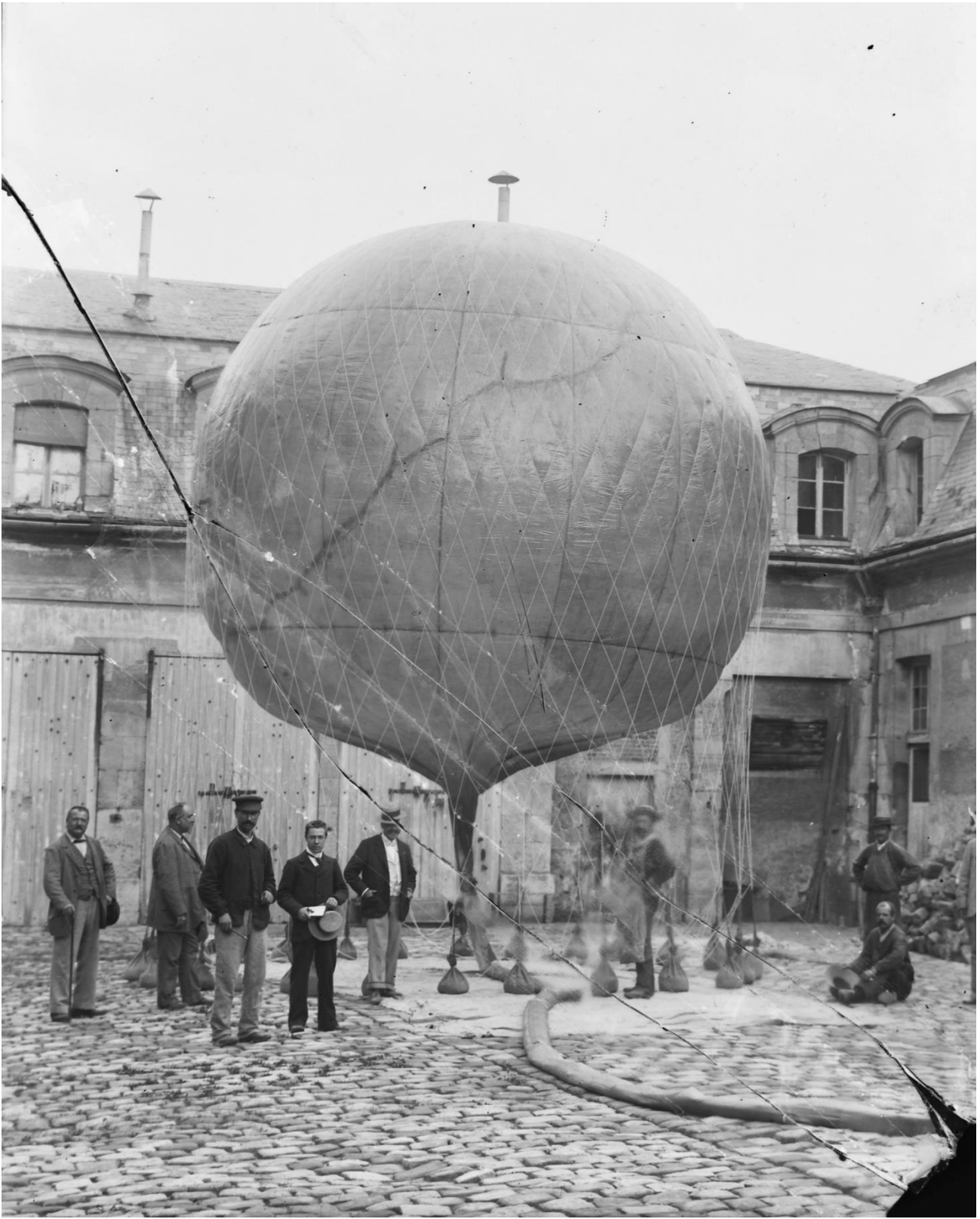

*Figure 47 : Aymar de la Baume Pluvinel's instrumented balloon, here shown in the courtyard of the outbuildings of the Château in Meudon, reached an altitude of 9000 m in 1899 (credit OP).*

The following adventures were the work of Audouin Dollfus, a famous aeronaut, just like his father Charles Dofffus. In 1956, Dollfus undertook the first observation of the Sun's surface using a large telescope attached to the balloon's basket, which rose to an altitude of 6000 m with its passenger (Figure 48). Why this attempt ? On the ground, images could hardly be better than the (already beautiful) granulation shots of the Janssen atlas, because the Earth's atmosphere is turbulent and blurs astronomical observations. If we want to improve the quality of the images, at this epoch, the only way was to climb the highest mountains, or better to use balloons, in order to reduce as best as possible the column of air above the observer.

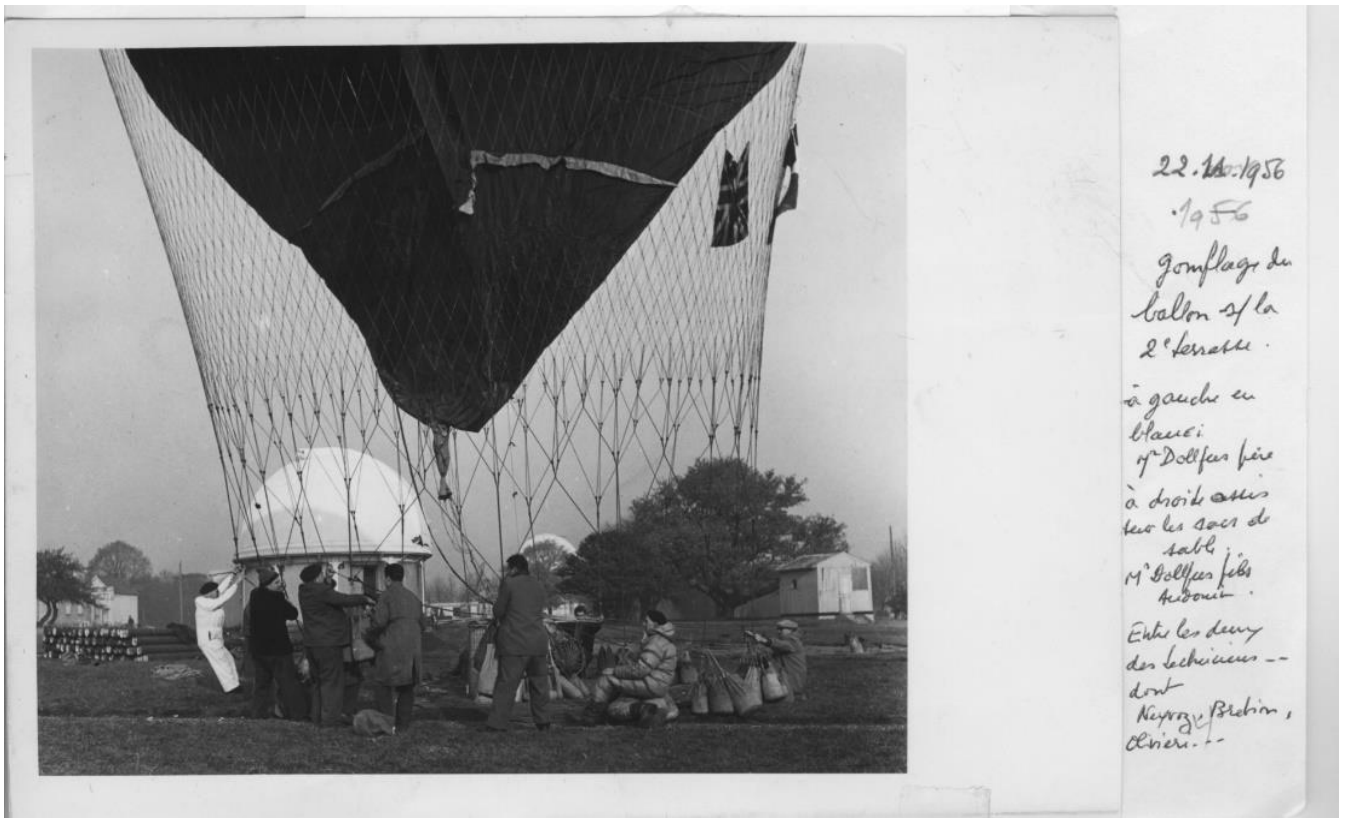

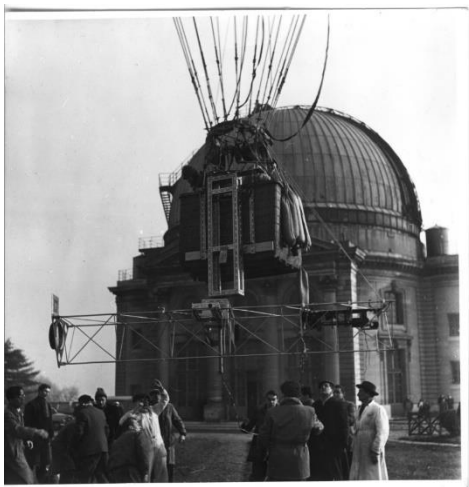

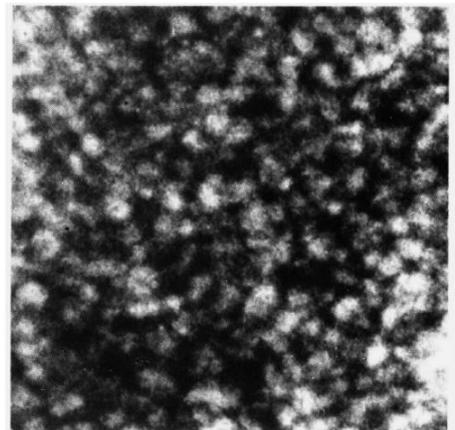

Fig. 226. – La première photographie astronomique réalisée hors de la surface terrestre : – la granulation solaire en ballon à l'altitude de 6 000 mètres, le 1er avril 1957.

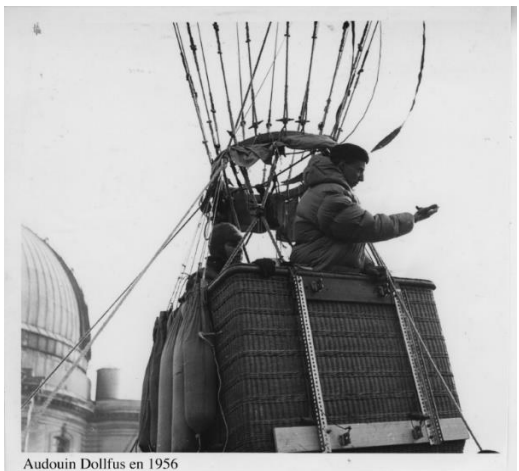

*Figure 48*

*Audouin Dollfus observes the solar granulation at the altitude of 6000 m. A 30 cm diameter refractor with a focal length of 3 m is supported by the basket and the instrument is controlled by the aeronaut. This experiment allowed Dollfus to bring back new photographs whose finesse was unequalled, thanks to the reduction of disturbances caused by the turbulence of atmospheric air. A performance ! Credit OP.*

The observation of the solar granulation, at the altitude of 6000 m, was a great success. But Dollfus dreamed of an experiment at the altitude of 25000 m with an instrumented telescope controlled from a pressurized capsule carrying the operator. Although this project concerned nocturnal astronomy, we mention it because it is in the continuation of the solar observation project at 6000 m. For the very high altitude, Dofffus turned to the multi-balloon technology with a long cluster of 100 small balloons. He also tested expandable balloons (Figure 49) under the Hangar Y at Chalais-Meudon (a former airship hangar now restored). Dollfus flew in 1959 from Villacoublay and reached the altitude of 14000 m with the capsule and the telescope. Even if it was not the expected 25000 m, it was already a famous record, related by an on-line film at the Institut National de l'Audiovisuel (Figure 50).

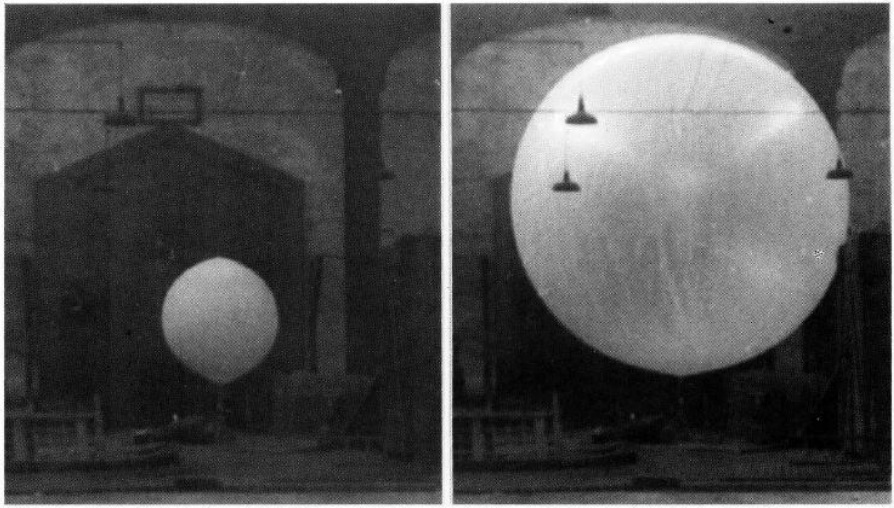

*Figure 49*

*Expandable balloons tested by Audouin Dollfus in the Hangar Y in Chalais-Meudon with the aim of reaching an altitude of 25000 m.*

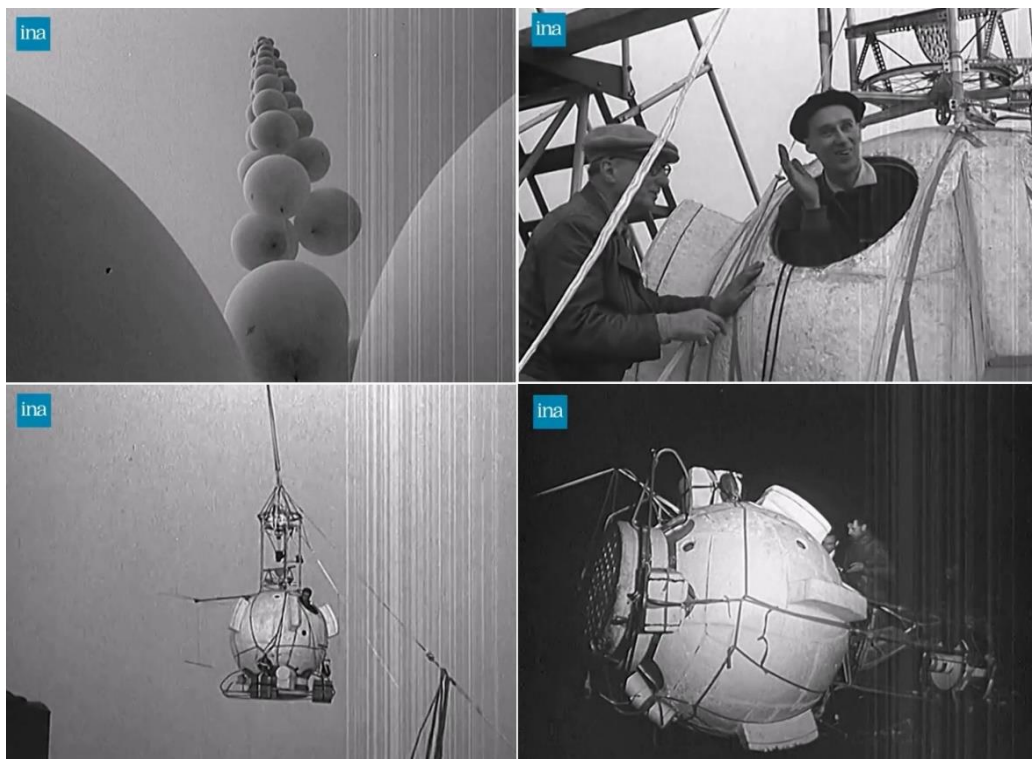

*Figure 50 : capsule and telescope of Audouin Dollfus, 1959 flight, film on-line on the INA website*

*Extract : https://www.lesia.obspm.fr/perso/jean-marie-malherbe/HY/Media4.mp4*

https://www.ina.fr/ina-eclaire-actu/video/afe85008273/audouin-dollfus-monte-a-13000-metres-en-ballon

Dollfus can thus be considered a pioneer of space astronomy in France. The adventure continued with the CNES agency for space after its creation in 1961. An automated and external occultation coronagraph was launched to observe the corona during several (and unmanned) flights at the altitude of 32000 m (Figure 51).

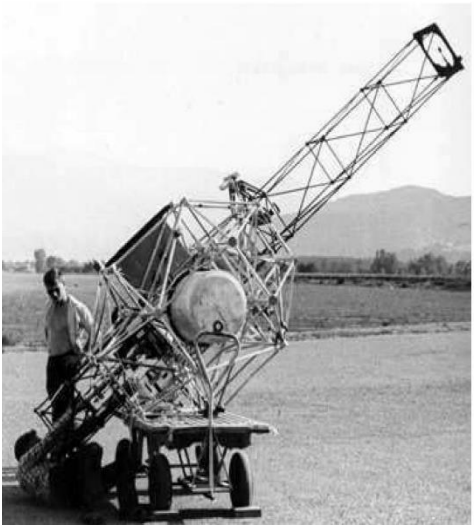
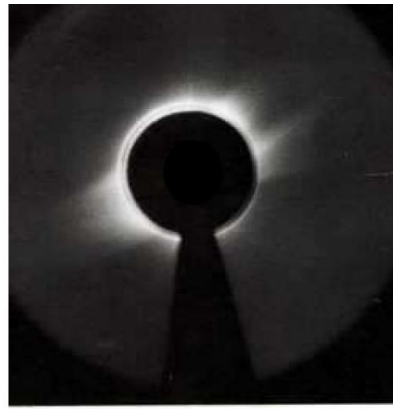

*Figure 51*

*Audouin Dollfus' external occultation coronagraph launched from Aire sur Adour, south of France, in 1973 with the CNES support. Credit OP.*

## 4 – After 1960: solar observations from space open a new window on the Sun

After 1960, large, powerful but small field of view, ground based instruments were built (Figure 52).

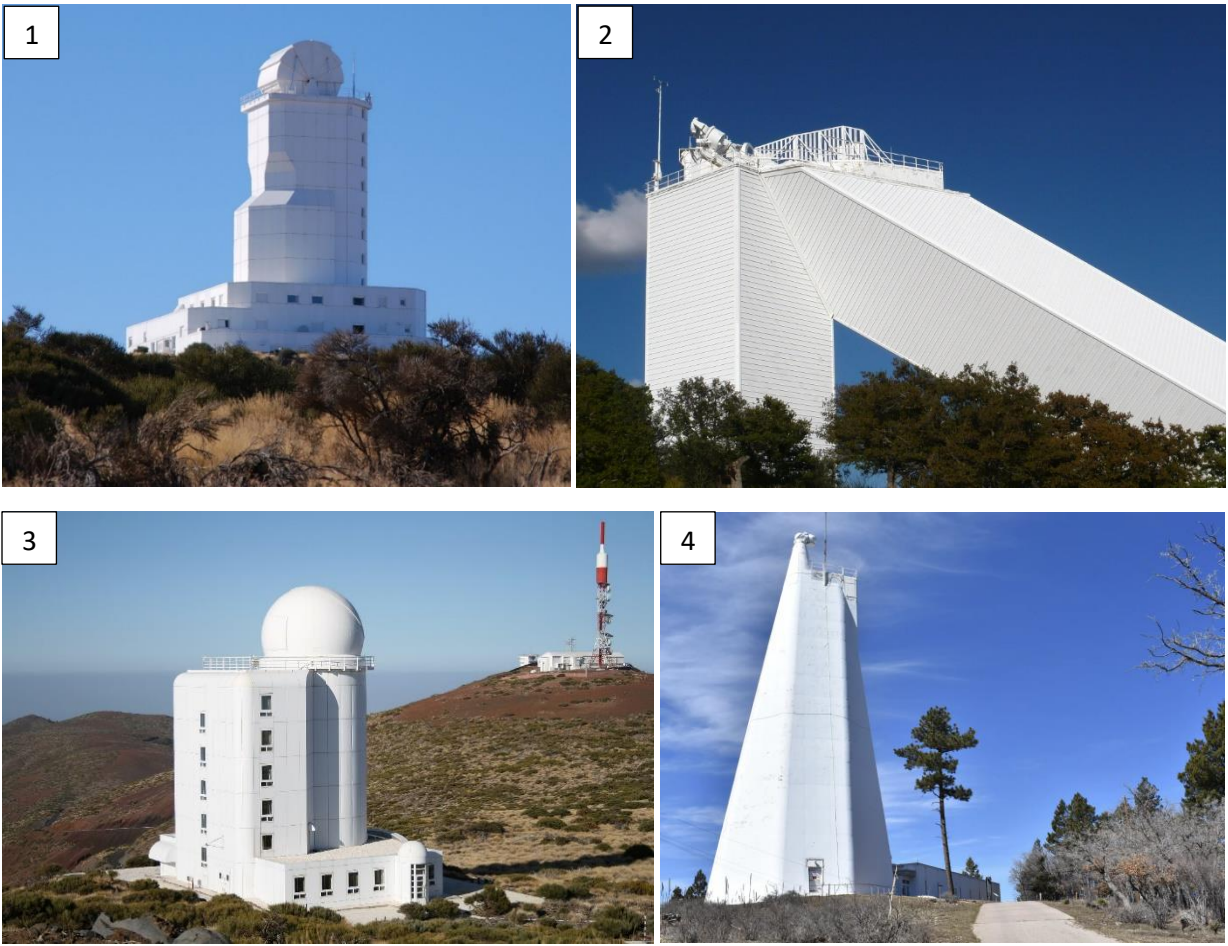

*Figure 52 : some examples of large solar telescopes built after 1960, (1) Vacuum Tower Telescope (Germany), (2) Kitt Peak (USA), (3) THEMIS (France), (4) Sacramento Peak (USA)*

These large telescopes often have a reduced field of view (a few arc minutes) and are designed to resolve small details of the surface (Figure 53), especially with adaptive optics at the turn of the year 2000. The largest solar telescope today is the American DKIST with a mirror of 4 m in diameter.

*Figure 53 : solar granulation observed by Janssen 1880 (1), at the Pic du Midi around 1970 (2), and by DKIST in Hawaii in 2020 (3)*

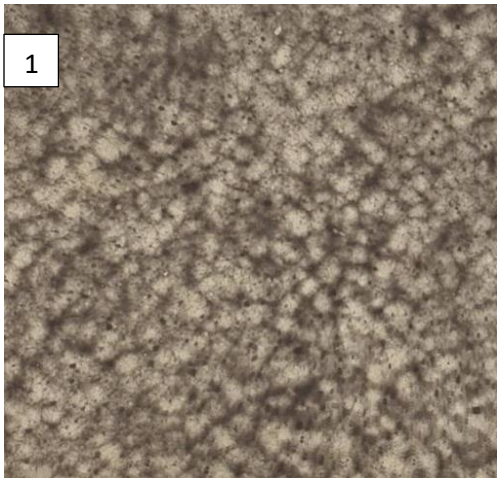
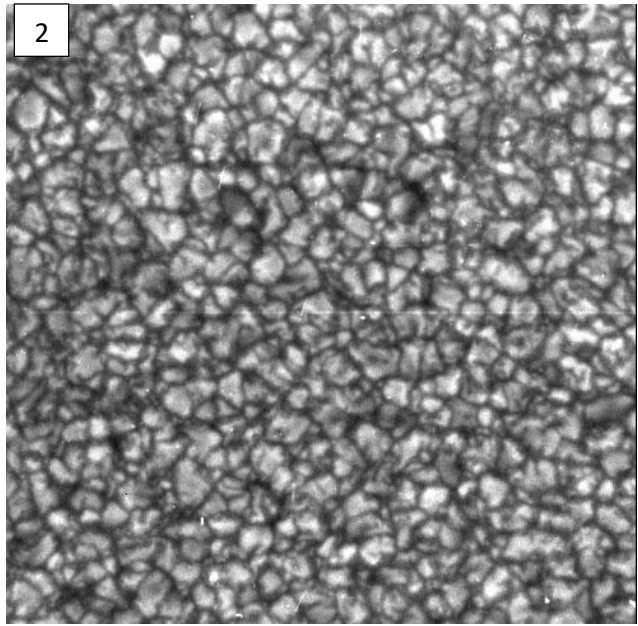
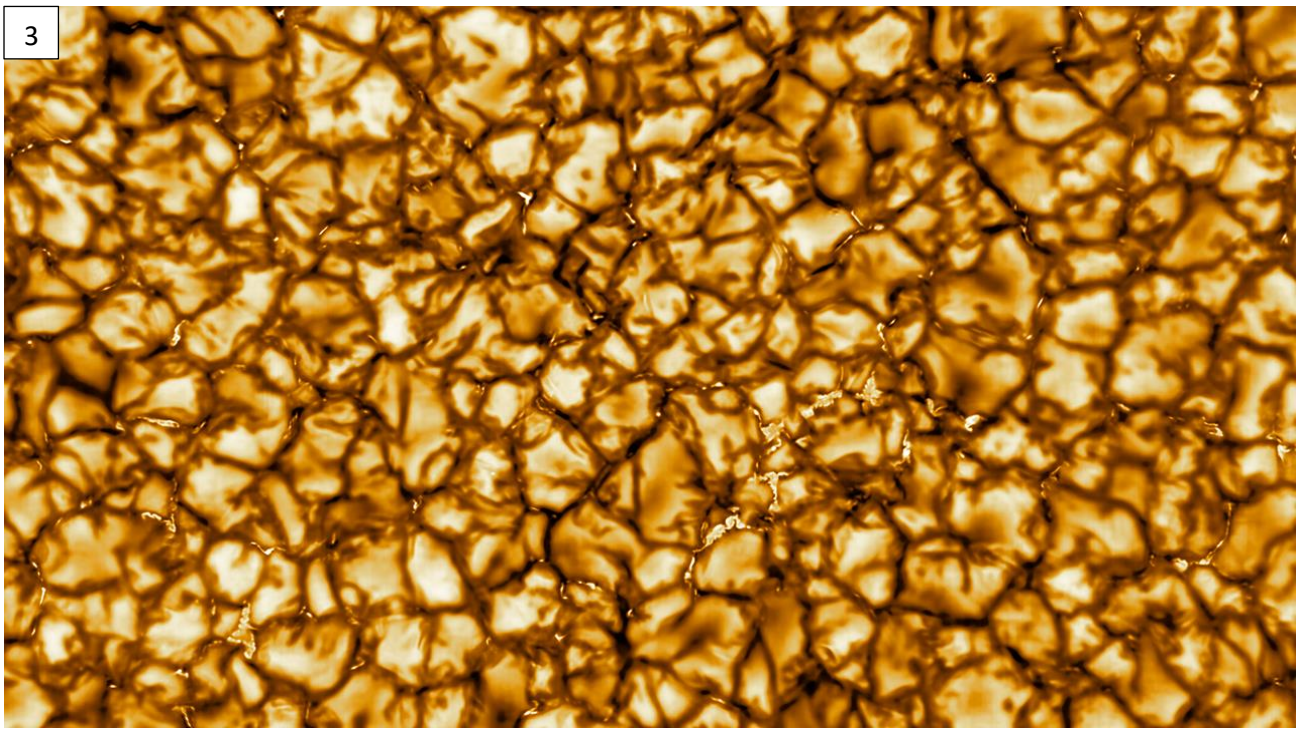

Space solar astronomy developed with short-lived rocket launches and orbiting satellites that opened a new window on the Sun, thanks to the observation of new radiations: Ultra Violet (UV), characteristic of hot plasmas in the corona (100000 to 10 million degrees, Figure 54), X-rays (Figure 55) and γ-rays as a signature of high-energy phenomena present in solar flares. An instrument (SOHO) was placed in 1996 at the L1 Lagrange point, 1.5 million km from Earth, to observe the Sun continuously; it is still in operation in 2024. A two-probe mission (STEREO A and B, 2006), orbiting around the Sun following the Earth's trajectory provides a stereoscopic view of our star (sometimes both hemispheres were seen at the same time). SDO delivers since 2010 high cadence images (45 s) in UV. Interplanetary probes approaching the Sun (Parker Solar Probe, PSP 2018, Solar Orbiter, SolO 2020) have the main objective of carrying out "in situ" measurements of physical parameters (densities, electro-magnetic field); PSP enters the corona at only 10 solar radii, while SolO leaves the plane of the Earth's orbit (the ecliptic) to better observe the solar poles.

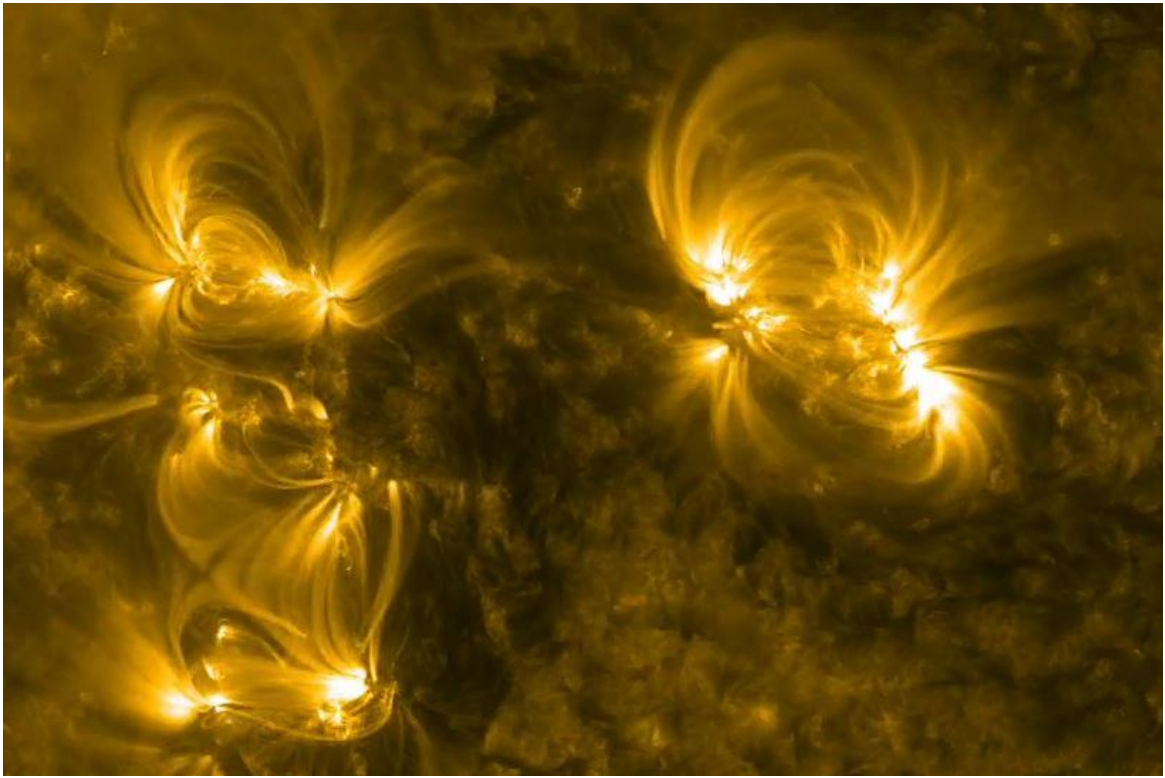

*Figure 54 : the Sun and its magnetic loops in UV, 1 million degrees (SDO/AIA), with animation:*

*https://www.lesia.obspm.fr/perso/jean-marie-malherbe/HY/Media5.mp4*

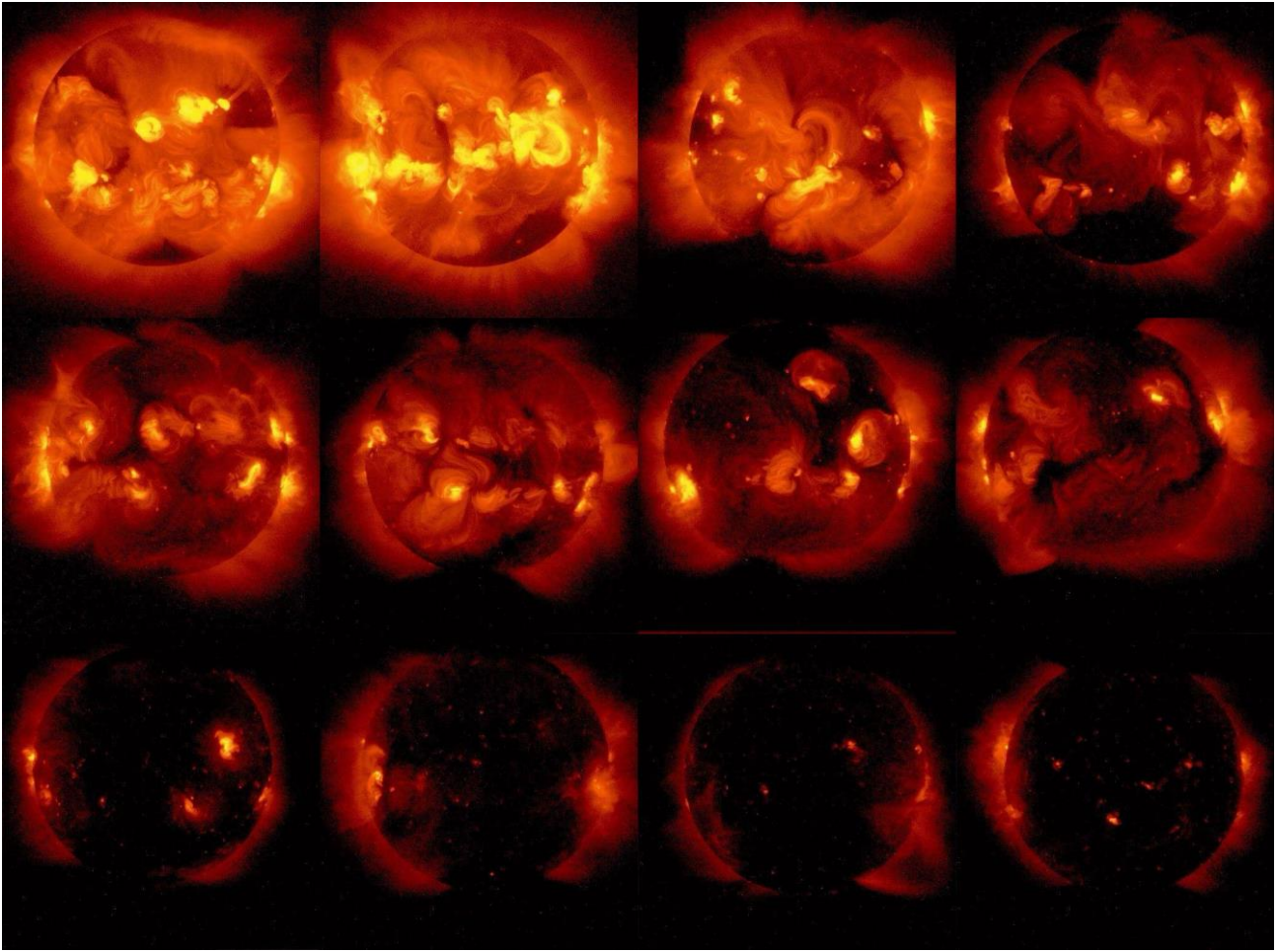

*Figure 55 : the X-ray solar cycle as seen by the Japanese satellite YOHKOH from 1990 to 1999. Credit JAXA.*

The American geostationary satellite Solar Dynamics Observatory, launched in 2010, which observes in Ultra Violet with more details than the older European SOHO, provides fantastic animated sequences of solar activity, including flares and coronal mass ejections, as shown in Figures 56, 57, 58, 59 and 60, thanks to its unprecedented frame rate. The figures are accompanied by associated animations.

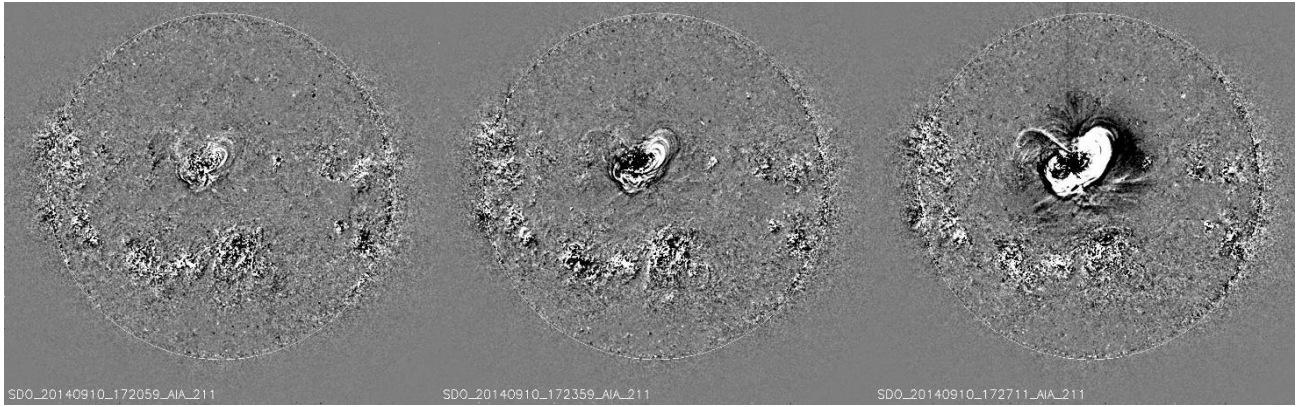

*Figure 56 : coronal shock wave triggered by an eruption sweeping the lower corona at the speed of 1000 km/s. SDO/AIA, animation: https://www.lesia.obspm.fr/perso/jean-marie-malherbe/HY/Media6.mp4*

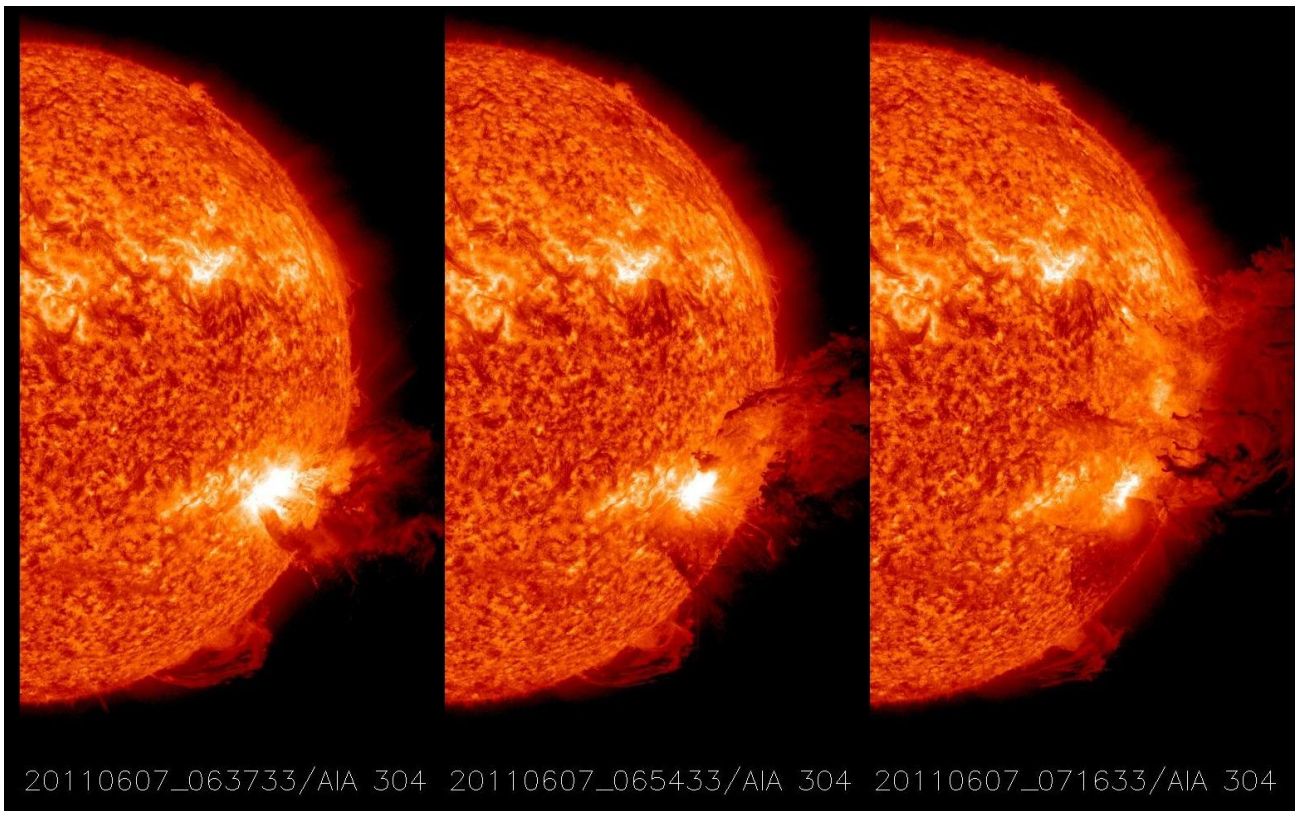

*Figure 57 : coronal mass ejection in ionized Helium, 100000 degrees. SDO/AIA, animation : https://www.lesia.obspm.fr/perso/jean-marie-malherbe/HY/Media7.mp4*

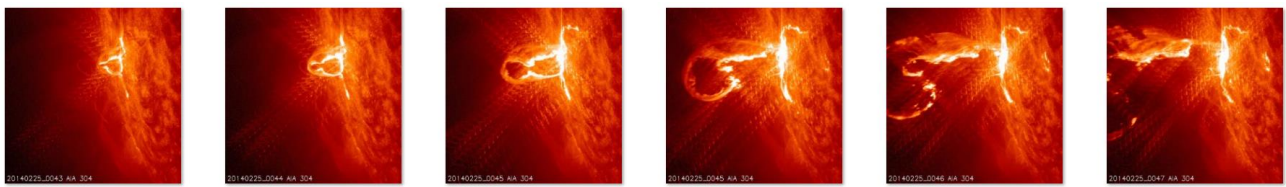

*Figure 58 : ejection of a magnetic structure visible in ionized Helium, 100000 degrees. SDO/AIA, animation : https://www.lesia.obspm.fr/perso/jean-marie-malherbe/HY/Media08a.gif*

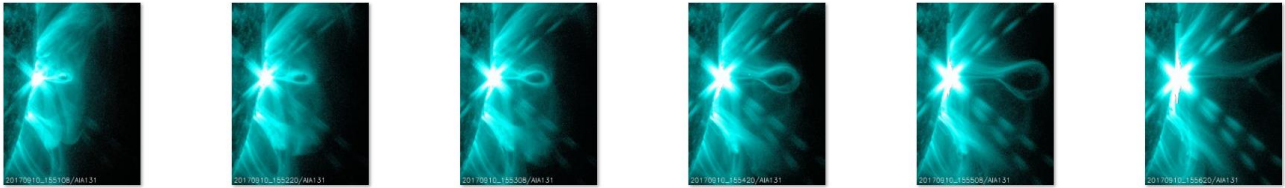

*Figure 59 : ejection of a plasma bubble, highly ionized Iron, 10 million degrees. SDO/AIA, animation :*
*https://www.lesia.obspm.fr/perso/jean-marie-malherbe/HY/Media08b.gif*

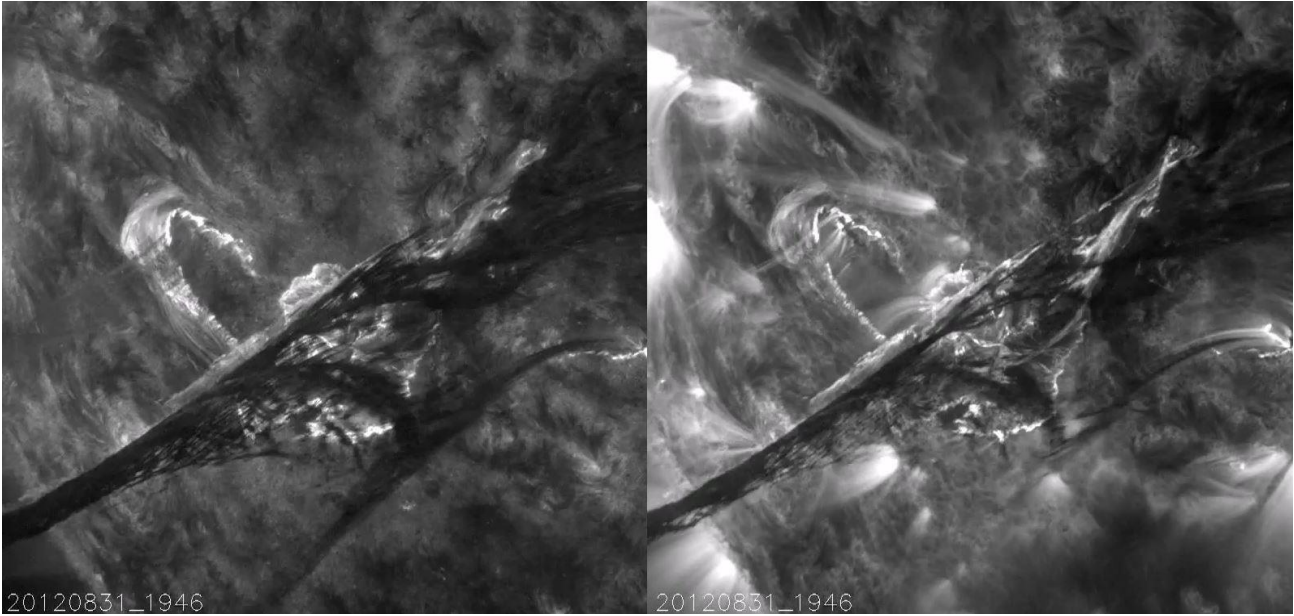

*Figure 60 : Ejection of a filament followed by a two-ribbon flare and formation of post-eruptive magnetic arches, two lines for two temperatures, 0.1 and 1 million degrees. SDO/AIA, animation :*
*https://www.lesia.obspm.fr/perso/jean-marie-malherbe/HY/Media8.mp4*

Figure 61 shows the trajectory of the European Solar Orbiter mission, which travels inside the orbit of Mercury and gradually leaves the plane of the ecliptic to observe the solar poles, which are not easily visible from Earth. The instrument performs remote observations and "in situ" measurements. Another mission, the American Parker Solar Probe (Figure 62), measures physical parameters directly in the solar corona, which was never done, much closer to the Sun than Solar Orbiter.

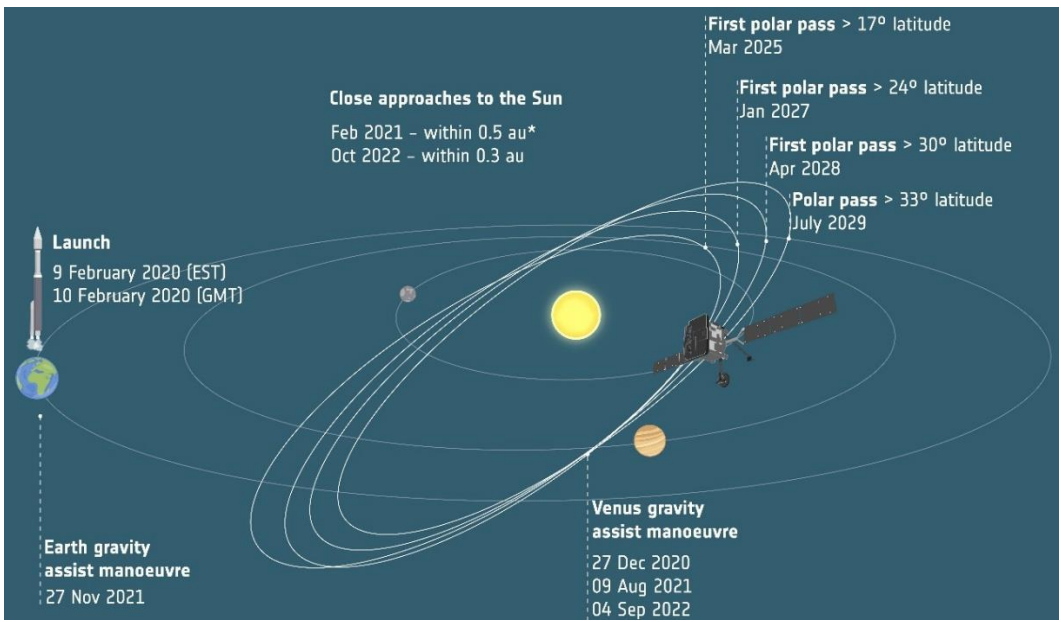

*Figure 61*

*Trajectory of the Solar Orbiter probe that leaves the plane of the ecliptic to better see the solar poles. Credit: ESA.*

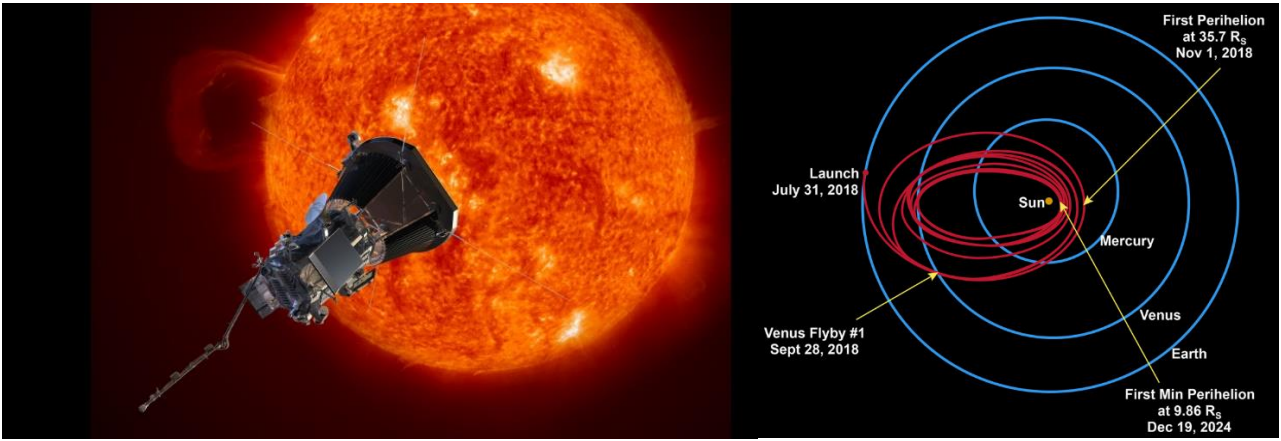

*Figure 62 : American Parker Solar Probe and its trajectory around the Sun. Credit: NASA.*

Let us conclude this brief overview of space-based solar observation with the wide-field vision provided by the LASCO coronagraphs carried by SOHO (ESA/NASA), whose longevity is exceptional (28 years). In space, you can see much farther than with ground-based coronagraphs. SOHO's field of view is 30 solar radii, in white light (Figure 63), which makes it possible to monitor the injection of solar material into the interplanetary medium and to estimate its possible impact at the Earth.

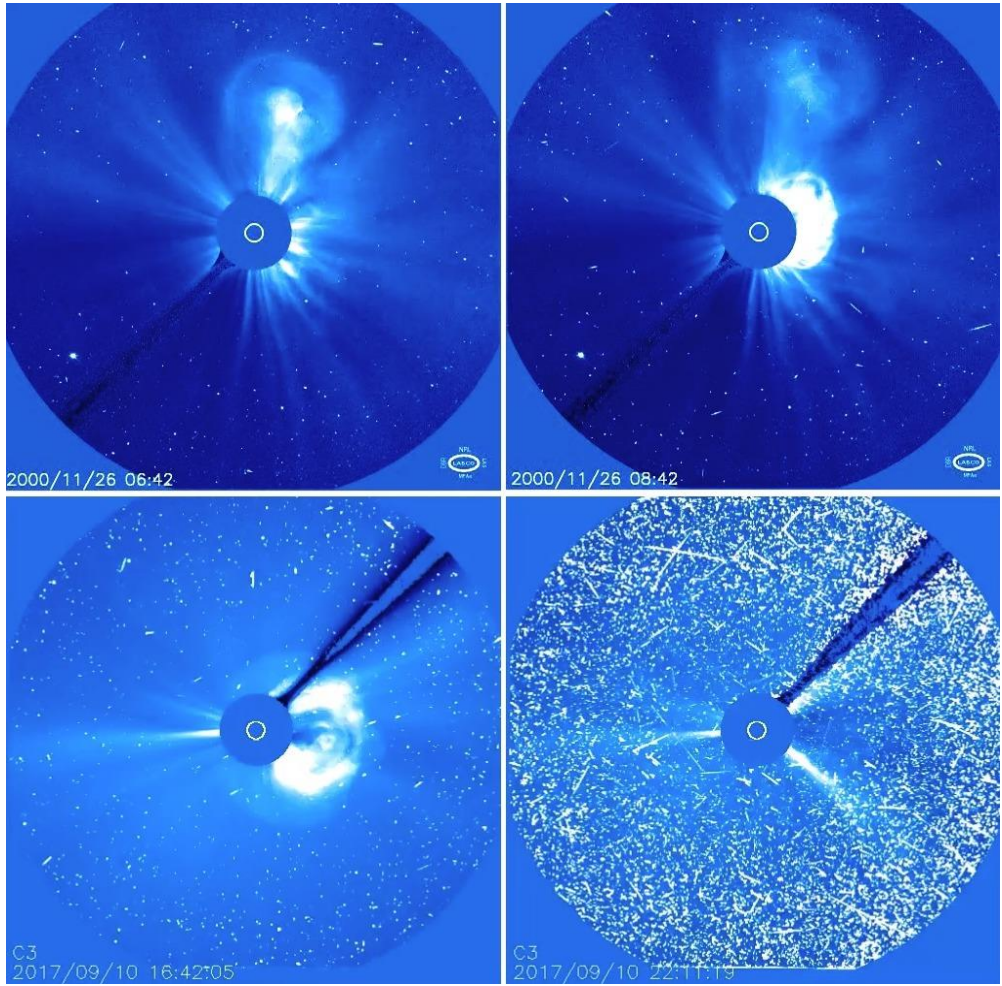

*Figure 63 : SOHO C3 white light coronagraph. Top: an event with no impact at the Earth. Bottom: an event with an impact at the Earth (please notice the trace of cosmic particles). Animations:*

*Top : https://www.lesia.obspm.fr/perso/jean-marie-malherbe/HY/Media9.mp4*

*Bottom : https://www.lesia.obspm.fr/perso/jean-marie-malherbe/HY/Media10.mp4*

The Sun, as shown by the C2 and C3 coronagraphs carried by SOHO at the L1 Lagrange point, can therefore have an impact at the Earth, by injecting charged particles transported by the solar wind that bombard the Earth's atmosphere. The most beautiful natural phenomenon is that of the polar auroras (Figure 64), which exist on all planets with a magnetic field (which traps particles) and a gaseous atmosphere or surface.

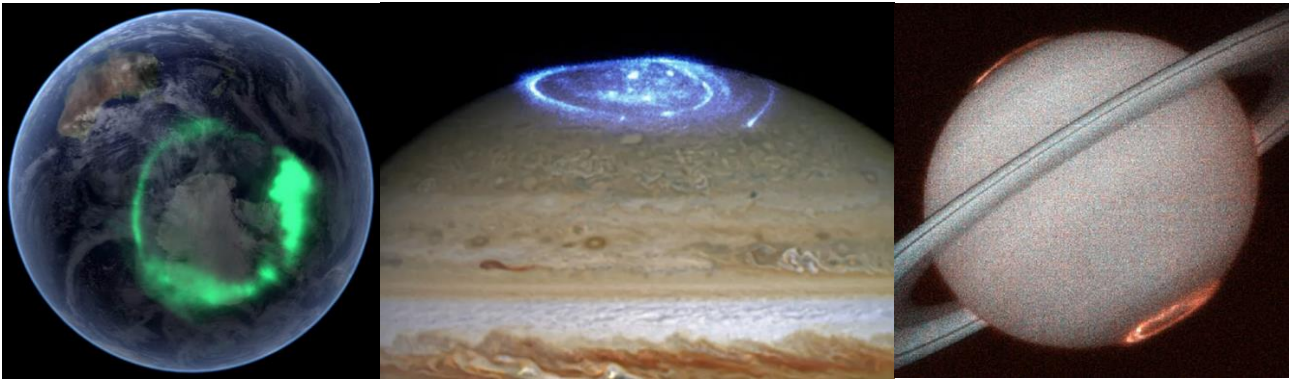

*Figure 64 : polar auroras on the planets Earth, Jupiter and Saturn. Credit NASA.*

## 5 – Supercomputers to the rescue of telescopes: the birth of a new "instrument" dedicated to numerical modelling

The development of computers from the 1970s onwards made it possible to move to digital data processing. Until the 1980s, those of optical telescopes were digitized by micro-densitometers and recorded on magnetic tapes. The radio data were plotted and then written on tapes. In orbiting instruments, digital sensors were used early to transmit data to the ground by telemetry. One saw soon an explosion in the volume of data produced in astronomy, so that large computers became essential tools, as indispensable as telescopes. This led to new needs for numerical calculations, because observations and modelling compete together. The CNRS in France then created a shared computing centre within the National Institute of Astronomy and Geophysics (INAG), the forerunner of the National Institute of Sciences of the Universe (INSU). It was equipped with an IBM 360/65 located in Meudon (figures 65 and 66) and operated until the end of the 70s. At this date, the centre was dismantled to make way for interactive computing (VAX under VMS from 1980, then subsequently UNIX systems). Laboratories belonging to INAG could connect to this national means via dedicated lines and terminals, consisting locally of a printer, a card reader, sometimes connected to a front-end computer such as the IBM 7040.

| ORDINATEUR | CAPACITÉ MÉMOIRE | POSITION GÉOGRAPHIQUE |
|---|---|---|
| I.B.M. 7 040 | 32 K* | Observatoire de Nice |
| Télémécanique Solar | 80 K (16 bits) | Observatoire de Nice |
| I.B.M. 360 /65 | 512 K en rapide<br>2 048 K en lente | Observatoire de Meudon |
| I.B.M. 370 /168 | 3 000 K (plus mémoire virtuelle) | C.I.R.C.É. Orsay |
| I.B.M. 370 /168 | 4 000 K (plus mémoire virtuelle) | C.I.R.C.É. Orsay |

\* Le K est une mesure de la capacité d'une mémoire électronique qui vaut $2^{10} = 1\,024$ mots ; chaque mot étant lui-même formé d'un certain nombre de bits (unité élémentaire d'information, qui ne peut prendre que deux valeurs distinctes, 0 ou 1). Les mots sont en général de 8, 12, 16 ou 32 bits.

*Table 1 : Features of IBM machines accessible to French astronomers in the mid-1970s*

The IBM 360/65 had 512 KB of fast memory (1 microsecond cycle), and 2048 KB of second-tier memory, which was slower (8 micro seconds), on 24 bits of addressing. There was no virtual memory. It performed about 0.6 million instructions per second. The most demanding calculations, for example to solve, in solar physics, the partial differential equations of fluid dynamics coupled with those of electromagnetism, were carried out on the large IBM 370/168 of the CIRCE ("Centre Inter Régional de Calcul Electronique" of the CNRS in Orsay, a multidisciplinary computer centre that opened in 1969), faster, equipped with a virtual memory system, with 4000 KB of main memory (see Table 1 above).

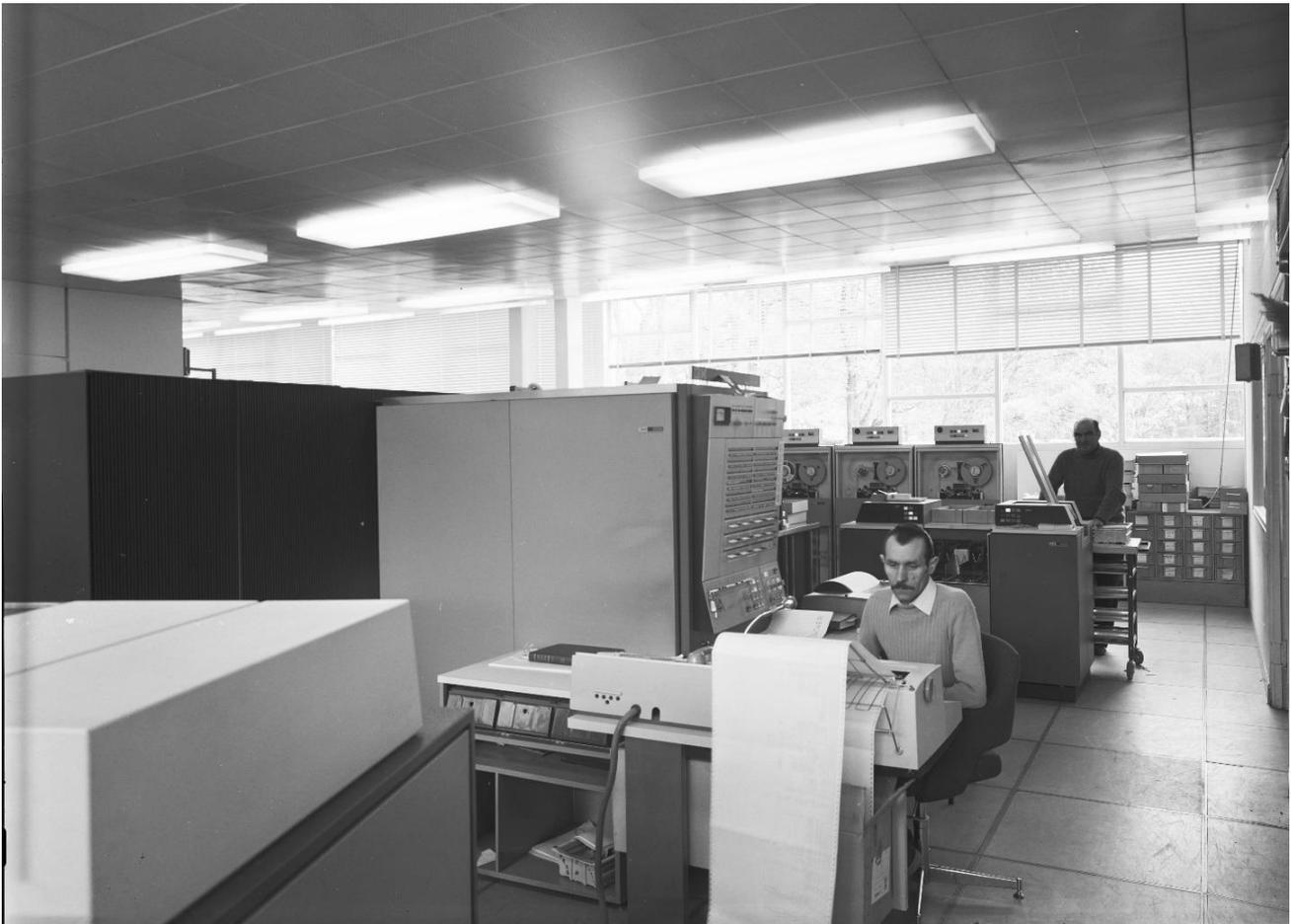

*Figure 65 : the IBM 360/65 of INAG at Meudon Observatory in the 70s (credit OP)*

At the time of the IBM 360/65, in the 70s, a room in Meudon housed many card punches used both to memorize FORTRAN code (one instruction per card) or astronomical data. 800/1600 bits per inch magnetic tape devices were also used to store data (20 to 40 MB per tape depending on density). All the work was done in batch mode, i.e. within a queue, in deferred time. There was a desk, which I knew in 1979, where you dropped your packet or box of punched cards. Results were later returned in the form of printer "listings", or even "Benson" drawings with an ink pen. This mode of operation, without any interactivity, would be discouraging today, but what a progress at that epoch ! Any programming error resulted in a rejection of the program at the compilation step, and you had to restart all: punching cards, submitting for new compilation, waiting ! Developing a complex program was time-consuming, but it was unavoidable.

The INAG computing centre was shut down at the end of the 1970s following the development of interactive computing in laboratories (dominated by American computers, first with 16-bit PDP 11 machines during the seventies, followed by 32-bit VAX computers with virtual memory after 1980, in particular the mythical VAX 11/780). In addition, the needs for high-performance computing (numerical simulations) were increasingly and covered in France by the CNRS computing centre (CIRCE), with machines (mainly IBM mainframes) that researchers could access remotely via terminals or specialised links.

A vector computing centre for research (CCVR equipped with the mythical CRAY 1) was then created and it merged with the CIRCE to form a joint centre, called IDRIS in 1993, which has continued to specialise in high-performance computing (HPC), with massively parallelised machines (from companies such as HPE, IBM or Bull) since the 2000s.

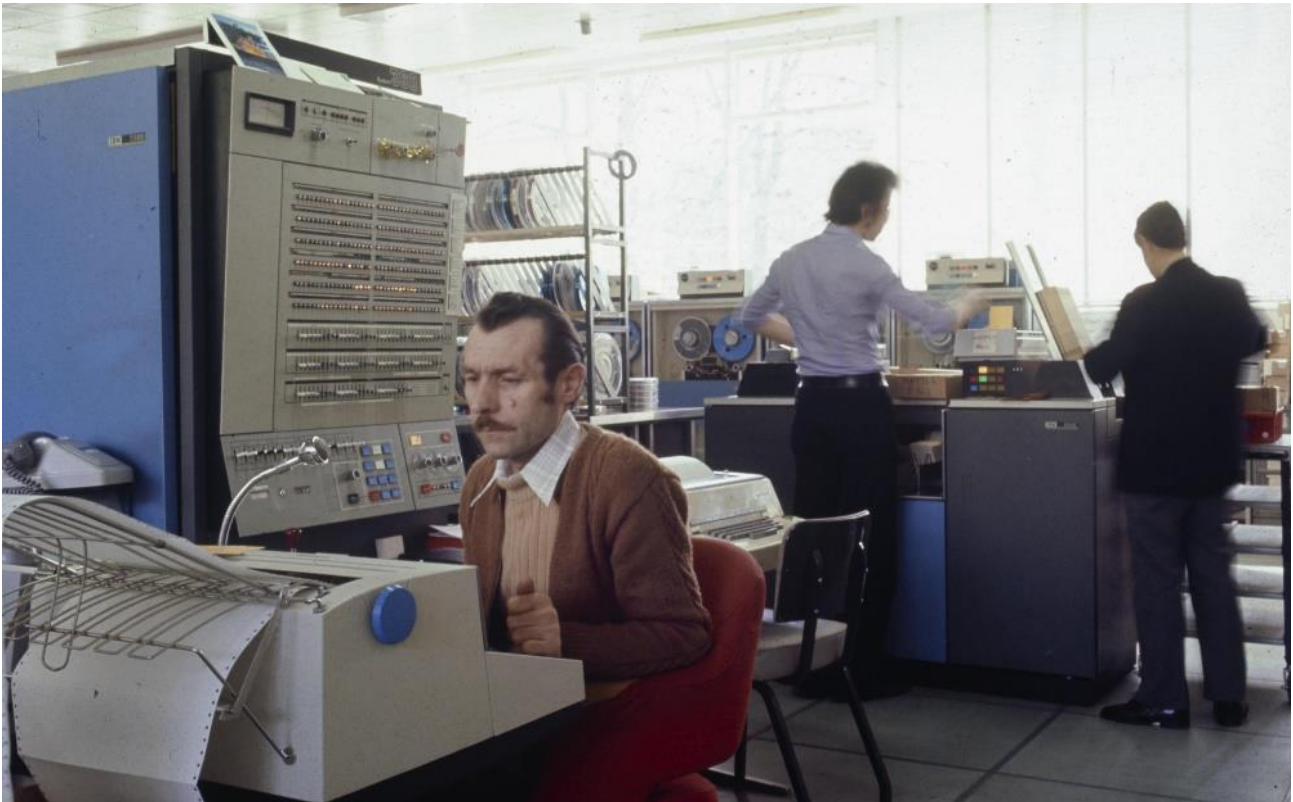

*Figure 66 : The IBM 360/65 of Meudon and its control panel. In the background, you can see the magnetic tapes and the punched card reader, with a large packet of cards being read. Credit: OP.*

| FAMOUS SUPERCOMPUTERS | | | | | |
|---|---|---|---|---|---|
| Date | Name | FLOPS | Memory | Storage | |
| 1961 | IBM 7030 | 1 Mflops | 2 Mo | | *Mflops Bar* |
| 1975 | Cray 1 | 160 Mflops | 8 Mo | 300 Mo | |
| 1985 | Cray 2 | 2 Gflops | 2 Go | | *Gflops Bar* |
| 1997 | IBM ASCI RED | 1 Tflops | 1 To | | *Tflops Bar* |
| 2008 | IBM RoadRunner | 1 Pflops | 100 To | | *Pflops Bar* |
| 2022 | HPE Frontier | 1 Eflops | 10 Po | 700 Po | *Eflops Bar* |

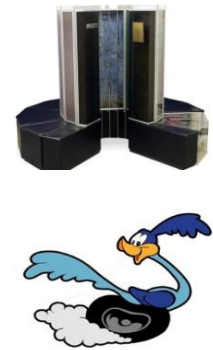

*Table 2 : List of some famous supercomputers*

Table 2 and Figure 67 show the evolution of the computer power in FLOPS (number of floating-point operations per second) of the fastest (and most expensive) supercomputers since 1960, used for numerical modelling in the world. The MFLOPS bar ($10^6$ FLOPS) was reached by the "stretch" IBM7030 in 1961 and then the CDC6600 and 7600 (from Control Data Corporation). That of the GFLOPS ($10^9$ FLOPS) by the Cray2 in 1985. That of the TFLOPS ($10^{12}$ FLOPS) by ASCI RED (IBM) around 1995. The PFLOPS ($10^{15}$ FLOPS) by Road Runner (IBM) in 2008. And finally, the EFLOPS bar ($10^{18}$ FLOPS) was reached by FRONTIER (HPE company) in 2022. The exceptional computing power available since 2005 is due to multi-core processors and massive parallelism (several million cores on exascale machines), and the integration of GPU accelerators (Graphic Processing Units) to support conventional CPUs (Central Processing Units) in vector or matrix mode. For example, on recent HPE machines, each AMD CPU (64 cores) is paired with 4 GPUs (each one with 220 cores).

Between the laboratory machines and the national centres (as those of table 2), there are intermediate facilities at the scale of one or more universities, such as the parallel machine on the Meudon campus (MesoPSL) which serves the PSL University and currently offers a considerable power, close to 300 TFLOPS with its 7500 cores.

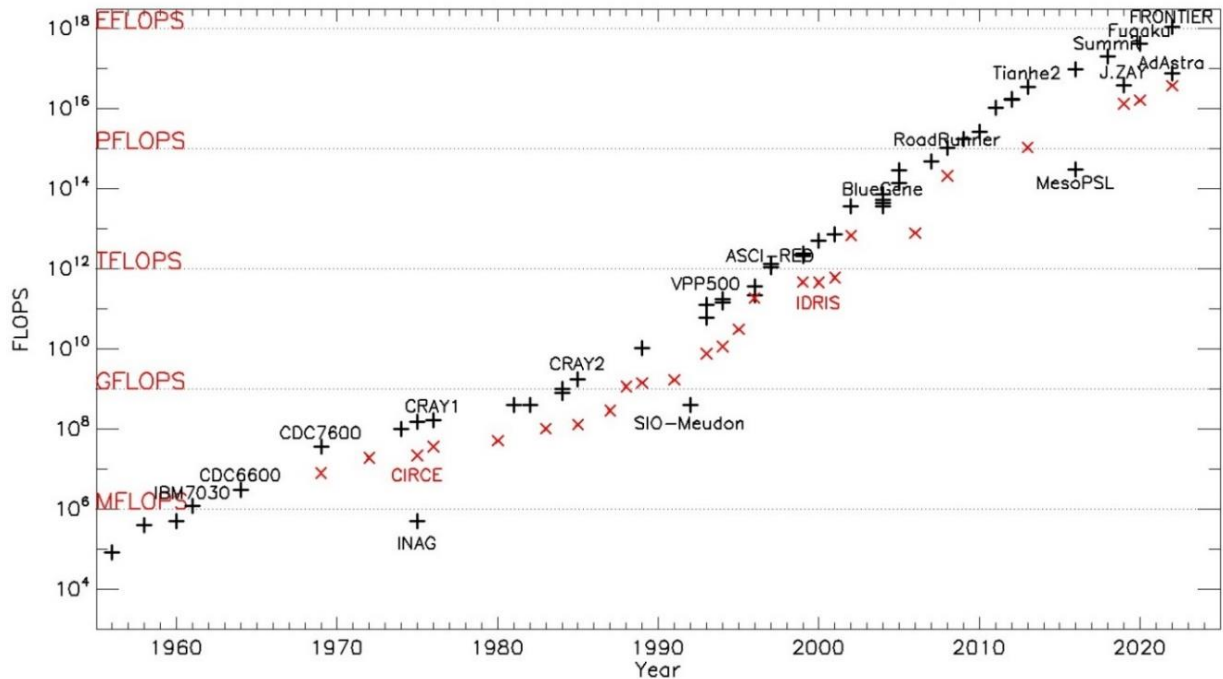

*Figure 67: Evolution of the computing power (FLOPS) of the most powerful supercomputers for numerical modelling between 1960 and 2022 (logarithmic scale). In log scale, the progression remains an affine function of time thanks to the appearance (2005) of multi-core processors and massively parallel computing using GPUs. The red crosses indicate the computing power of the CNRS, France (CIRCE since 1969 and IDRIS after 1993). The computing capability increased by **a billion** in 45 years. Author's document.*

## 6 – REFERENCES

A. Dollfus, **1959**, "observations astronomiques en ballon libre, suite et fin", *L'Astronomie*, 73, 411.

## on the beginnings of radio astronomy in Meudon

Laffineur, M., Houtgast, J., **1949**, "observations solaires sur la longueur d'onde de 55 cm", *Annales d'Astrophysique*, 12, 137.

Laffineur, M., **1951**, "radioastronomie", *l'Astronomie*, 65, 379.